\documentclass[twocolumn]{aastex63}

\usepackage{amsmath}
\usepackage{txfonts}
\usepackage{xcolor}
\usepackage{graphicx}
\usepackage{gensymb}
\usepackage{booktabs}
\usepackage{lipsum}
\usepackage{CJKutf8}
\usepackage{bm}

\AtBeginDocument{\mathcode`v=\varv}

\newcommand\ab{\bm{a}}

\newcommand\rb{\bm{r}}

\newcommand\mub{\bm{\mu}}
\newcommand\sigmab{\bm{\sigma}}
\newcommand\thetab{\bm{\theta}}
\newcommand{\HI}{\ifmmode \mathrm{\ion{H}{1}} \else \ion{H}{1} \fi}
\newcommand{\Nh}{\ifmmode N_{{\mathrm{H}} \, \mathrm{I}} \else $N_{{\mathrm{H}} \, \mathrm{I}}$\fi} 
\newcommand{\mh}{\ifmmode M_{{\mathrm{H}} \, \mathrm{I}} \else $M_{{\mathrm{H}} \, \mathrm{I}}$\fi} 
\newcommand{\nh}{\ifmmode n_{{\mathrm{H}} \, \mathrm{I}} \else $n_{{\mathrm{H}} \, \mathrm{I}}$\fi} 
\newcommand{\ROHSA}{{\tt ROHSA}}
\newcommand{\env}{region}

\def\GHz{\ifmmode $\,GHz$\else \,GHz\fi}
\def\MJysr{\ifmmode \,$MJy\,sr\mo$\else \,MJy\,sr\mo\fi}
\def\microns{\ifmmode \,\mu$m$\else \,$\mu$m\fi}

\def\kms{\ifmmode $\,km\,s$^{-1}\else \,km\,s$^{-1}$\fi}
    
\defcitealias{marchal_rohsa:_2019}{M19}
\newcommand{\cib}{C\,I\,B}

\received{October 20, 2020}
\revised{May 31, 2021}
\accepted{June 23, 2021}
\submitjournal{ApJ}

\shorttitle{Resolving the formation of cold \HI\ filaments in the HVC complex C}
\shortauthors{Marchal, Martin, and Gong}
\graphicspath{{./}{figures/}}

\begin{document}

\title{Resolving the formation of cold \HI\ filaments in the high velocity cloud complex C}

\correspondingauthor{Antoine Marchal}
\email{amarchal@cita.utoronto.ca}

\author[0000-0002-5501-232X]{Antoine Marchal}
\affiliation{Canadian Institute for Theoretical Astrophysics, University of Toronto, 60 St. George Street, 
    Toronto, ON M5S 3H8, Canada}

\author[0000-0002-5236-3896]{Peter G. Martin}
\affiliation{Canadian Institute for Theoretical Astrophysics, University of Toronto, 60 St. George Street, 
    Toronto, ON M5S 3H8, Canada}

\author[0000-0003-1613-6263]{Munan Gong (
\begin{CJK*}{UTF8}{gbsn}
龚慕南
\end{CJK*}
)}
\affiliation{Max-Planck Institute for Extraterrestrial Physics, 
    Garching by Munich, 85748, Germany}

\begin{abstract}
  {The physical properties of galactic halo gas have a profound impact
  on the life cycle of galaxies. As gas travels through a galactic halo, it undergoes dynamical interactions, influencing its impact on star formation and the chemical evolution of the galactic disk. In the Milky-Way halo, considerable effort has been made to understand the spatial distribution of neutral gas, which are mostly in the form of large complexes. However, the internal variations of their physical properties remains unclear.}  
  {In this study, we investigate the thermal and dynamical state of the neutral gas in high velocity clouds (HVCs). High-resolution observations (1\farcm 1) of the 21\,cm line emission in the EN field of the DHIGLS \HI\ survey are used to analyze the physical properties of the bright concentration \cib\ located at an edge of a large HVC complex, complex C.} 
  {We use the Gaussian decomposition code \ROHSA\ to model the multiphase content of \cib, and perform a power spectrum analysis  to analyze its multi-scale structure. Physical properties of 
  some 200
  structures extracted using dendrograms are examined. Each phase exhibits different thermal and turbulent properties.}
  {We identify two distinct regions, one of which has a prominent protrusion extending from the edge of complex C that exhibits an ongoing phase transition from warm diffuse gas to cold dense gas and filaments. The scale at which the warm gas becomes unstable and undergoes a thermal condensation is about $15$\,pc, corresponding to a cooling time about $1.5$\,Myr.
  Our study characterizes the statistical properties of turbulence in the fluid of a HVC for the first time. 
  We find that a transition from subsonic to trans-sonic turbulence is associated with the thermal condensation, going from large to small scales.}
  {A large scale perspective of complex C suggests that hydrodynamic instabilities are involved in creating the structured concentration \cib\ and the phase transition therein. However, the details of the dynamical and thermal processes remain unclear and will require further investigation, through both observations and numerical simulations.}
\end{abstract}



\section{Introduction}
\label{sec:introduction}

In their \HI\ survey of neutral high-velocity gas (HVC) in the Galactic halo, in the EN field of DHIGLS,\footnote{DRAO \HI\ Intermediate
Galactic Latitude Survey: \url{https://www.cita.utoronto.ca/DHIGLS/}} \citet{blagrave_dhigls:_2017} remarked on an intricate pattern of coherent narrow ribbons of emission associated with narrow line widths, reminiscent of the cold neutral medium (CNM) in the interstellar medium (ISM) in the Milky Way.
In this paper, we investigate quantitatively the multiphase structure of this HVC gas, its thermal and dynamical state, and the origin of the structured concentration itself.

The EN field (or simply EN) is a sub-field at an edge of
complex C\footnote{Among HVC complexes, complex C has the largest sky coverage ($\sim1600$\,deg$^2$) and, using a distance $D=10\pm2.5$\,kpc, the largest mass of atomic gas, $\mu_m\,(4.9^{+2.9}_{-2.2})\times10^6$\,$M_{\sun}$ \citep[][ see also \citep{wakker_2007}]{thom_2008}, where $\mu_m=1.4$ accounting for helium.}
\citep{hulsbosch_1966}.
Complex C was mapped in three parts (I, II, III) by \cite{hulsbosch_1968}. With higher resolution (10\arcmin), \cite{giovanelli_1973} identified bright concentrations within more diffuse gas. The EN field coincides with their concentration called \cib. Higher spectral resolution observations 
documented by \cite{cram_1976} confirmed that the HVC concentrations have narrower line widths than the diffuse gas, and this was interpreted as evidence for two different thermal phases. In particular, their Gaussian decomposition of the line profile \#26 within a 20\arcmin\ beam toward \cib\ revealed a narrow component with a full width half maximum (FWHM) of 7.58 \kms.
Building on this pioneering work, and benefiting from the high resolution (1\farcm 1) EN data mapping the entire \cib\ concentration, our spectral decomposition below further quantifies that the narrow components noted by \cite{blagrave_dhigls:_2017} have a typical FWHM of 4.2~\kms\ (velocity dispersion $\sigma = $ FWHM/2.355 about 1.8~\kms).

\subsection{The HVC context}
The reservoir of gas in halos is key to understanding the life cycle of galaxies. Halo gas exists in several forms: hot plasma \citep[$T_k\gtrsim10^6$\,K,][]{kerp_1999,wang_2005}, largely-ionized warm and warm-hot gas \citep[$T_k\sim10^4$-10$^5$\,K,][]{weiner_williams_1996,tufte_1998,putman_2003a,gaensler_2008}, and neutral gas \citep[$T_k\lesssim10^4$\,K,][]{muller_hydrogene_1963,giovanelli_1973,putman_2002}.
The hot and warm ionized phases are difficult to detect due to their very diffuse nature, but the neutral phase of the Milky-Way halo provides a major probe of its dynamical state and content \citep{kalberla_1999}.

The physical properties of the neutral Milky-Way halo gas are needed to assess
the fuel available for star formation \citep{putman_2012}, and 
the impact on galactic chemical evolution \citep{chiappini_2001} due to its low metallicity \citep{wakker_1999,gibson_2001,tripp_2003,collins_2003,collins_2007}.
HVC gas contributes to the mass influx through the Galactic halo, $\sim0.14$\,$M_{\sun}$\,yr$^{-1}$ from complex C alone \citep{thom_2008}.
The neutral gas also indirectly probes 
the properties of the warm/hot ionized Galactic halo and 
the origin and evolution of the gas throughout the halo.

From the broadening of the observed \HI\ lines, the internal structure of HVCs is turbulent \citep{bruns_deep_2001}, but the statistical properties of the energy cascade in the multiphase medium remain largely unexplored. The external dynamics can be probed through the interactions with the surrounding halo gas, for example how the morphology and velocity of HVC gas is affected near interfaces \citep{bruns_2000}.

Empirically, some HVCs exhibit a multiphase structure \citep{giovanelli_1973,cram_1976,giovanelli_1976,cohen_1979,wakker_schwarz_1991,bruns_deep_2001,kalberla_2006} as might be expected for colder structures in pressure equilibrium with warmer more diffuse gas \citep{wolfire_neutral_1995,wolfire_multiphase_1995}, in this case the diffuse halo. However, multiphase structure is not universal and varies between HVCs \citep{kalberla_2006,hsu_2011}.

In the ISM, thermal instability (TI) is thought to be the main process that leads to thermal condensation of the neutral phase, and therefore its multiphase structure \citep{field_thermal_1965,wolfire_neutral_1995,hennebelle_dynamical_1999,hennebelle_dynamical_2000,audit_thermal_2005,marchal_2021}.
However, for the condensation mode of thermal instability to grow freely, the cooling time must be shorter than the dynamical time.
For HVCs, both time scales are different than those in the disk and less constrained observationally.  But the thermal state of HVCs seems likely to be linked intimately to the dynamical state.

\subsection{Our goals}
This motives our investigation of both thermal and dynamical aspects. Using the high resolution EN observations, we quantify the multiphase structure
of the concentration \cib\ in HVC complex C in detail. Furthermore, the EN data cover a projected edge of the complex, providing insight into the dynamics of the interaction and origin of the structured concentration produced.

The paper is organized as follows. 
In Sect.~\ref{sec:data-processing} we present the data used in this work and the Gaussian decomposition performed to model its multiphase content.
A power spectrum analysis is presented in Sect.~\ref{sec:sps}. 
In Sect.~\ref{sec:phase-segmentation} we analyze the physical properties of structures from EN, including their scaling laws. 
Thermal equilibrium and thermal instability are discussed in Sect.~\ref{sec:discussion}.
Sect.~\ref{sec:context} examines the origin of the concentration and its relationship to the triggering of the thermal instability in the large-scale context of complex C.
A summary is provided in Sect.~\ref{sec:summary}.

\begin{figure}[!t]
  \centering
  \includegraphics[width=\linewidth]{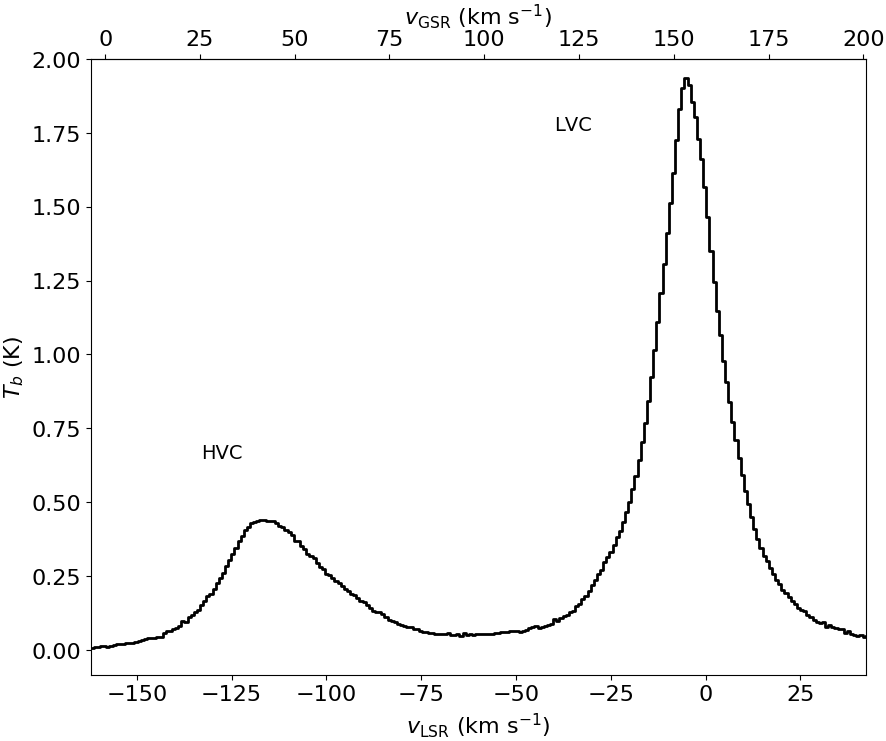}
  \caption{Mean \HI\ spectrum of the EN dataset from DHIGLS, showing both HVC and LVC peaks well separated by an interval with little emission.
  Bottom and top axis give velocities in the LSR and GSR, respectively.}
  \label{fig:mean_spectrum_EN}
\end{figure}

\begin{figure*}[!t]
  \centering
  \includegraphics[width=0.49\linewidth]{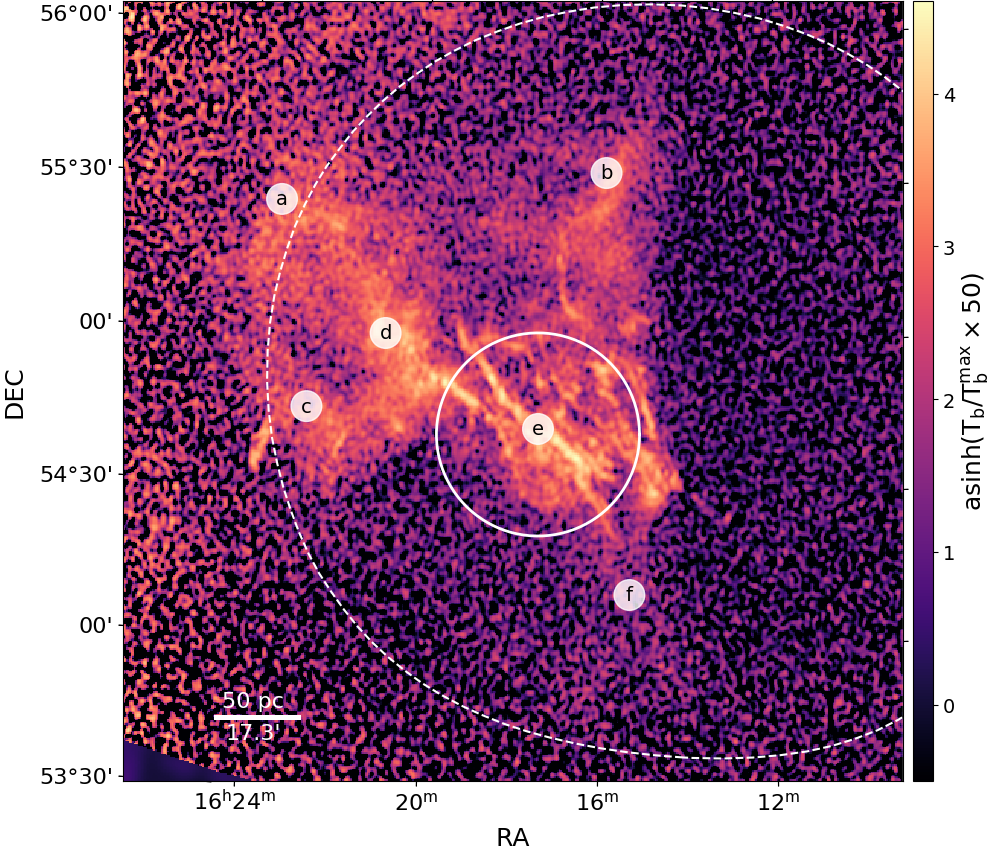}
  \includegraphics[width=0.48\linewidth]{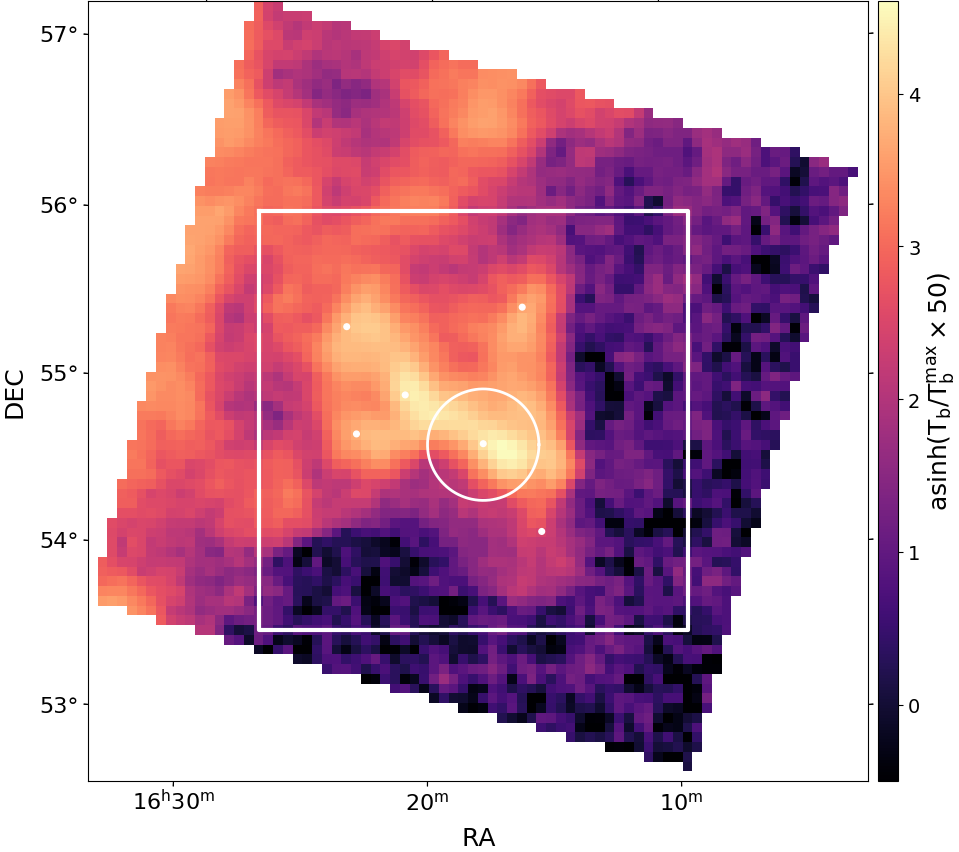}
  \caption{
Left: Channel map from the DHIGLS EN data at 1\farcm 1 resolution for $v_{\rm LSR} = -121$ \kms.
The area shown is the 6.55 square degrees analysed. Because of the coverage of ST pointings making up the final EN mosaic, the noise increases from a minimum of 0.37\,K outward, by a factor two at the open white dashed contour.
An asinh non-linear scaling of the brightness has been used to bring this out.
White circles labeled (a-f) indicate the positions of the six spectra shown in Fig.~\ref{fig:mosaic_spectra_all}.
The 20\arcmin-radius white circle centred on position e corresponds to the beam for profile \#26 of \cite{cram_1976} in which two-component multiphase structure was identified.
The physical scale of 50\,pc (17\farcm 3) assumes a distance of 10\,kpc. 
Right: Channel map at lower (10\arcmin) resolution and wider coverage from the GHIGLS N1 data. The bright concentration identified as \cib\ by \cite{giovanelli_1973} is clearly seen at an edge in this broader context. The white box shows the area selected from EN in the left panel.
}

\label{fig:NHI_EN_TOT}
\end{figure*}

\section{\HI\ spectral data and decomposition}
\label{sec:data-processing}
\subsection{Data}
\label{subsec:data}

The 14.6 square degree EN dataset used in this paper, located at $(\alpha,\, \delta) = (16^{{\mathrm h}}$14$^{{\mathrm m}},\, 54\degree 49')$ or $(l,\, b) = (84\fdg 5,\, 44\fdg 3)$
was part of the DHIGLS \HI\ survey \citep{blagrave_dhigls:_2017} with the Synthesis Telescope (ST) at the Dominion Radio Astrophysical Observatory.
The 256-channel spectrometer, spacing $\Delta v=0.824$\,\kms and velocity resolution 1.32\,\kms, was centered at $v_c=-60$\,\kms\ relative to the Local Standard of Rest (LSR). 
The spatial resolution of the ST interferometric data was about 1\farcm 1.
EN is embedded in the N1 field of the GHIGLS\footnote{GBT \HI\ Intermediate
Galactic Latitude Survey: \url{https://www.cita.utoronto.ca/GHIGLS/}} \HI\ survey \citep{martin_ghigls:_2015} with the Green Bank Telescope (GBT), with spatial resolution about $9\farcm 4$. 
The DHIGLS EN product has the full range of spatial frequencies, obtained by
a rigorous combination of the ST interferometric and GBT single dish data
\citep[see Sect.~5 in][]{blagrave_dhigls:_2017}. The pixel size is 18\arcsec.

The mean \HI\ spectrum of the EN dataset is shown in Fig.~\ref{fig:mean_spectrum_EN}. 
A velocity range of at least $\Delta v \sim$ 20 \kms\ with little emission cleanly separates the emission of HVC gas associated with complex C and a low velocity component (LVC) associated with the Milky-Way disk.
This important gap coupled with a relatively high fraction of HVC emission limit the confusion between different Galactic environments and, 
along with the dynamic structure already seen in the N1 data, is what motivated the deep observations
with the ST \citep{blagrave_dhigls:_2017}.
The top axis of the figure gives the velocity with respect to the Galactic Standard of Rest (GSR):
\begin{equation}
  v_{\rm GSR} = v_{\rm LSR} + 220 \sin(l) \cos(b) \, ,
  \label{eq:defGSR}
\end{equation}
the radial component of the velocity relative to the Sun in a reference frame in which the Galaxy rotates, i.e., removing the effect of the rotation of the LSR about the Galactic Center.
This is often thought to be more directly relevant to assessing the kinematics of the HVC gas \citep{woerden_high-velocity_2004}.
For a tiny patch like covered by the EN data, the conversion is a simple translation but in the wide-angle context of an entire complex (Sect.~\ref{sec:context}) it is of more interest.

Figure~\ref{fig:NHI_EN_TOT} presents channel maps of HVC emission for $v_{\rm LSR} = -121$ \kms, at which the bright concentration \cib\ was first identified by \cite{giovanelli_1973}.
On the left is the 6.55 square degree field from the EN data that we analysed\footnote{Specifically, counting pixels from (0,0) at the lower left of the full EN data set, this is the 512 pixel square with lower left corner at coordinate $(89, 169)$.} and on the right is a corresponding map from the N1 data with the GBT. Both the extent of the bright concentration and the finer spatial structure now available are clearly seen.

As we shall see in much more detail below, most of the diffuse emission is in the upper left triangle of the map; the lower right is relatively void except for a prominent protrusion (``finger") extending from the main body of emission. The boundary between these triangular areas corresponds to the local ``edge'' of the much larger complex C and is oriented at position angle about 120\degree\ or $-60\degree$. 
The wider scale context of the concentration \cib\ is discussed in Sect.~\ref{sec:context} and can be seen in Figures~\ref{fig:NHI_EBHIS_nomask_csc_label}, \ref{fig:CV_GSR_EBHIS}, and \ref{fig:NHI_complex_C_zoom_edge_complex_A_M} (top right).

The physical scale of 50\,pc shown is 17\farcm 3 at an assumed distance of 10 kpc at this position in complex C.\footnote{Complex C is of large angular extent. Among the probes used by \cite{thom_2008} to bracket the distance, no HVC absorption was present in the line of sight to SDSS J153915.24+575731.7 (S441), thus setting a lower limit of $10.2\pm2.6$ kpc at its position, $(l,\, b) = (91\fdg 2,\, 47\fdg 5)$, only $\sim 6\degree$ from \cib.}
The range of spatial scales accessible (4\,pc $\lesssim l \lesssim$ 300\,pc) makes \cib\ a unique laboratory for probing the multi-scale and multiphase properties of neutral gas in HVCs. 

\subsection{Gaussian decomposition}
\label{subsec:gaussian-decomposition}

\subsubsection{Model and optimization}
\label{subsubsec:model}
We performed multiphase separations of EN and N1 spectra of \cib\ using the publicly available code \ROHSA,\footnote{\url{https://github.com/antoinemarchal/ROHSA}} a multi-Gaussian decomposition algorithm originally developed for just such analyses. As described by \citet[][hereafter M19]{marchal_rohsa:_2019}, \ROHSA\ is based on a regularized nonlinear least-squares criterion that takes into account the spatial coherence of the emission across a field at coordinates $\rb$ and the multiphase nature of the gas. 

The model $\tilde T_b\big(v_z, \thetab(\rb)\big)$ used to fit the measured brightness temperature $T_b(v_z, \rb)$ at radial velocity $v_z$ and coordinates $\rb$ is 
\begin{equation}
  \tilde T_b\big(v_z, \thetab(\rb)\big) = \sum_{n=1}^{N} G\big(v_z, \thetab_n(\rb)\big) \, ,
  \label{eq::model_gauss}
\end{equation}
where each of the $N$ Gaussians
\begin{equation}
  G\big(v_z, \thetab_n(\rb)\big) = \ab_n(\rb) \exp
  \left( - \frac{\big(v_z - \mub_n(\rb)\big)^2}{2 \sigmab_n(\rb)^2} \right)
\end{equation}
is parametrized by three 2D spatial fields across $\rb$: $\thetab_n(\rb)  = \big(\ab_n(\rb),\, \mub_n(\rb),\, \sigmab_n(\rb)\big)$, with amplitude $\ab_{n} \geq \bm{0}$, mean velocity $\mub_{n}$, and standard deviation
$\sigmab_{n}$.

The initialization of each Gaussian is accomplished in \ROHSA\ by a multi-resolution procedure from coarse to fine grid \citepalias[see Sect.~2.4.3 in][]{marchal_rohsa:_2019}. The parameters $\hat{\thetab}$ are optimized by minimizing a cost function that goes beyond the standard $\chi^2$. As described in \citetalias{marchal_rohsa:_2019}, to penalize variations at the smallest spatial frequencies, the cost function includes Laplacian filtering of each of the three parameter maps $\big(\ab_n,\, \mub_n,\, \sigmab_n\big)$, with cost controlled by three hyper-parameters. A fourth penalty term, for minimizing the variance of $\sigmab_n$ across the whole field, is added to enable the multiphase separation. 
As examined and recommended in \citetalias{marchal_rohsa:_2019}, the magnitudes of these four hyper-parameters, $\lambda_{\ab},\, \lambda_{\mub},\, \lambda_{\sigmab}$, and $\lambda'_{\sigmab}$, are chosen empirically so that the solution converges toward a noise-dominated residual and a signal that is encoded with a minimum number of Gaussian components.

\subsubsection{Decomposition of EN data from DHIGLS}
\label{subsubsec:DHIGLS-decomposition}

We decomposed the spectra for EN from DHIGLS for the area shown in Fig.~\ref{fig:NHI_EN_TOT} (left) and HVC spectral range ($-162.21 \leq v_{\rm LSR}$ [\kms] $\leq -60.82$) using 
$\lambda_{\ab}=1000$, $\lambda_{\mub}= 100$, $\lambda_{\sigmab}=1000$, and $\lambda'_{\sigmab}=10$.
Because of the relative simplicity of the HVC spectra, as opposed to complex LVC emission from the Milky-Way disk, only a small number of Gaussians, $N=6$, is needed and only four of the six 
encode emission associated with the HVC; the other two deal with noise at the extreme of the spectral range toward intermediate velocities.

\begin{figure}
  \centering
  \includegraphics[width=\linewidth]{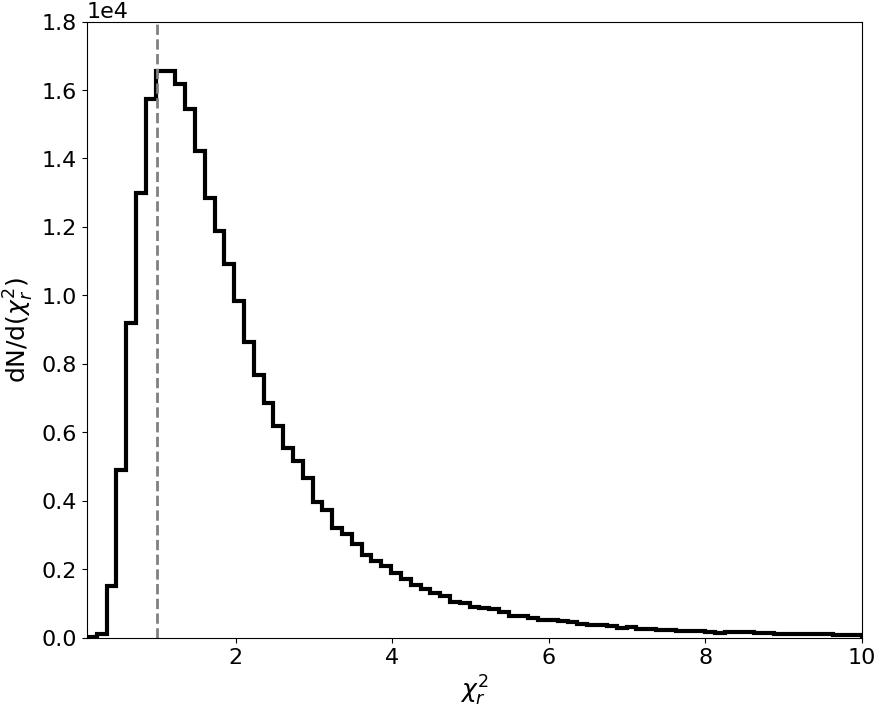}
  \caption{Probability distribution function of the reduced chi-square $\chi^2_r$ obtained with {\tt ROHSA} on EN. Vertical line indicates $\chi^2_r=1$.}
  \label{fig:PDF_rchi2_linear}
\end{figure}

The simple model fits the data well. A map of the reduced chi-squared $\chi^2_r$ of the decomposition shows no structure, only  random fluctuations.
Figure~\ref{fig:PDF_rchi2_linear} shows the one-dimensional probability distribution function (PDF) of $\chi^2_r$, which peaks near the expected value of 1 denoted by the vertical line.

\begin{figure*}[!t]
  \centering
  \includegraphics[width=\linewidth]{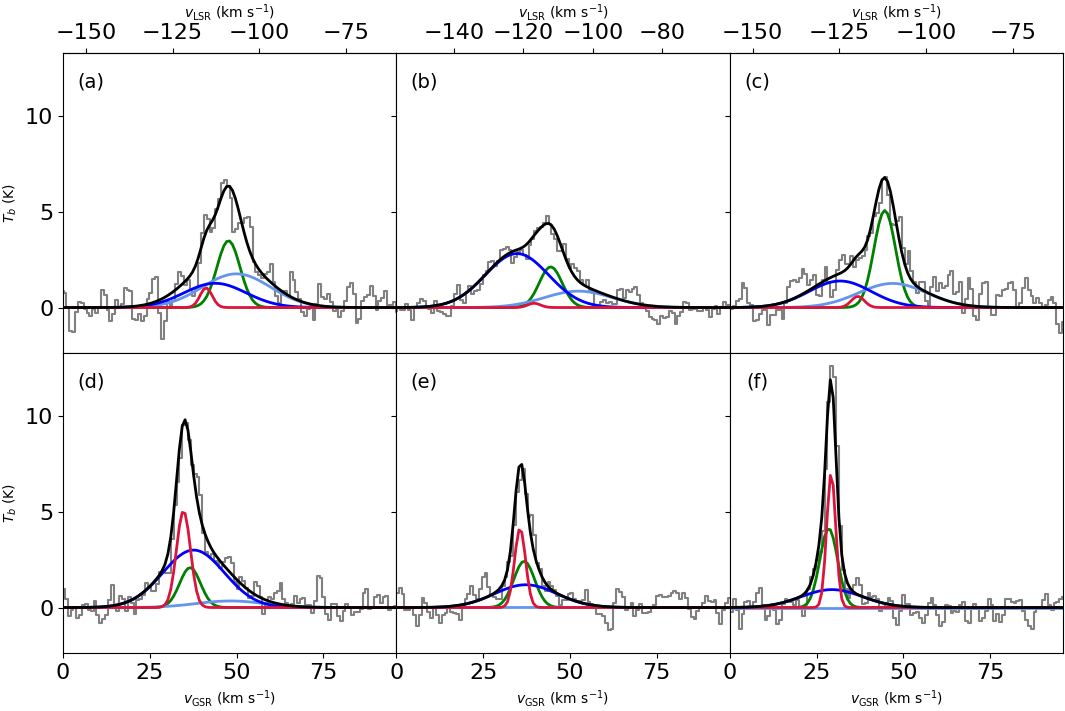}
  \caption{Example Gaussian decomposition by \ROHSA\
   for six lines of sight toward \cib, those annotated (a--f) in Fig.~\ref{fig:NHI_EN_TOT} (left). The original signal is shown in gray and total signal encoded by \ROHSA\ in black. The four individual Gaussians are color coded: $G_1$ (light blue), $G_2$ (blue), $G_3$ (green), and $G_4$ (red).}
  \label{fig:mosaic_spectra_all}
\end{figure*}

Across the field, the spectra are quite varied.  Figure~\ref{fig:mosaic_spectra_all} illustrates the different decompositions obtained for six lines of sight marked in Fig.~\ref{fig:NHI_EN_TOT}.  Some spectra, particularly in the lower row, contain narrow Gaussian components, indicating the presence of colder gas.  Each of these spectra also has emission spread broadly over many channels.  This can be fit by broader components, indicating warmer gas, similar to that in the pioneering work of \citet{giovanelli_1973} and \citet{cram_1976}.

\begin{figure}[!t]
  \centering
  \includegraphics[width=\linewidth]{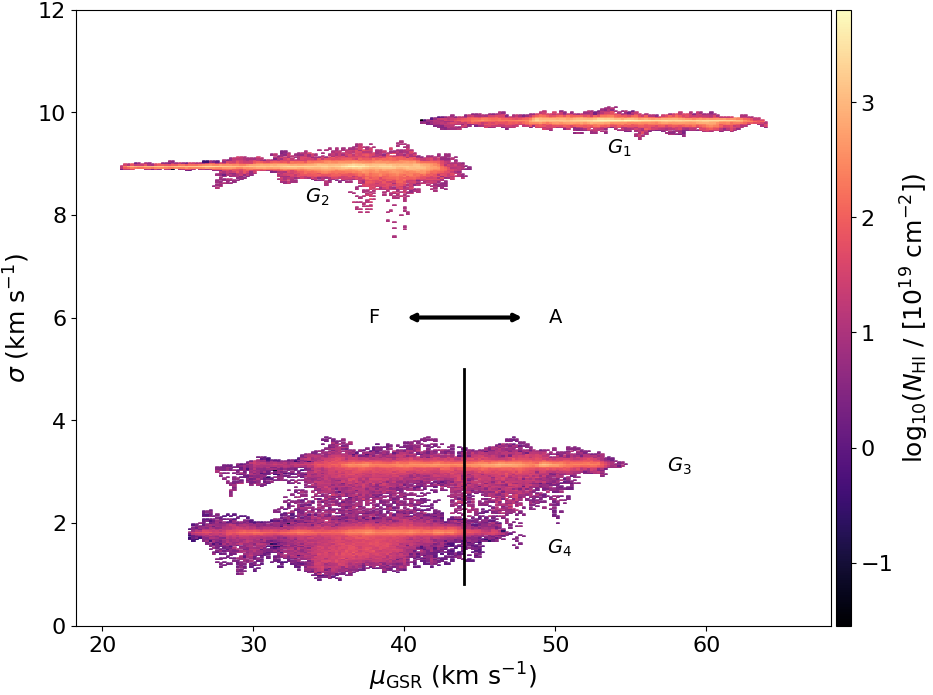}
  \caption{Two-dimensional probability distribution function
  $\sigma - \mu$} weighted by the column density of each Gaussian along each line of sight. For the two lower clusters $G_3$ and $G_4$, vertical black line shows the velocity split between \env\ A ($v_{\rm GSR}\geq44\kms$) and \env\ F.
  \label{fig:heatmap}
\end{figure}

\begin{deluxetable}{lccccc}
\tablecaption{Mean kinematic properties (in \kms) of Gaussians modeling DHIGLS/EN and GHIGLS/N1 data toward \cib}
\label{table:mean_var_DHIGLS_GHIGLS}
\tablewidth{0pt}
\tablehead{
\nocolhead{Field} & \colhead{Component} & \colhead{$G_{1}$} & \colhead{$G_{2}$} & \colhead{$G_{3}$} & \colhead{$G_{4}$} \\
\nocolhead{Field}   & \nocolhead{Name}    & \colhead{WNM$_{\rm A}$} & \colhead{WNM$_{\rm F}$} & \colhead{LNM}  & \colhead{CNM} \\
\colhead{Field}  
}

\startdata
{} & $\langle\sigmab_n\rangle$ & 9.8 & 8.9 & 3.1 & 1.8 \\
{EN} & $\langle\mub_n^{\rm{GSR}}\rangle$ & 55 & 34.4 & 44.5 & 36.8 \\
{} & $\langle\mub_n^{\rm{LSR}}\rangle$ & $-103.8$ & $-124.6$ & $-114.8$ & $-121.3$  \\
\hline
{} & $\langle\sigmab_n\rangle$ & 9.6 & 9.4 & 7.2 & 2.6 \\
{N1} & $\langle\mub_n^{\rm{GSR}}\rangle$ & 57.8 & 25.5 & 41.1 & 33.6 \\
\enddata
\end{deluxetable}

It is clear from the original spectra that mean velocities ($\mu$) and dispersions ($\sigma$) of the fitted components vary with position. Figure~\ref{fig:heatmap} summarizes the outcome in a two-dimensional PDF of $\sigma - \mu$ weighted by the column density of each Gaussian along each line of sight. In the regularized decomposition obtained with \ROHSA, these properties are clustered, each cluster corresponding to one of the four Gaussian components. 
In particular, the separation vertically into clusters of broader and narrow components results from the hyper-parameter $\lambda'_{\sigmab}$ \citepalias{marchal_rohsa:_2019}.
Table~\ref{table:mean_var_DHIGLS_GHIGLS} summarizes the spatially-averaged mean dispersion $\langle\sigmab_n(\rb)\rangle$ of each component $n$ and also the mean velocity $\langle\mub_n(\rb)\rangle$, both GSR and LSR.

Clusters $G_1$ and $G_2$ have similarly large velocity dispersions (separated by only about 1 \kms), but cover two distinct velocity ranges and are statistically uncorrelated in their spatial distributions (see Sect.~\ref{subsec:cross_correlation_coefficients}). 
Hereafter, these components for two physically distinct regions in this field will be called more memorably WNM$_{\rm A}$ and WNM$_{\rm F}$, respectively, adopting the standard abbreviation WNM for a Warm Neutral Medium ``phase", with subscripts A (``Arc") and F (``Filaments") motivated by their different morphologies as described in Sect.~\ref{subsec:general-description}.

This motivated us to identify the unstable (lukewarm) and cold phases, LNM$_{\rm (A,F)}$ and CNM$_{\rm (A,F)}$, that are associated with these two regions. To accomplish this within this simple decomposition, the G$_3$ and G$_4$ clusters each needs to be divided with respect to velocity $\mub_n(\rb)$.
The location of this split was determined by examining the spatial distribution of the emission encoded in the Gaussians, looking for spatial correlations with WNM$_{\rm A}$ and WNM$_{\rm F}$. The resulting split adopted is at $v_{\rm GSR} = 44\kms$, marked by the black vertical line in Fig.~\ref{fig:heatmap}. Not surprisingly, this is close to the velocity extremes where the G$_1$ and G$_2$ clusters (WNM$_{\rm A}$ and WNM$_{\rm F}$) separate. 
Hereafter, gas in clusters G$_3$ and G$_4$ with $v_{\rm GSR}\geq44\kms$ (to the right in the figure) will be called LNM$_{\rm A}$ and CNM$_{\rm A}$, respectively, with the complementary gas being LNM$_{\rm F}$ and CNM$_{\rm F}$.

\begin{figure*}[!t]
  \centering
  \includegraphics[width=0.7\linewidth]{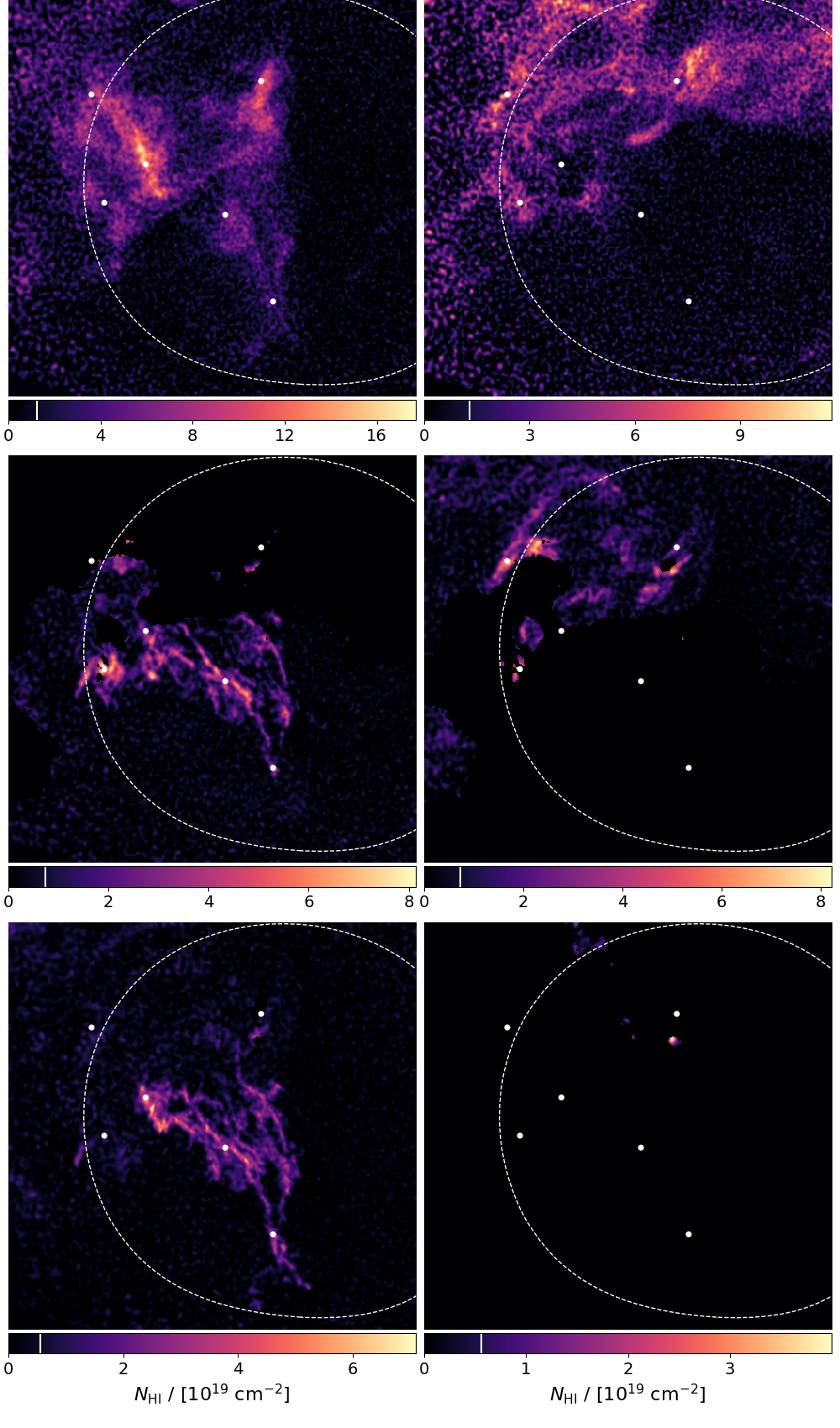}
  \caption{Column density maps of the three HVC phase components identified in two regions of \cib\ from EN data. 
          (left): WNM$_{\rm F}$ (top), LNM$_{\rm F}$ (middle), and CNM$_{\rm F}$ (bottom); (right): WNM$_{\rm A}$ (top), LNM$_{\rm A}$ (middle), and CNM$_{\rm A}$ (bottom). 
          Note that color bars have different scales.
          Coordinates (not shown here) are the same as in Fig.~\ref{fig:NHI_EN_TOT} (left).
          The white dots indicate the positions for the six spectra shown in Fig.~\ref{fig:mosaic_spectra_all}. 
          The white dashed contours indicate where the noise has increased by a factor two relative to the central minimum.
          The white marks on color bars indicate the column density sensitivity limits (3$\sigma$) within the white dashed contours (see Table~\ref{table:detection_limits}), showing that the features seen with this color representation are well detected (see also Fig.~\ref{fig:mosaic_field_0_1_dendrograms} for another representation).
          }
  \label{fig:mosaic_field_0_1}
\end{figure*}

\begin{figure*}[!t]
  \centering
  \includegraphics[width=0.45\linewidth]{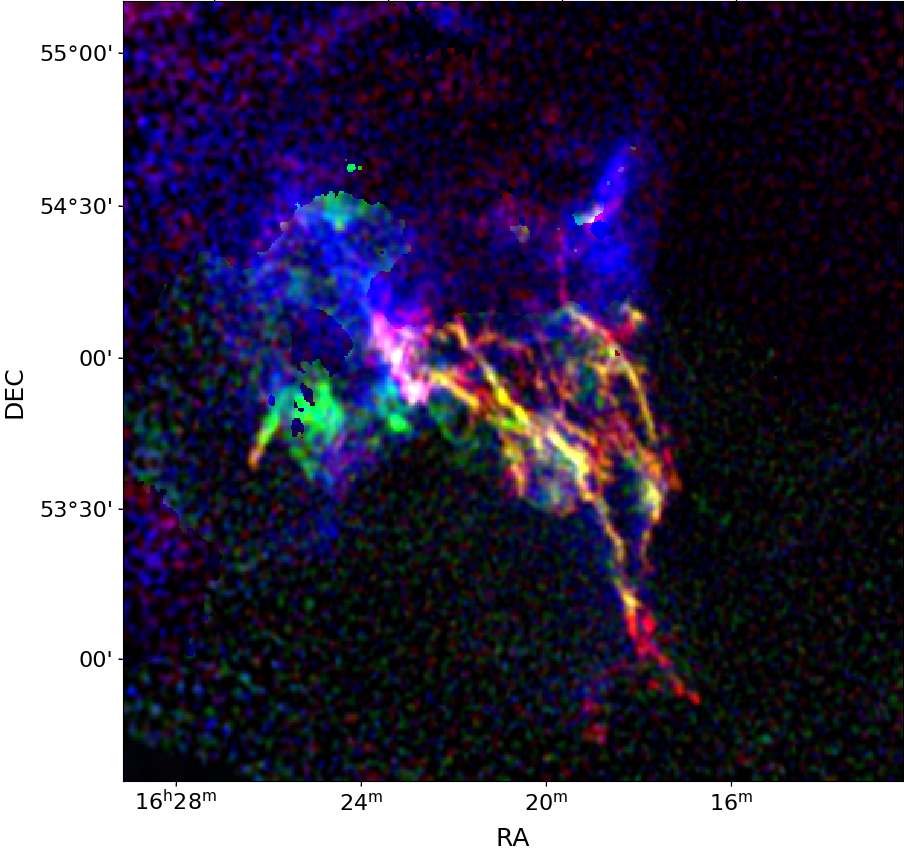}
  \includegraphics[width=0.45\linewidth]{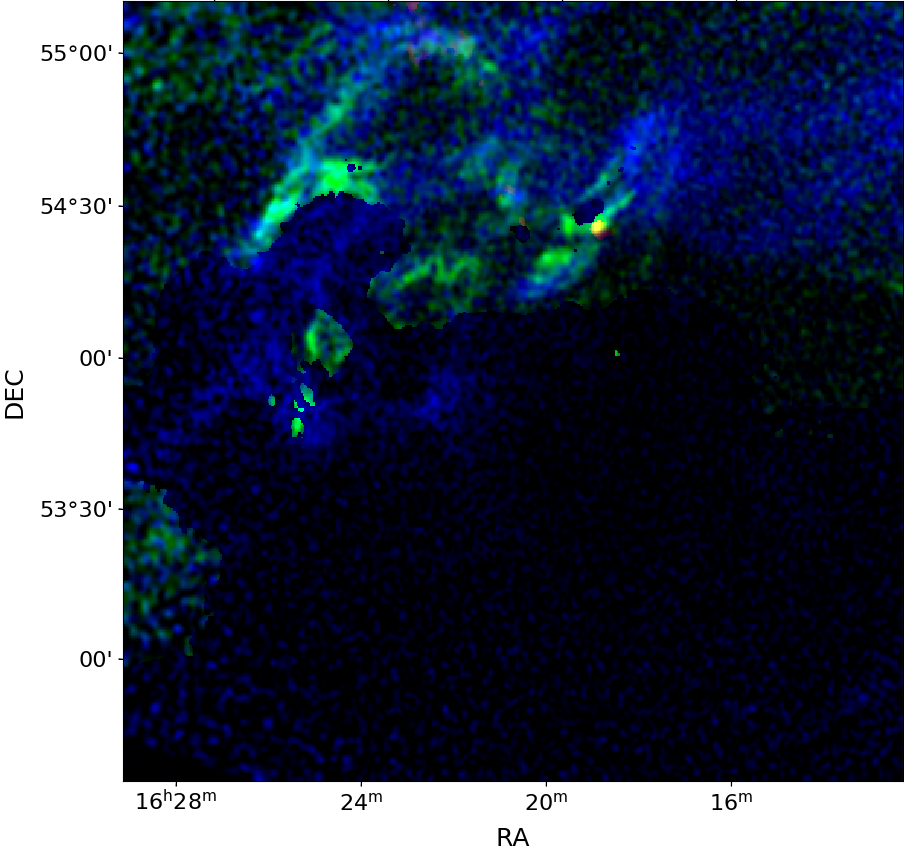}
  \caption{Left: Three-phase view of \env\ F in \cib\ (EN) from column density maps of WNM$_{\rm F}$ (blue), LNM$_{\rm F}$ (green), and CNM$_{\rm F}$ (red). 
  Maps were scaled with relative weightings of 1, 2, 3 for WNM$_{\rm F}$, LNM$_{\rm F}$ and CNM$_{\rm F}$, respectively, to highlight the interrelationship, rather than simply their relative column densities. Right: Same for \env\ A.}
  \label{fig:RGB_WCS}
\end{figure*}

\subsection{Sensitivity limits}
\label{sec:senslim}

Noise in the spectral data varies over the field, though only by a factor two within the white dashed contour in Fig.~\ref{fig:NHI_EN_TOT}. In this low-noise central area, the average noise ($3\sigma$) is 1.53\,K.
Based on this, the corresponding column density sensitivity limit for each phase of regions F and A was evaluated from 
\begin{equation}
    \label{eq:senslim}
    \Nh^{\rm lim} \simeq 0.414 \,  \sqrt{\langle \sigma(\rb) \rangle} \times 10^{19} \, \rm{cm}^{-2}\,,
\end{equation}
where $\langle \sigma(\rb) \rangle$ is the average dispersion (\kms) of the relevant Gaussians from Table~\ref{table:mean_var_DHIGLS_GHIGLS}.
Values are tabulated in Table~\ref{table:detection_limits}.
These are marked on the color bars in Figs.~\ref{fig:mosaic_field_0_1} and \ref{fig:mosaic_field_0_1_dendrograms} and can be seen to be reasonable estimates.

\begin{deluxetable}{lcccccc}
    \tablecaption{Column density sensitivity limits (3$\sigma$) by phase (in 10$^{19}$\,cm$^{-2}$)}
    \label{table:detection_limits}
    \tablewidth{0pt}
    \tablehead{\nocolhead{}  & \colhead{WNM$_{\rm F}$} & \colhead{LNM$_{\rm F}$} & \colhead{CNM$_{\rm F}$} & \colhead{WNM$_{\rm A}$} & \colhead{LNM$_{\rm A}$} & \colhead{CNM$_{\rm A}$}}
    \startdata
    \Nh$^{\rm lim}$ & 1.24 & 0.73 & 0.56 & 1.30 & 0.73 & 0.56
    \enddata
\end{deluxetable}

\subsubsection{The impact of spatial resolution}
\label{subsubsec:decomposition-resolution}
To evaluate the impact of spatial resolution, we performed a decomposition of N1 HVC spectra from the GHIGLS survey in the same velocity range as for the EN data, as detailed in Appendix~\ref{app:GHIGLS}. Again \ROHSA\ converges toward four components and a phase separation is still detectable.
As seen in Table~\ref{table:mean_var_DHIGLS_GHIGLS}, the mean kinematic properties of Gaussians are fairly similar.
The most notable difference is the larger velocity dispersion of $G_3$ at the lower spatial resolution of the N1 data (9\farcm 4 compared to 1\farcm 1), which would be classified as warm gas. 
This foreshadows the finding in Sect.~\ref{sec:sps} that emission in narrow lines (by LNM and CNM gas) has more significant fluctuations on small spatial scales and so is more affected by beam smearing (mixing physically distinct structures inside one beam leads to an unresolved crowding along the velocity axis if their respective velocities are not exactly aligned). 
In the following, only the higher resolution results are used to analyze the physical properties of the gas toward the concentration \cib.

\subsection{Multiphase and multi-scale structure}
\label{subsec:general-description}
Figure~\ref{fig:mosaic_field_0_1} shows HVC column density maps of the six phases modeled in \cib.\footnote{For comparison, maps from analysis of lower resolution DHIGLS/N1 are shown in Fig.~\ref{fig:mosaic_field_N1} in Appendix~\ref{app:GHIGLS}).}
Column density detection limits (3$\sigma$, see Table~\ref{table:detection_limits}) are shown by white marks on the color bars.
In \env\ F (right panels), WNM$_{\rm F}$, LNM$_{\rm F}$, and CNM$_{\rm F}$ are clearly associated spatially, underlying the split of the $\sigma - \mu$ clusters, especially, LNM ($G_3$), in Fig.~\ref{fig:heatmap}.
By contrast, \env\ A shows a dearth of cold gas and emission only in the upper left triangle of the field with a continuous ``arc'' parallel to the edge in the total column density maps described in Sect.~\ref{subsec:data}, Fig.~\ref{fig:NHI_EN_TOT}.

The finger protruding beyond the edge is traced by WNM$_{\rm F}$, LNM$_{\rm F}$, and CNM$_{\rm F}$ and includes very striking elongated filaments along the structure.
The close relationship of these filaments to each phase is brought out in Fig.~\ref{fig:RGB_WCS} (left), which displays the phases simultaneously, \Nh\ maps of CNM$_{\rm F}$, LNM$_{\rm F}$, and WNM$_{\rm F}$ being represented by RGB, respectively).\footnote{Fig.~\ref{fig:RGB_WCS}, based on the phase decomposition here, is complementary to the RGB image in figure 27 of \citet{blagrave_dhigls:_2017}, which encodes kinematic information of the CNM gas by using three channel maps.} 
Correlations between phases within each environment and between regions will be quantified statistically in Sect.~\ref{subsec:cross_correlation_coefficients}.

The other characteristic feature of each \Nh\ map in Fig.~\ref{fig:mosaic_field_0_1} is the multi-scale structure. In both regions F and A, lukewarm filaments are visible, often within warmer envelopes.
Environment F is furthermore populated with smaller cold filaments within warmer envelopes.
Building in this, we focus our analysis on the correlated phases WNM$_{\rm F}$, LNM$_{\rm F}$, and CNM$_{\rm F}$ from \env\ F, which are suggestive of an ongoing phase transition where all three phases are present simultaneously.
The dearth of cold gas in \env\ A will be discussed in Sect.~\ref{subsec:edgeC}.

\section{Power spectrum analysis of \Nh\ maps from environment F} 
\label{sec:sps}

To investigate the multi-scale properties, here focusing on \env\ F, we make use of three statistical tools: the power spectrum, the cross power spectrum, and the cross correlation coefficients. For completeness, the results tabulated in Table~\ref{table:index} for \env\ F have their equivalent for \env\ A in Table~\ref{table:index_0}.

\subsection{Power spectrum}
\label{subsec:power-spectrum}

All spatial (angular) power spectra $P(k)$ presented here are obtained following the methodology described in \cite{martin_ghigls:_2015} and \cite{blagrave_dhigls:_2017}. Each power spectrum $P(k)$ is the azimuthal average of the modulus of the Fourier transform of the corresponding field, and is modelled as 
\begin{equation}
    P(k) = B(k) \times P_0 k^{\gamma} + N(k) \, ,
    \label{eq:model-Pk}
\end{equation}
where, $P_0$ is the amplitude of the power spectrum, $\gamma$ is the scaling exponent, $B(k)$ describes the cutoff of the spectrum at high $k$ due to the beam of the instrument, assumed to be a 2D Gaussian of FWHM = 1\farcm 1, and $N(k)$ is the noise estimated by taking the power spectrum of empty channel maps of the PPV cube. 
The finite images were apodized using a cosine function to minimize systematic edge effects from the implementation of the Fourier transform.

\begin{figure}[!t]
  \centering
  \includegraphics[width=\linewidth]{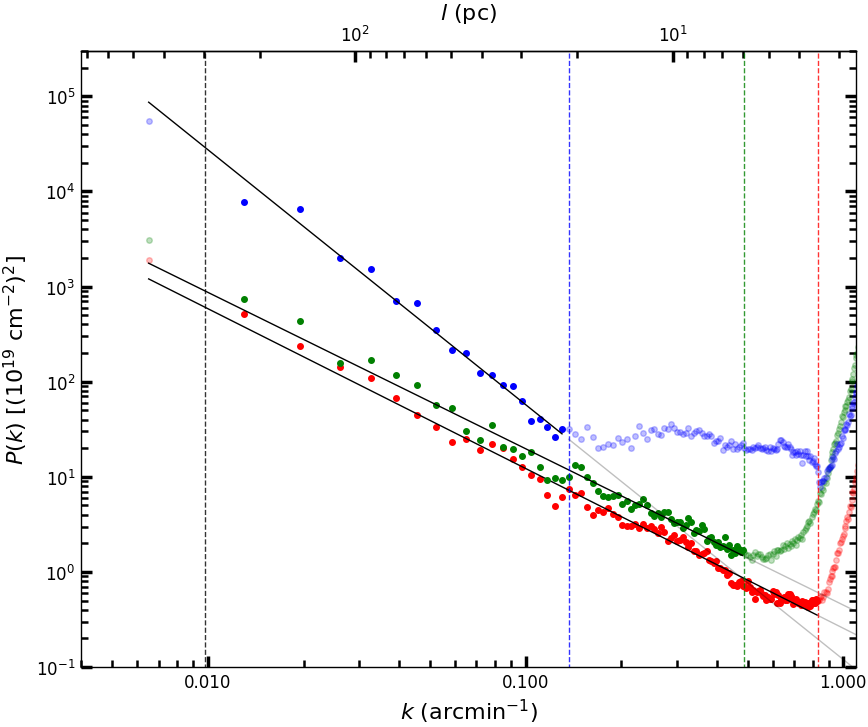}
  \caption{Spatial power spectra and models of the column density maps of WNM$_{\rm F}$ (blue), LNM$_{\rm F}$ (green), and CNM$_{\rm F}$ (red). Corrected (de-convolved) power spectra $P(k)$/$B(k)$ are shown with colored dots. Solid lines show the respective model fits between the vertical dashed lines, chosen to select $k$ ranges 
  avoiding noise or systematics (high- and low-$k$ limits, respectively). Exponents are in Table~\ref{table:index} on the diagonal.}
  \label{fig:SPS_NHI_CNM_LNM_WNM}
\end{figure}

\begin{deluxetable}{lcccc}
\tablecaption{Power spectrum analysis of the multiphase \env\ F}
\label{table:index}
\tablewidth{0pt}
\tablehead{
\nocolhead{} & \colhead{WNM$_{\rm F}$} & \colhead{LNM$_{\rm F}$} & \colhead{CNM$_{\rm F}$}
}
\startdata
Exponent \\
WNM$_{\rm F}$ & -2.68$\pm$0.10 & -2.51$\pm$0.18 & -2.87$\pm$0.37\\
LNM$_{\rm F}$ & & -1.64$\pm$0.03 & -2.00$\pm$0.03 \\
CNM$_{\rm F}$ & & & -1.68$\pm$0.02 \\
\hline
Correlation \\
WNM$_{\rm F}$ & 1 & 0.27$\pm$0.14 & 0.38$\pm$0.14 \\
LNM$_{\rm F}$ & & 1 & 0.53$\pm$0.09 \\
CNM$_{\rm F}$ & & & 1 \\
\enddata
\end{deluxetable}

\begin{deluxetable}{lcccc}
\tablecaption{Power spectrum analysis of the multiphase \env\ A}
\label{table:index_0}
\tablewidth{0pt}
\tablehead{
\nocolhead{} & \colhead{WNM$_{\rm A}$} & \colhead{LNM$_{\rm A}$} & \colhead{CNM$_{\rm A}$}
}
\startdata
Exponent \\
WNM$_{\rm A}$ & -2.37$\pm$0.10 & -2.91$\pm$0.50 & -2.28$\pm$0.42 \\
LNM$_{\rm A}$ & & -1.67$\pm$0.04 & -2.16$\pm$0.11 \\
CNM$_{\rm A}$ & & & -0.53$\pm$0.03 \\
\hline
Correlation \\
WNM$_{\rm A}$ & 1 & 0.32$\pm$0.15 & 0.10$\pm$0.12 \\
LNM$_{\rm A}$ & & 1 & 0.10$\pm$0.12 \\
CNM$_{\rm A}$ & & & 1 \\
\enddata
\end{deluxetable}

\begin{figure}[!t]
  \centering
  \includegraphics[width=\linewidth]{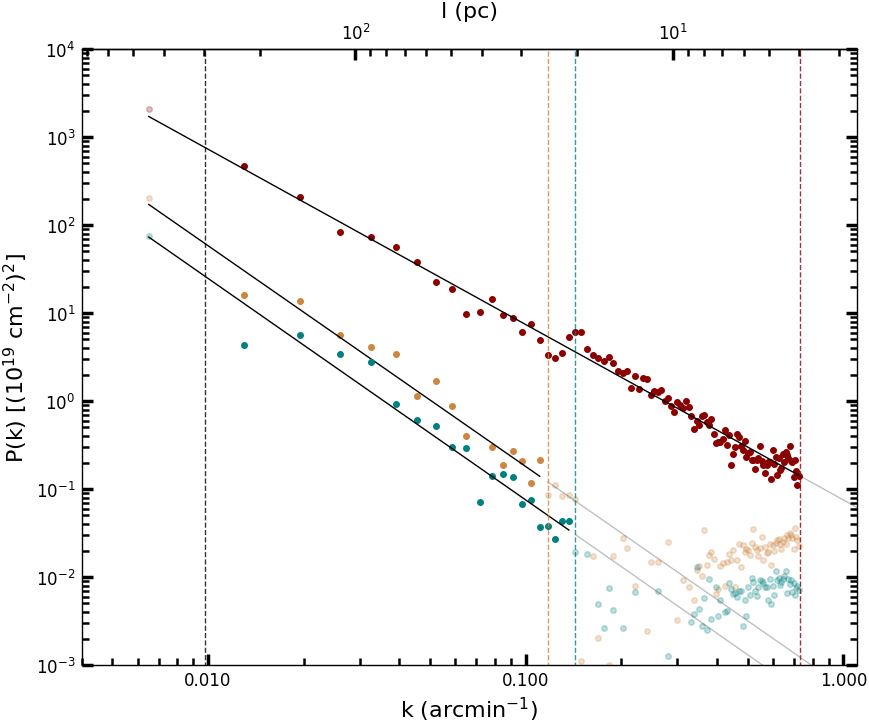}
  \caption{Cross power spectra and models of WNM$_{\rm F}\times$LNM$_{\rm F}$ (teal), WNM$_{\rm F}\times$CNM$_{\rm F}$ (peru), and LNM$_{\rm F}\times$CNM$_{\rm F}$ (dark-red). Corrected cross power spectra $P^{i,j}(k)$/$B(k)$ are shown with colored dots. 
  Solid lines show the respective model fits between the vertical dashed lines.
  Exponents are in Table~\ref{table:index}, off the diagonal.
}
  \label{fig:Cross_SPS_NHI_COMBINED}
\end{figure}

Figure~\ref{fig:SPS_NHI_CNM_LNM_WNM} shows the beam-corrected (de-convolved) spatial power spectra of WNM$_{\rm F}$, LNM$_{\rm F}$, and CNM$_{\rm F}$. LNM$_{\rm F}$ and CNM$_{\rm F}$ are well described by power laws in the spatial range 4\,pc $\lesssim l \lesssim$ 300\,pc. At low spatial frequencies ($ k \lesssim$ 0.15), the power spectrum of WNM$_{\rm F}$ follows a steeper power law.  However, at higher $k$, on scales $l \lesssim$ 20\,pc, the power spectrum flattens, just as it does for the total \Nh\ of the HVC toward \cib\ presented by \citet[][figure 22]{blagrave_dhigls:_2017} using the EN data.  This is caused by noise.
Unlike CNM$_{\rm F}$ and LNM$_{\rm F}$, the WNM covers parts of the field where the intrinsic noise of the observation is the highest, notably visible in the upper left of WNM$_{\rm F}$ and WNM$_{\rm F}$ in Fig~\ref{fig:mosaic_field_0_1}, outside the white dashed contour that indicates where the noise has increased by a factor two relative to the central minimum.
Furthermore, compared to CNM$_{\rm F}$ and LNM$_{\rm F}$, the WNM Gaussian components WNM$_{\rm F}$ and WNM$_{\rm F}$ span significantly more channels (i.e., have large velocity dispersions, see Fig.~\ref{fig:mosaic_spectra_all}), so that again the warm phase is more sensitive to noise (Eq.~\ref{eq:senslim}).

Therefore, different spatial ranges are used to fit the power spectrum exponents, as denoted by the vertical dashed lines on the right. 
These exponents are reported on the diagonal of Table~\ref{table:index}. The exponents for LNM$_{\rm F}$ and CNM$_{\rm F}$ are significantly less negative (the spectra are flatter) than for WNM$_{\rm F}$, quantifying that the unstable and cold phases have relatively more structure on small scales.
A similar trend is observed in Table~\ref{table:index_0} for \env\ A.

\subsection{Cross power spectrum}
\label{subsec:cross_power_spectrum}

The cross power spectrum $P^{i,j}$(k) is the azimuthal average of the Fourier transform of image $i$ times the conjugate of the Fourier transform of $j$.  It is also modeled using Eq.~\ref{eq:model-Pk}. Figure~\ref{fig:Cross_SPS_NHI_COMBINED} shows the beam corrected spatial cross power spectra of WNM$_{\rm F}\times$LNM$_{\rm F}$, WNM$_{\rm F}\times$CNM$_{\rm F}$, and LNM$_{\rm F}\times$CNM$_{\rm F}$. 
The cross power spectrum LNM$_{\rm F}\times$CNM$_{\rm F}$ is well constrained due to the broad spatial coverage 4\,pc $\lesssim l \lesssim$ 300\,pc available. On the other hand, WNM$_{\rm F}\times$LNM$_{\rm F}$ and WNM$_{\rm F}\times$CNM$_{\rm F}$ are dominated by noise on scales $l \lesssim$ 20\,pc. Exponents from the model fits are reported in Table~\ref{table:index} in the off-diagonal elements. 
The cross power spectrum of LNM$_{\rm F}\times$CNM$_{\rm F}$ is significantly flatter than WNM$_{\rm F}\times$CNM$_{\rm F}$ and WNM$_{\rm F}\times$LNM$_{\rm F}$, considering the uncertainties.

The cross power amplitude of LNM$_{\rm F}\times$CNM$_{\rm F}$ is much higher than for the other two cross powers, but similar to the amplitudes for these cooler components in Fig.~\ref{fig:SPS_NHI_CNM_LNM_WNM}, revealing a strong correlation between these two maps. The exponent for LNM$_{\rm F}\times$CNM$_{\rm F}$ is somewhat steeper than for the power spectrum of CNM$_{\rm F}$, indicating a relative de-correlation at the smallest scales.

\begin{figure}[!t]
  \centering
  \includegraphics[width=0.98\linewidth]{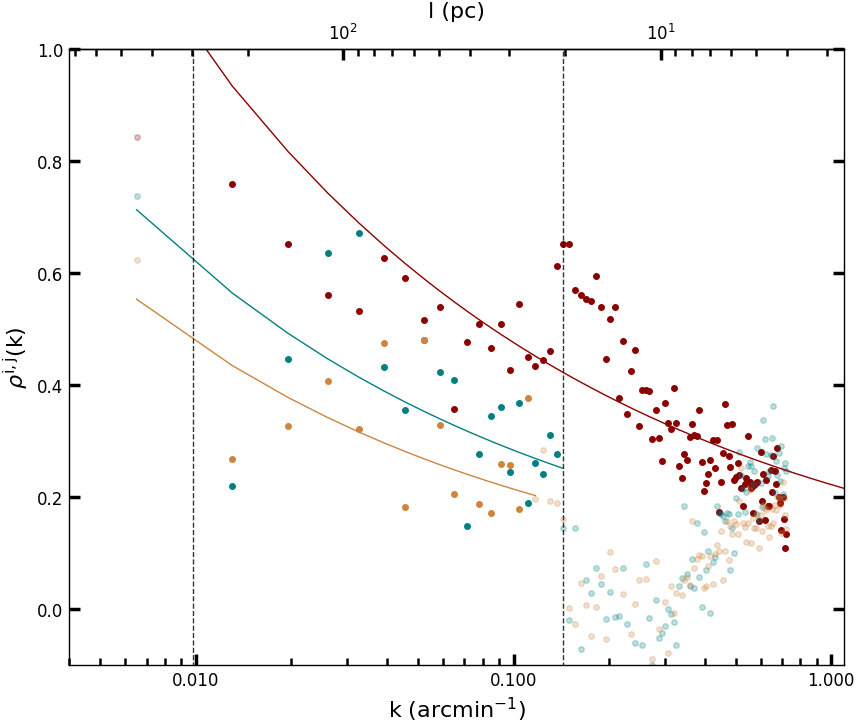}
  \caption{Cross correlation coefficients $\rho^{i,j}(k)$ of LNM$_{\rm F}\times$CNM$_{\rm F}$ (dark-red), WNM$_{\rm F}\times$CNM$_{\rm F}$ (peru), and WNM$_{\rm F}\times$LNM$_{\rm F}$ (teal). 
  Calculated coefficients are shown with colored dots. 
  Solid lines are based on the spectral fits shown in Figs.~\ref{fig:SPS_NHI_CNM_LNM_WNM} and \ref{fig:Cross_SPS_NHI_COMBINED}. Vertical dashed lines shows high- and low-$k$ limits within which the mean and standard deviation of the correlation coefficient for each pair are computed (see Table~\ref{table:index}).
  }
  \label{fig:Cross_coeff_NHI}
\end{figure}

\subsection{Cross correlation coefficients}
\label{subsec:cross_correlation_coefficients}

For further statistical quantification of the interrelationship between phases,
we combined the power spectra and the cross power spectra results to obtain the cross correlation coefficients $\rho^{i,j}(k)$ for each pair of column density maps, formally

\begin{equation}
    \rho^{i,j}(k) = \frac{P^{i,j}(k)}{\sqrt{P^{i,i}(k) \, P^{j,j}(k)}} \, .
\end{equation}

Figure~\ref{fig:Cross_coeff_NHI} shows the cross correlation coefficients of LNM$_{\rm F}\times$CNM$_{\rm F}$, WNM$_{\rm F}\times$CNM$_{\rm F}$, and WNM$_{\rm F}\times$LNM$_{\rm F}$. 
Solid lines show the models $\tilde \rho^{i,j}(k)$ obtained using the spectral fits shown in Figs.~\ref{fig:SPS_NHI_CNM_LNM_WNM} and \ref{fig:Cross_SPS_NHI_COMBINED}.  
Note again the de-correlation toward smaller spatial scales.

In addition, for each pair we calculated the mean and standard deviation (not uncertainty) of the cross correlation coefficients in the spatial range $20$\,pc $< l < 300$\,pc.
These results are tabulated in Table~\ref{table:index} (bottom section). 
The correlation of LNM$_{\rm F}\times$CNM$_{\rm F}$ is the highest and as seen in Fig.~\ref{fig:Cross_coeff_NHI} the other two are not negligible. But in \env\ A, only WNM$_{\rm A}\times$LNM$_{\rm A}$ has a hint of correlation.

Finally, for WNM$_{\rm A}\times$WNM$_{\rm F}$ we find a cross correlation coefficient of $-0.01\pm0.13$, which confirms that regions A and F are uncorrelated statistically and should be analyzed independently.

\section{Properties of structures from segmentation of \Nh\ maps in region F}
\label{sec:phase-segmentation}

As can be appreciated from Fig.~\ref{fig:mosaic_field_0_1} (left), the surface coverage of the EN data is large enough to explore substructures within each phase of \env\ F of \cib. The largest scale ($\sim400$\,pc) allows us to segment even the warm phase and the smallest scale ($\sim2$\,pc) allows us to quantify the finer structures seen in the colder phase. 
For completeness, results of a similar analysis for \env\ A are given in Appendix~\ref{app:properties_env_0}.

\subsection{Hierarchical clustering of \Nh\ maps from region F using dendrograms}
\label{sec:dendrogram}

To perform the clustering analysis, we made use of the \textit{astrodendro} python package that follows the changing topology of the isosurfaces as a function of their contour levels \citep{rosolowsky_2008}. Our choice of this specific method was motivated by the potentially multi-scale nature of the observed phase transition, a direct consequence of turbulence. 
Also, as noted by \cite{goodman_2009}, a dendrogram is almost entirely data driven and is weakly sensitive to the chosen user-parameters. 

A detailed description is provided in Appendix~\ref{app:dendrogram} where visualizations of the extracted structures are shown in the left panels of Fig.~\ref{fig:mosaic_field_0_1_dendrograms} (see right panels for \env\ A). We obtained $N_c = 21,$ 60, and 73 structures for WNM$_{\rm F}$, LNM$_{\rm F}$, and CNM$_{\rm F}$, respectively.

In separate subsections below, we have evaluated a number of properties of the structures -- physical, thermodynamic, and turbulent -- for each phase and for the total ensemble.
The values have a considerable range, so that their PDFs are best displayed logarithmically. Although the PDFs are not precisely log-normal, they are well summarized by the mean and standard deviation of the log of the parameter.  The value of the parameter corresponding to this mean is reported in Table~\ref{table:turbulence_1}, along with the standard deviation (hereafter ``spread", e.g., 2 for a 0.3 dex standard deviation) in parenthesis.

\subsubsection{Evaluation of uncertainties}
To evaluate the uncertainties of physical properties derived from dendrograms, we have repeated the segmentation on \Nh\ maps using two series of decompositions obtained with {\tt ROHSA}. 
For the first series, composed of 50 runs, in the spectral data for each line of sight we injected Gaussian random noise with the same dispersion as the noise in the original data. The hyper-parameters were kept the same. For the second series, composed of 80 runs, we kept the original data cube (i.e., each run has just the original noise) but randomly perturbed the four {\tt ROHSA} hyper-parameters in a $\pm 10$\% interval around the original values. 
From the catalogs obtained by segmentation for each run, properties of the structures -- physical, thermodynamic, and turbulent -- were computed.

The new structures extracted were cross-matched with structures in the original catalogs using their spatial coordinates. 
The uncertainties of the properties of a structure were estimated by calculating the standard deviation over the cross-matched members for each series. Finally, for each property the contributions from the two series were summed in quadrature to yield the total uncertainty (the contribution from the first series was generally slightly higher than from the second).

These uncertainties were then taken into account in producing
the PDFs presented below. In each PDF, the bin  size (logarithmically constant) was chosen so that it was larger than the relative uncertainty of the quantity being analyzed in each bin.

\subsection{Physical properties}
\label{physical-properties}

\begin{deluxetable*}{lccccccc}
\tablecaption{Properties of structures in \env\ F}
\label{table:turbulence_1}
\tablewidth{0pt}
\tablehead{
\nocolhead{\textbf{Quantity}} & \colhead{\textbf{Symbol}} & \colhead{\textbf{WNM$_{\rm F}$}} & \colhead{\textbf{LNM$_{\rm F}$}} & \colhead{\textbf{CNM$_{\rm F}$}} & \colhead{\textbf{Total$_{\rm F}$}} & \colhead{\textbf{Units}} }
\startdata
 & $N_c$ & 21 & 60 & 73 & 154\\
\hline
\textit{\textbf{Physical}} & & & & & &                     
\\
Size & $L$ & 28\,(1.6) & 7.3\,(1.4) & 6.4\,(1.4) & 8.2\,(1.8) & pc \\
Aspect ratio & $r$ & 2.0\,(1.4) & 2.1\,(1.5) & 2.1\,(1.5) & 2.1\,(1.5) & \\
Mass & $\mh$ & 296\,(3.4) & 18\,(2.3) & 12\,(1.9) & 22\,(3.8) & M$_{\odot}$  \\
Average number of H atoms per unit volume & $n$ & 0.41\,(1.9) & 1.4\,(1.8) & 1.3\,(1.9) & 2.1\,(2.1) & cm$^{-3}$ \\
\hline
\textit{\textbf{Thermodynamic}} & & & & & &  \\
Doppler velocity dispersion & $\sigma_{T_b}$ & 8.9\,(1.0) & 3.0\,(1.1) & 1.6\,(1.1) & 2.6\,(1.8) & \kms \\
Turbulent velocity dispersion & $\sigma_{v_{z}}$ & 0.79\,(2.4) & 0.94\,(1.6) & 0.74\,(1.8) & 0.82\,(1.8) & \kms  \\
Thermal velocity dispersion & $\sigma_{\rm th}$ & 8.8\,(1.0) & 2.7\,(1.1) & 1.3\,(1.4) & 2.3\,(2.0) & \kms \\
Kinetic temperature & $T_k$ & 9.5\,(1.0) & 0.90\,(1.2) & 0.20\,(2.1) & 0.62\,(4.0) & 10$^{3}$\,K \\
Sound speed & $C_s$ & 9.6\,(1.0) & 3.0\,(1.1) & 1.4\,(1.4) & 2.5\,(2.0) & \kms \\
Thermal crossing time & $t_{\rm cross}^{\rm th}$ & 2.8\,(1.6) & 2.4\,(1.5) & 4.5\,(1.8) & 3.3\,(1.8) & Myr \\
Turbulent crossing time & $t_{\rm cross}^{v_z}$ & 34\,(2.1) & 7.6\,(1.5) & 8.5\,(1.7) & 9.8\,(2.1) & Myr \\
Thermal pressure & $P_{\rm th}/k_B$ & 3.8\,(1.9) & 1.2\,(1.8) & 0.25\,(2.8) & 0.70\,(3.7) & 10$^{3}$\,K\,cm$^{-3}$ \\
Turbulent pressure & $P_{\rm v_z}/k_B$ & 0.1\,(7.6) & 0.5\,(2.3) & 0.3\,(3.1) & 0.3\,(3.7) & 10$^{3}$\,K\,cm$^{-3}$ \\
Total pressure & $P_{\rm tot}/k_B$ & 4.5\,(1.9) & 2.0\,(1.6) & 0.7\,(1.8) & 1.4\,(2.5)  & 10$^{3}$\,K\,cm$^{-3}$ \\
\hline
\textit{\textbf{Turbulent cascade}} &                            & &&&&                             \\
Turbulent sonic Mach number & $\mathcal{M}_s$ & 0.14\,(2.4) & 0.55\,(1.7) & 0.86\,(2.2) & 0.56\,(2.5) & \\
Mean free path  & $\lambda$ & 0.80\,(1.9) & 0.24\,(1.8) & 0.25\,(1.9) & 0.29\,(2.1) & 10$^{-3}$\,pc \\
Kinematic molecular viscosity & $\nu$ & 1.0\,(1.9) & 0.09\,(1.8) & 0.05\,(2.0) & 0.09\,(3.3) & 10$^{21}$\,cm$^2$\,s$^{-1}$ \\
Knudsen number  & $K_n$ & 2.9\,(1.8) & 3.2\,(1.5) & 4.0\,(1.6) & 3.5\,(1.6) & 10$^{-5}$ \\
Reynolds number & $Re$  & 0.62\,(3.9) & 2.1\,(1.9) & 2.7\,(2.4) & 2.0\,(2.7) & 10$^4$   \\
Dissipation scale & $\eta$ & 4.0\,(2.5) & 0.42\,(1.6) & 0.30\,(1.9) & 0.49(2.9) & 10$^{-2}$\,pc \\
Dissipation time & $t_{\eta}$ & 4.9\,(3.9) & 0.59\,(1.9) & 0.61\,(2.3) & 0.80\,(3.1) & 10$^{-1}$\,Myr \\
Convective time & $t_{L}$ & 38\,(3.9) & 8.5\,(1.9) & 10\,(2.3) & 11.3\,(3.1) & Myr \\
Traversal time & $\tau_L$ & 39\,(2.1) & 8.6\,(1.5) & 10.1\,(1.6) & 11.4\,(2.0) & Myr \\
Energy transfer rate & $\epsilon$ & 0.22\,(11) & 1.4\,(3.7) & 0.6\,(4.1) & 0.7\,(5.3) & 10$^{-5}$\,L$_{\odot}$\,M$_{\odot}^{-1}$ \\
\enddata
\tablecomments{Values correspond to the mean and spread from the logarithmic PDF. 
}
\end{deluxetable*}
The sky coordinates of a structure in a particular frame 
(taken to be the International Celestial Reference System for the terminology here) are the mean positions of all pixels, weighted by column density:
\begin{equation}
    (\bar x, \bar y) = \sum_{\rb} (x(\rb), y(\rb)) \, \Nh(\rb) \, / \,  \sum_{\rb} \Nh(\rb) \, ,
\end{equation}
where $x$ and $y$ denote the Right Ascension and Declination.

The geometry of each structure is modeled by an on-sky ellipse.
Following \citet{hennebelle_structure_2007} and \citet{miville-deschenes_structure_2017}, the inertia matrix of the projected emission is
\begin{equation}
    I = \left[\begin{array}{ll}
    \sigma_{x}^{2} & \sigma_{xy}^{2} \\
    \sigma_{xy}^{2} & \sigma_{y}^{2}
    \end{array}\right] \, ,
\end{equation}
where the matrix coefficients are 
\begin{equation}
   \sigma_{xy}^2 = \sum_{\rb} \left(x(\rb) - \bar x \right) \left(y(\rb) - \bar y \right) \Nh(\rb) \, / \, \sum_{\rb} \Nh(\rb) \, .
\end{equation}

\subsubsection{Size}
\label{subsubsec:size}

Using the eigenvalues $\lambda_{\rm min}$ and $\lambda_{\rm maj}$ of $I$ and the pixel area $S_A = (D \delta)^2$ (the scaling with distance $D$ can be tracked),
the lengths of the semi-major and semi-minor axes are $L_{{\rm min, \, maj}} = \sqrt{8 \ln(2) \, S_A \,  \lambda_{{\rm min, \, maj}}}$.
Making the assumption that the depth along the line of sight is more likely to be $L_{\rm min}$, the volume of the ellipsoid is $V = (4\pi/3)\, L_{\rm min}^2 L_{\rm maj}$.
Finally, our size estimate is simply $L = V^{1/3}$. 

\begin{figure}[!t]
  \centering
  \includegraphics[width=0.95\linewidth]{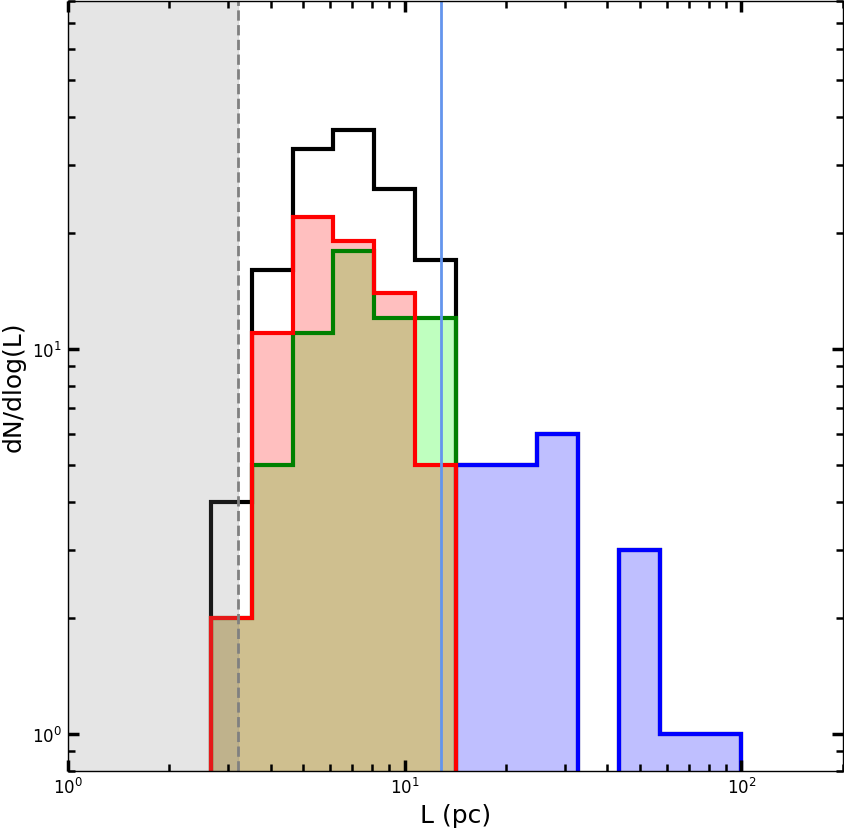}
  \caption{Probability distribution function of the typical size $L$ of structures extracted from WNM$_{\rm F}$ (blue), LNM$_{\rm F}$ (green), and CNM$_{\rm F}$ (red). The total shown in black. The gray dashed line shows physical size of 1\farcm 1 beam at 10 kpc. The blue vertical line shows physical size of 4\farcm4 beam (i.e., the convolved WNM$_{\rm F}$ map used for the clustering) at the same distance.
  } 
  \label{fig:dend_length}
\end{figure}

Figure~\ref{fig:dend_length} shows the PDFs of the size, with color coding for each phase. The PDF for all phases combined is also shown in black (note the logarithmic scale vertically). Sizes range from $\sim$ 3\,pc (the spatial resolution of the maps for LNM$_{\rm F}$ and CNM$_{\rm F}$, see vertical gray dashed line) up to $\sim$ 100\,pc. 

Inspection of Fig.~\ref{fig:dend_length} suggests that the PDF of $L$ for CNM$_{\rm F}$ and LNM$_{\rm F}$ might be impacted by the resolution, 1\farcm 1 or 3.2\,pc at the assumed distance. Perhaps smaller structures would be identified with observations of higher resolution and higher signal to noise. In none of the statistics, here and below, have we applied any correction for this potential bias. Similarly, because the segmentation for WNM$_{\rm F}$ was carried out on maps convolved to 4\farcm4\ (a factor of four in resolution, see vertical blue line for the corresponding physical size), that phase too is potentially impacted. 
Even so, there is some evidence that the typical size of structures decreases from the warmer to the cooler phases, as expected from the thermal condensation.

\subsubsection{Aspect ratio}
\label{subsubsec:ratio}

\begin{figure}[!t]
  \centering
  \includegraphics[width=0.95\linewidth]{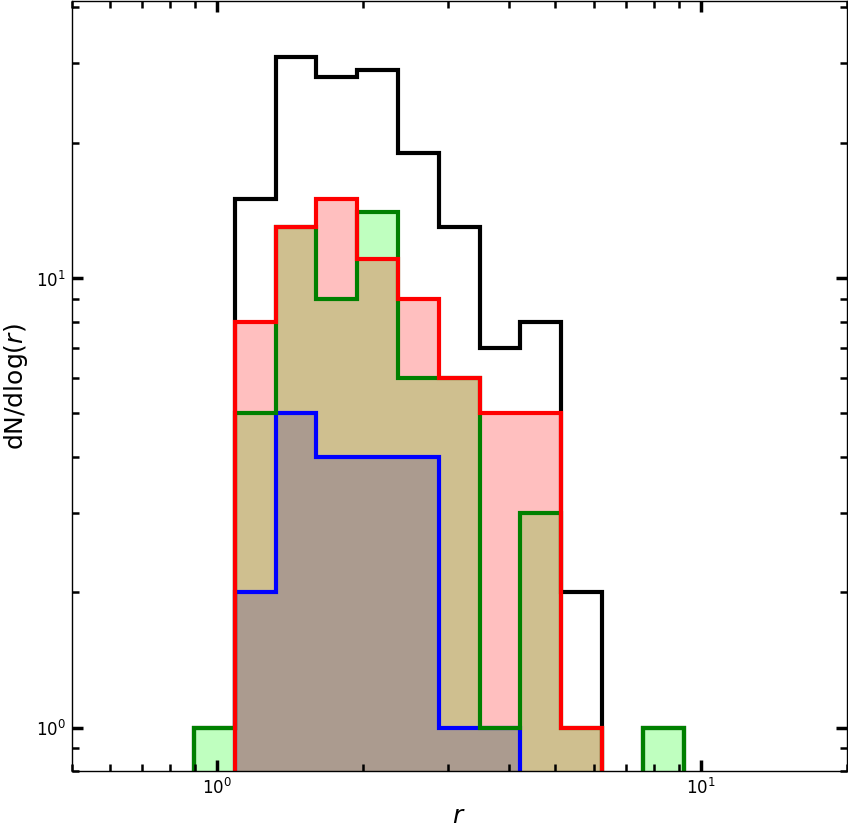}
  \caption{Probability distribution function of the aspect ratio $r$ of structures extracted from WNM$_{\rm F}$ (blue), LNM$_{\rm F}$ (green), and CNM$_{\rm F}$ (red). The total is shown in black.} 
  \label{fig:dend_aspect_ratio}
\end{figure}

Figure~\ref{fig:dend_aspect_ratio} shows the PDFs of the aspect ratio $r = L_{\rm maj} / L_{\rm min}$. In each phase, $r$ is generally higher than 1.5 and can reach values up to 6 in CNM$_{\rm F}$ and LNM$_{\rm F}$. 
The shape of the distributions looks similar among phases and the mean of all distributions is close to 2 (see Table~\ref{table:turbulence_1}), showing that elongated structures are seen not just in the cold phase but rather are a multiphase property.

\subsubsection{Orientation}
\label{subsubsec:orientation}

\begin{figure}[!t]
  \centering
  \includegraphics[width=0.95\linewidth]{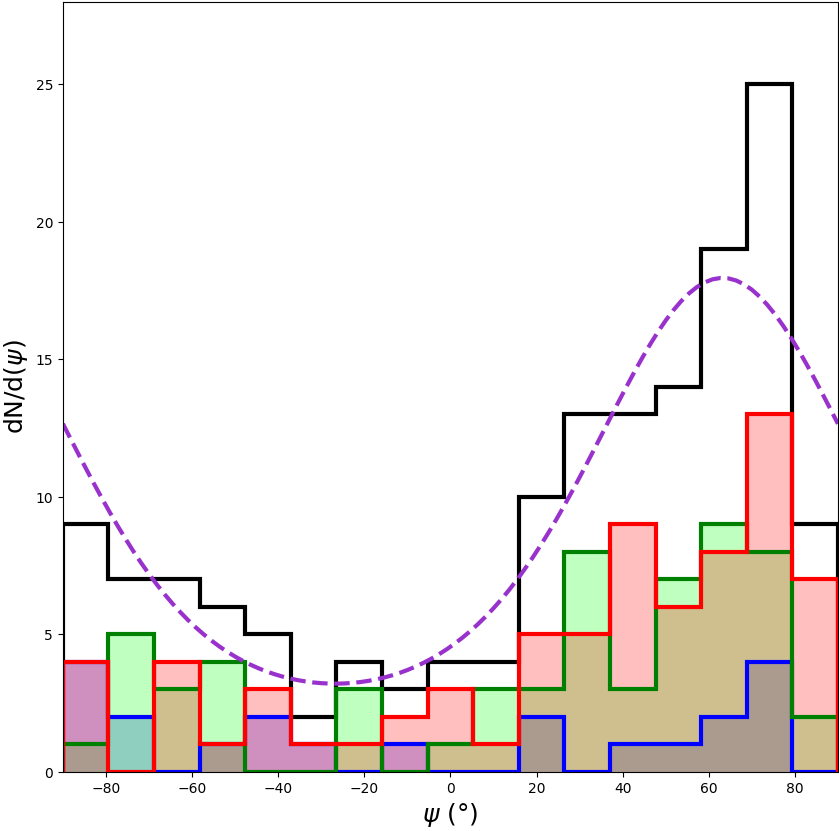}
  \caption{Probability distribution function of the position angle $\psi$ of structures extracted from WNM$_{\rm F}$ (blue), LNM$_{\rm F}$ (green), and CNM$_{\rm F}$ (red). The total is shown in black. 
  The purple dashed line shows the MLE fit of the Von Mises distribution to the (unbinned) orientations of all structures.
  }
  \label{fig:dend_orientation_1}
\end{figure}

The position angle $\psi$ of each structure is 
\begin{equation}
    \psi = tan^{-1} \left( \frac{v_2}{v_1} \right) \, ,
\end{equation}
where 
$(v_1, \, v_2)$ is the eigenvector corresponding to $\lambda_{m\rm ax}$. 

Figure~\ref{fig:dend_orientation_1} shows the PDF of $\psi$ for each phase and the whole sample. CNM$_{\rm F}$ and LNM$_{\rm F}$ clearly show a distribution dominated by positive position angles; the evidence for WNM$_{\rm F}$ is less pronounced. Maximum Likelihood Estimates (MLE) of the location (orientation) and dispersion of a Von Mises distribution fit to the orientations of all of the structures 
are $+63\degree$ and $31\degree$, respectively. The purple dashed line shows the model.

In \env\ A, we find an orientation of $-43\degree$ and a dispersion of $38\degree$ (see Fig.~\ref{fig:dend_orientation_0} in Appendix~\ref{app:properties_env_0}).
That the mean orientations of structures in regions F and A are roughly orthogonal can be appreciated visually from Figs.~\ref{fig:mosaic_field_0_1} and \ref{fig:RGB_WCS}.

\subsubsection{Mass}
\label{subsubsec:mass}

\begin{figure}[!t]
  \centering
  \includegraphics[width=0.95\linewidth]{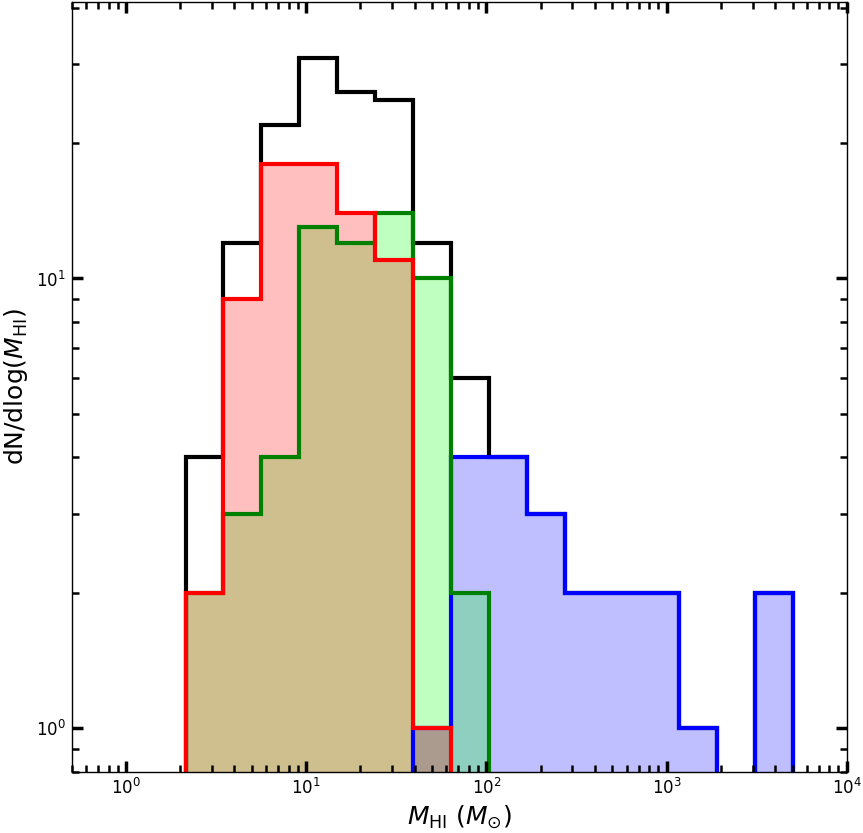}
  \caption{Probability distribution function of the mass $\mh$ of structures extracted from WNM$_{\rm F}$ (blue), LNM$_{\rm F}$ (green), and CNM$_{\rm F}$ (red). The total is shown in black.} 
  \label{fig:dend_mass}
\end{figure}

The mass of each structure derived from the column density is 
\begin{equation}
    \mh = S_A \, \mu_m \, m_{\rm H} \, \sum_p \Nh^p \, . 
    \label{eq:mass}
\end{equation}
Here $m_{\rm H}$ is the mass of the hydrogen atom and $\mu_m = 1.4$ accounts for the atomic Galactic composition, so that $\mh$ is actually the total mass of the neutral gas.

Figure~\ref{fig:dend_mass} shows the PDFs of $\mh$. Masses increase systematically from CNM$_{\rm F}$ to LNM$_{\rm F}$ and WNM$_{\rm F}$, with values ranging from $\sim$ 2 $M_{\odot}$ in the cold phase up to $\sim$ 3000 $M_{\odot}$ in the warm phase. Note that CNM structures would reach even lower masses if they were segmented more finely at higher spatial resolution.

The total masses of CNM$_{\rm F}$, LNM$_{\rm F}$, and WNM$_{\rm F}$ are 
$(1, 1.5, {\rm and\,} 15) \times 10^3$\,$M_{\odot}$,
respectively. 
The corresponding mass fractions within \env\ F are 0.06, 0.08, and 0.86.
Finally, for perspective, the total mass of the neutral gas of these structures, 
$18 \times 10^3$\,$M_{\odot}$, is just 0.3\% 
of the total atomic mass of complex C \citep{thom_2008}.

\begin{figure}[!t]
  \centering
  \includegraphics[width=0.95\linewidth]{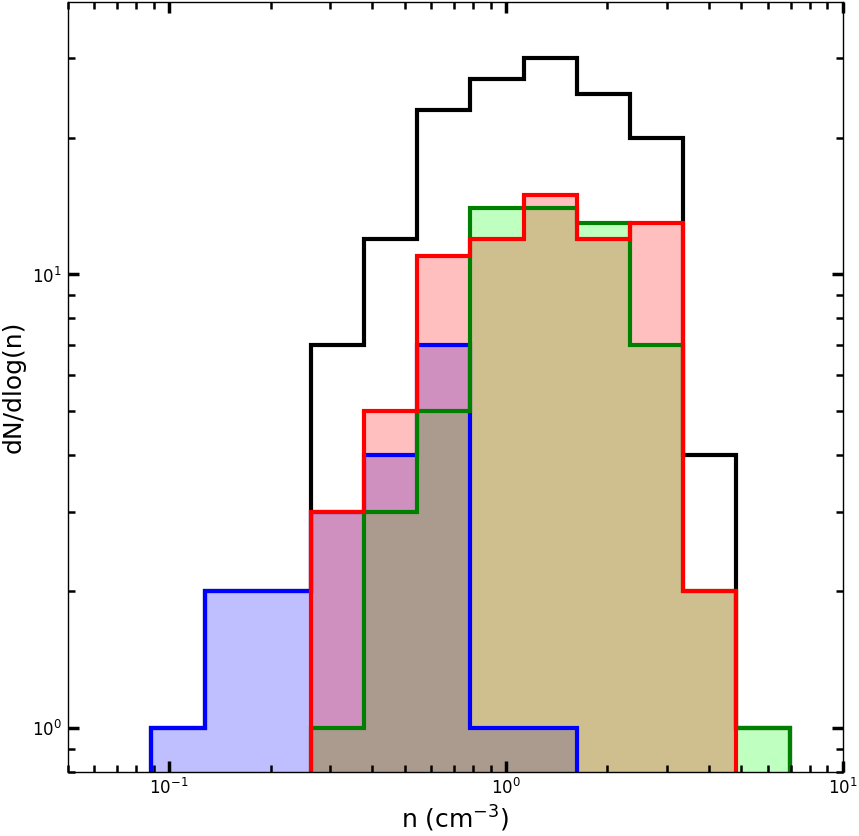}
  \caption{Probability distribution function of the average number density of H atoms $n$ of structures extracted from WNM$_{\rm F}$ (blue), LNM$_{\rm F}$ (green), and CNM$_{\rm F}$ (red). The total is shown in black.}
  \label{fig:dend_density}
\end{figure}

\subsubsection{Average number density of H atoms}
\label{subsubsec:volume-density}

For each structure, the average number of H atoms per unit volume is
\begin{equation}
    n = \frac{1}{\mu_m m_{\rm H}} \frac{\mh}{V} \, ,
    \label{eq:density}
\end{equation}
(or equivalently the total column density \Nh\ times $S_A$ divided by $(4\pi/3)\, L_{\rm min}^2 L_{\rm maj}$)
which scales as $D^{-1}$.
Figure~\ref{fig:dend_density} shows the PDFs of $n$.
From the condensation mode of thermal instability we expect to observe an increase of the 
$n$ from the warm to the cold phase. There is some evidence for this, and it is possible that $n$ for CNM$_{\rm F}$ is underestimated because of the finite resolution.

\subsection{Thermodynamic properties}
\label{subsec:thermodynamic-properties}

\subsubsection{Separation of thermal and non-thermal motions}
\label{subsec:sepmo}

The observed Doppler dispersions $\sigma_{T_b}$ of the structures (Table~\ref{table:turbulence_1}) measure the total velocity dispersion of gas along the line of sight. This is often modeled as a quadratic sum of a thermal component and a non-thermal component
\begin{equation}
    \sigma_{T_b} = \sqrt{\sigma_{\rm th}^2 + \sigma_{v_z}^2} \, .
    \label{eq:quadratic-disp}
\end{equation}
Separation of the two components is not possible using data for a single line of sight, but can be done statistically for an ensemble.
According to studies by \citet[][and references therein]{ossenkopf_interstellar_2006}, 
for a turbulent medium the statistics of the 3D velocity field can be recovered from its 2D projection if density fluctuations are small compared to the mean density of the fluid ($\sigma_{\rho/\rho_0}\lesssim 0.5$).

To verify that sufficiently small density contrast is the case here, the methodology proposed by \cite{brunt_method_2010} and applied to 21\,cm line emission data in \cite{marchal_2021} was used to calculate, for each structure, the density contrast of its three-dimensional (3D) density field $\sigma_{\rho/\rho_0}$ from the column density contrast of its projection along the line-of-sight $\sigma_{N/N_0}$. \citet{brunt_method_2010}, assuming that the statistical properties of $\rho$ are isotropic and using Parseval's Theorem, have shown that the ratio of these contrasts is
\begin{align}
    R &= \left( \frac{\sigma_{N/N_0}}{\sigma_{\rho/\rho_0}} \right)^2 \\
    &= \frac{\left( \sum\limits_{k_x=-L_{\rm min}/2+1}^{L_{\rm min}/2} 
        \sum\limits_{k_y=-L_{\rm maj}/2+1}^{L_{\rm maj}/2}
        P^{3D}_{\rho}(k) \right) - P^{3D}_{\rho}(0)}{\left( \sum\limits_{k_x=-L_{\rm min}/2+1}^{L_{\rm min}/2} 
        \sum\limits_{k_y=-L_{\rm maj}/2+1}^{L_{\rm maj}/2}
        \sum\limits_{k_z=-L_{\rm min}/2+1}^{L_{\rm min}/2}
        P^{3D}_{\rho}(k) \right) - P^{3D}_{\rho}(0)} \, ,
    \label{eq:R}
\end{align}
where $P^{3D}_{\rho}(k)$ is the azimuthally-averaged power spectrum of $\rho$.

For a structure of size $L_{\rm min} \times L_{\rm maj}$, two parameters control the ratio $R$: the slope of $P^{3D}_{\rho}(k)$ and the depth of the structure. 
For this model we consider $P^{3D}_{\rho}(k)\propto k^{-11/3}$, representative of a sub/trans-sonic turbulence. The depth over which velocity fluctuations are averaged is assumed to be $L_{\rm min}$, so that $R$ depends on the aspect ratio $r$ of each structure. For the whole sample, we find a mean value $\sigma_{\rho/\rho_0}\sim0.2$ and a standard deviation of 0.1.
This result is not sensitive to a variation of $\pm1/3$ for the slope of  $P^{3D}_{\rho}(k)$.

Having verified that density fluctuations are sufficiently small, using the same formalism (i.e., the same coefficient $R$ for each structure), we can infer the non-thermal velocity dispersion $\sigma_{v_z}$ of the three-dimensional (3D) velocity field from the observed dispersion of its projection along the line of sight, $\sigma_{\left<v_z\right>_z}$.
We find a similar velocity dispersion $\sigma_{v_z}$ across phases (Table~\ref{table:turbulence_1}), with a value about 0.8 \kms\ overall. This is a significant part of the Doppler dispersion only for the CNM, as reflected in the thermal velocity dispersions $\sigma_{\rm th}$ recorded, derived using Eq.~\ref{eq:quadratic-disp}.
Formally, nothing prevents the inferred value of $\sigma_{\rm th}$ from being less than the channel spacing of the observations. However, this occurs for only 9 out of the 73 structures in CNM$_{\rm F}$
and for these there is considerable uncertainty.

Properties derived below that are dependent on these separated velocity dispersions are less directly data-driven than the physical quantities in Sect.~\ref{physical-properties}. Nevertheless, tabulating these provides a global view of the properties of the fluid constituting the concentration \cib, which is useful for interpreting our results in the context of condensation via thermal instability.

\subsubsection{Velocity dispersions and related properties}
\label{subsubsec:related-properties}

The inferred kinetic temperature is $T_k = \sigma_{\rm th}^2 \, m_{\rm H}  / k_B$, where $k_B$ the Boltzmann constant and as used here $\sigma_{\rm th}$ describes the motions of H atoms.
The adiabatic sound speed is $C_s = \sqrt{ \gamma k_B T_k/\mu_m m_{\rm H} } \equiv \sqrt{\gamma/\mu_m} \, \sigma_{\rm th}$, with the adiabatic index $\gamma=5/3$ for atomic gas.
In combinations with the size of the structure, the thermal crossing time is $t_{\rm cross}^{\rm th} = L/C_s$
and the turbulent crossing time is $t_{\rm cross}^{v_z} = L/\sigma_{v_z}$.

\begin{figure}[!t]
  \centering
  \includegraphics[width=0.95\linewidth]{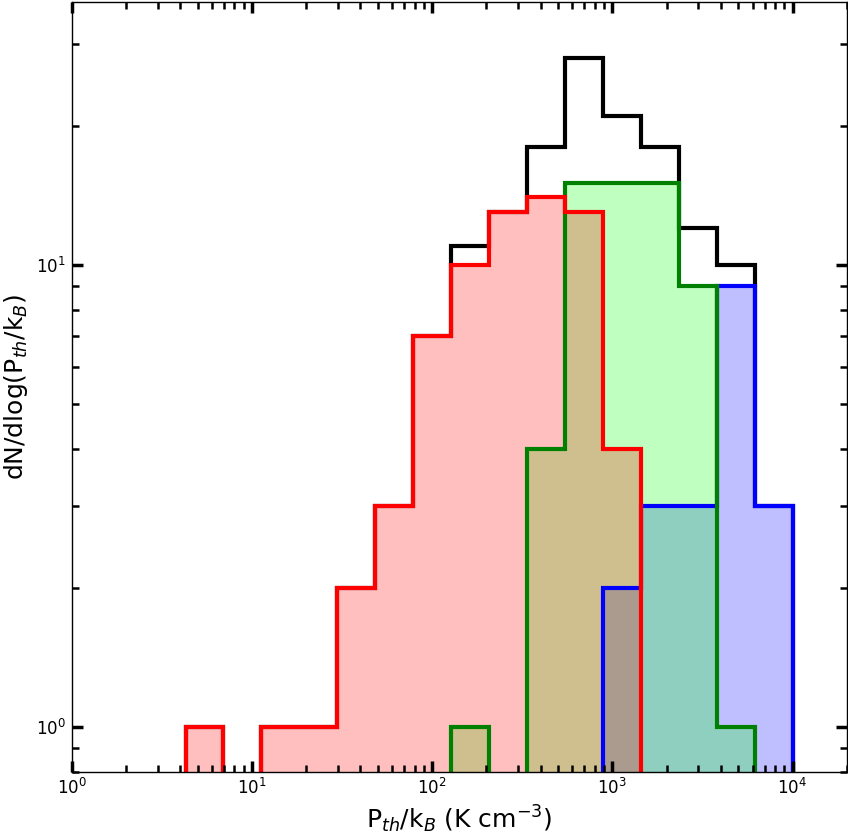}
  \caption{Probability distribution function of the thermal pressure $P_{\rm th}/k_B$ of structures extracted from WNM$_{\rm F}$ (blue), LNM$_{\rm F}$ (green), and CNM$_{\rm F}$ (red). The total is shown in black.} 
  \label{fig:dend_pressure}
\end{figure}

\subsubsection{Thermal pressure}
\label{subsubsec:pressure}

Given the kinetic temperature and the average number density of \HI of a structure, the thermal pressure is 
\begin{equation}
    P_{\rm th}/k_B = \mu_P\, n \, T_k  \equiv \mu_P\, n\, \sigma_{\rm th}^2 m_{\rm H} / k_B \, ,
    \label{eq:thermal_pressure}
\end{equation}
with $\mu_P = 1.1$ accounting for He.
Figure~\ref{fig:dend_pressure} shows the PDFs of $P_{\rm th}/k_B$ for the phases. 
There is an apparent decrease of mean thermal pressure from WNM$_{\rm F}$ to LNM$_{\rm F}$ and CNM$_{\rm F}$ (see also Table~\ref{table:turbulence_1}).
As mentioned in Sect.~\ref{subsubsec:volume-density}, because of finite resolution the average number of H atoms per unit volume of CNM structures might be higher than that above, which would increase the thermal pressure. A secondary compensating effect might be less beam smearing, possibly leading to a lower estimate of the thermal broadening and therefore a lower thermal pressure.

\subsection{Turbulent and total pressure}
The turbulent pressure can be calculated using the deduced turbulent velocity dispersion:
\begin{equation}
    P_{\rm t}/k_B = 3 \, \mu_P\, n \, \sigma_{\rm v_z}^2 m_{\rm H} / k_B \,.
    \label{eq:turb_pressure}
\end{equation}
where the factor 3 account for the dimensionality of $\sigma_{\rm v_z}$ (from 1D to 3D).
As expected from the typical relative sizes of $\sigma_{\rm v_z}$ and $\sigma_{\rm th}$ (Table~\ref{table:astrodendro}), this pressure is of most interest in the CNM phase.
The total pressure is just $P_{\rm tot}/k_B = (P_{\rm t} + P_{\rm th})/k_B$. Table~\ref{table:astrodendro} gives the mean and spread of each quantity calculated for the ensemble.

\subsection{Properties of the turbulent cascade}
\label{subsec:turbulent-properties}

Turbulent fluids are commonly characterized by properties like the Mach number. These can be derived from the velocity dispersions in combinations with the size $L$ and average number of H atoms per unit volume $n$, and so again can be considered as different recastings not as directly data-driven.  They are nevertheless useful in connecting to the literature and understanding numerical simulations. Values of such properties are given in Table~\ref{table:turbulence_1} for structures in different phases. These are motivated and discussed briefly here.

\begin{figure}[!t]
  \centering
  \includegraphics[width=0.95\linewidth]{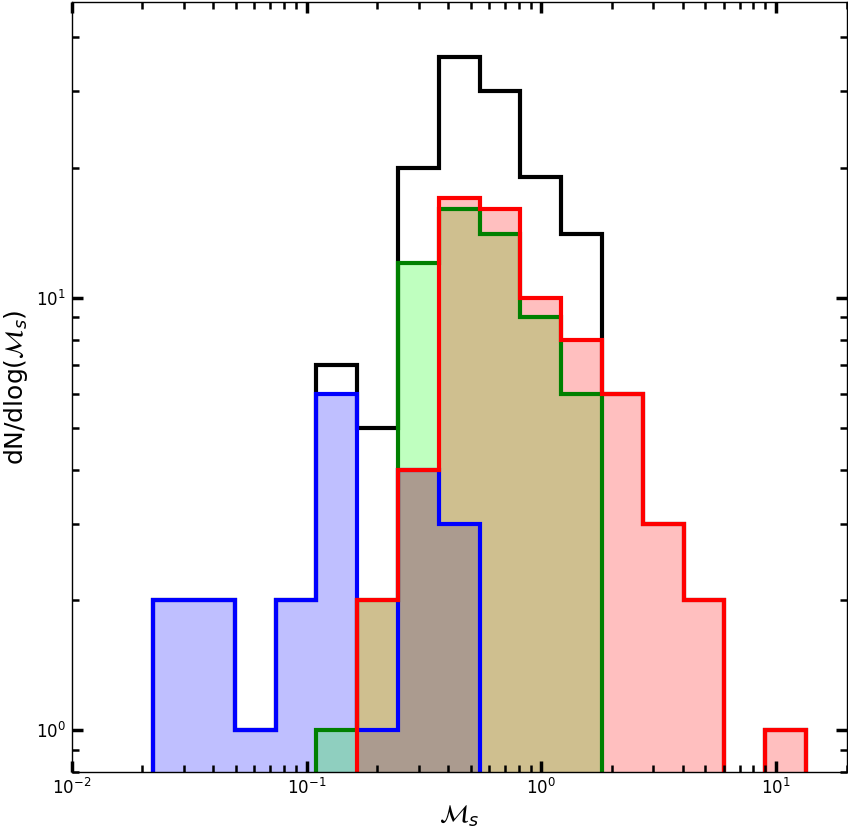}
  \caption{Left: Probability distribution function of the sonic Mach number $\mathcal{M}_s$ of structures extracted from WNM$_{\rm F}$ (blue), LNM$_{\rm F}$ (green), and CNM$_{\rm F}$ (red). The total is shown in black.}
  \label{fig:dend_Mach}
\end{figure}

\begin{figure}[!t]
  \centering
  \includegraphics[width=\linewidth]{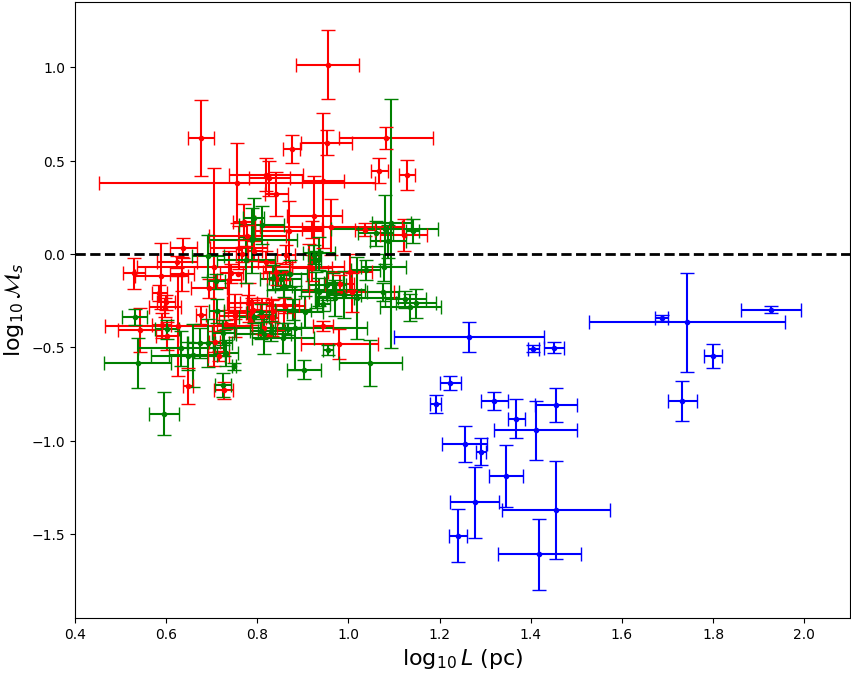}
  \caption{Scatter plot of $\log_{10}\mathcal{M}_s$ vs $\log_{10}L$. Values for structures from CNM$_{\rm F}$, LNM$_{\rm F}$, and WNM$_{\rm F}$ are shown in red, green, and blue, respectively. The black dashed line indicates $\mathcal{M}_s=1$.}
  \label{fig:dend_Mach_size}
\end{figure}

\begin{figure}[!t]
  \centering
  \includegraphics[width=\linewidth]{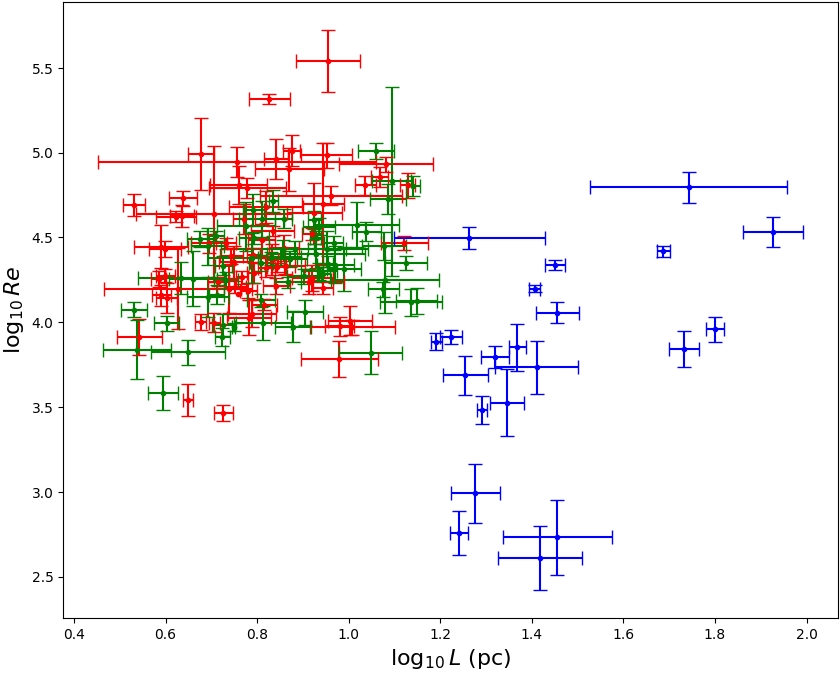}
  \caption{Scatter plot of $\log_{10}Re$ vs $\log_{10}L$. Values for structures from CNM$_{\rm F}$, LNM$_{\rm F}$, and WNM$_{\rm F}$ are shown in red, green, and blue, respectively. The black dashed line indicates $\mathcal{M}_s=1$.}
  \label{fig:dend_Re_size}
\end{figure}

The turbulent sonic Mach number $\mathcal{M}_s$ of a structure is
\begin{equation}
    \mathcal{M}_s = \frac{\sqrt{3} \, \sigma_{v_z}}{C_s} \, .
    \label{eq:mach}
\end{equation}
Figure~\ref{fig:dend_Mach} shows the PDFs of $\mathcal{M}_s$.
Because $\sigma_{v_z}$ is fairly constant, $\mathcal{M}_s$ increases from the warm phase to the cold phase, from a sub-sonic regime ($0.14\,(2.4)$) to a trans-sonic regime ($0.75\,(2.1)$). This implies an inverse trend of $\mathcal{M}_s$ with size, as shown in as Fig.~\ref{fig:dend_Mach_size}.

For properties below involving a mean free path, $\lambda$, we assume $\lambda=1/(n\, \sigma$ with $\sigma = 1\times$10$^{-15}$ cm$^{2}$ for the hydrogen collision cross-section \citep{lequeux_milieu_2012}.
One such property is the kinematic viscosity
\begin{equation}
\label{eq:visc}
    \nu = \frac{1}{3} \lambda\, v_{\rm th} \, ,
\end{equation}
where the arithmetic mean speed $v_{\rm th} = \sqrt{8/\pi\gamma}C_s \equiv \sqrt{8/\pi\mu} \, \sigma_{\rm th}$.

Another such property is the Knudsen number $Kn=\lambda$/$L$. For an isothermal gas, as assumed here for each individual structure, the ratio of
$\mathcal{M}_{s}$ to the Knudsen number is linked directly to
the Reynolds number 
\begin{equation}
    Re = \sqrt{\frac{\pi}{2}} \frac{\mathcal{M}_{s}}{Kn} \, ,
    \label{eq:Reynolds-from-mach}
\end{equation}
which quantifies the relative influence of advection and diffusion in a turbulent fluid. 
Unlike $\mathcal{M}_{s}$, $Re$ is proportional to $n$, through the inverse dependence on $\lambda$.
The net dependence on size is shown in the scatter plot of $Re$ and $L$ in Fig.~\ref{fig:dend_Re_size}. We observe the same trend as for $\mathcal{M}_s$, i.e., $Re$ increases from large warm structures to small cold structures.
The typical values of $Re$ are of interest too.  If initial perturbations applied to a flow are not too small, turbulence appears at $Re\sim2000$ \citep{reynolds_experimental_1883}.\footnote{For minimal perturbations, the flow can remain laminar up to $Re\sim13000$.} It therefore seems likely that the turbulence in gas toward \cib\ is well developed for CNM$_{\rm F}$, LNM$_{\rm F}$, and even WNM$_{\rm F}$, except for four structures with $Re<1000$ (Fig.~\ref{fig:dend_Re_size}).

Other combinations give characteristic size and time scales for structures. 
The dissipation scale $\eta$, on which the smallest eddies dissipate the turbulent energy into heat through viscosity, is 
\begin{equation}
    \eta = L Re^{-3/4} \, ,
    \label{eq:disspation-scale-NEP}
\end{equation}
and the dissipation time scale $t_{\eta}$ is
\begin{equation}
    t_{\eta} = \frac{\eta^2}{\nu} \, .
    \label{eq:dissipation-time-NEP}
\end{equation} 
Note that $\eta$ and $t_{\eta}$ are also called the 
Kolmogorov length and time scales.
The convective time, also called the large eddy turnover time, is 
\begin{equation}
    t_L = t_{\eta} Re^{1/2} \, .
    \label{eq:turnover-time}
\end{equation}
The traversal time $\tau_L$ needed for an eddy of size $L$ to traverse the inertial range, in which viscous effects are essentially negligible, 
down to the Kolmogorov length scale $\eta$
is related to $Re$ through the relation
\begin{equation}
    \tau_L = \frac{t_L}{1 - Re^{-1/2}} \, .
    \label{eq:traverse-time}
\end{equation}
For high $Re$, i.e., fully developed turbulence, $\tau_L \sim t_L$.
Finally, combining the dissipation scale and the dissipation time, the energy transfer rate $\epsilon$ is
\begin{equation}
    \epsilon = \frac{\eta^2}{t_{\eta}^3} \, .
    \label{eq:energy-transfer-rate-NEP}
\end{equation}

\subsection{Scaling laws between properties of extracted structures} 
\label{sec:scaling-laws}

\subsubsection{Mass and density -- size relations}
\label{subsec:mass-size-relation}

\begin{figure}[!t]
  \centering
  \includegraphics[width=\linewidth]{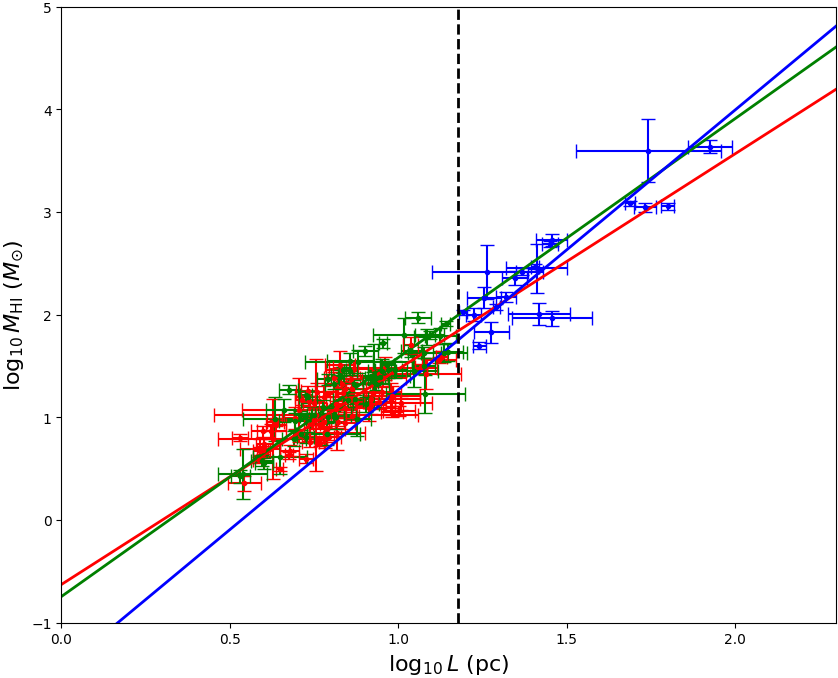}  
  \caption{Scatter plot of $\log_{10}\mh$ vs $\log_{10}L$. Values for structures from CNM$_{\rm F}$, LNM$_{\rm F}$, and WNM$_{\rm F}$ are shown in red, green, and blue, respectively. Pearson correlation coefficients are 0.90, 0.81, and 0.76, respectively. Solid lines show power law fits using the bisector estimator. Black dashed line marks our estimate of the cooling length scale, at which the warm gas undergoes a phase transition.}
  \label{fig:dend_mass_size}
\end{figure}

A scatter plot of \HI\ mass (proportional to total column density: Eq.~\ref{eq:mass}) and size is shown in Fig.~\ref{fig:dend_mass_size}, color coded by phase. Note the logarithmic scales, such that the slope in this diagram is the power-law exponent of the mass-size relation.  Using a bisector estimator of the slopes, the exponents are $2.7\pm0.2$, $2.2\pm0.2$, and $2.0\pm0.1$ for WNM$_{\rm F}$, LNM$_{\rm F}$, and CNM$_{\rm F}$, respectively.

Recognizing the uncertainties, we find that the exponent for WNM$_{\rm F}$ is higher than those of LNM$_{\rm F}$ and CNM$_{\rm F}$.
This trend is consistent with the ranking of the exponents found in Sect.~\ref{sec:sps} for the power spectra of the column density maps (Fig.~\ref{fig:SPS_NHI_CNM_LNM_WNM} and Table~\ref{table:index}), that for WNM$_{\rm F}$ being steeper.

By definition, $M\propto n \,L^3$. The deviations of the above exponents from 3 show that the average number of H atoms per unit volume $n$ is not constant even within a single phase.  Instead, $n$ varies inversely with $L$, with exponents
$-0.3\pm0.2$, $-0.8\pm0.2$, and $-1.0\pm0.1$ for WNM$_{\rm F}$, LNM$_{\rm F}$, and CNM$_{\rm F}$, respectively.
The increase of $n$ from large to small scales is more pronounced in the unstable and cold phases compared to the warm phase, a result of the thermal condensation.

Finally, we have made an empirical estimate of the typical cooling length scale at which the warm gas is non-linearly unstable \citep{audit_thermal_2005}. 
This should be larger than most LNM$_{\rm F}$ and CNM$_{\rm F}$ structures and so from Fig.~\ref{fig:dend_length} $\lambda_{\rm cool} \sim15$\,pc, as marked by the vertical black dashed line in Fig.~\ref{fig:dend_mass_size}.
This spatial scale is five times higher that the spatial resolution of the observation (also the size of the smallest cold structures) and this estimate is only weakly sensitive to the user parameters of the dendrograms. Note also that this result is driven by the largest structures extracted in LNM$_{\rm F}$ (i.e., 14.1\,pc) as opposed to the smallest structures found in WNM$_{\rm F}$. Therefore, it is not affected by the limited spatial resolution of the WNM$_{\rm F}$ map from the convolution applied to suppress noise (see Appendix~\ref{app:dendrogram}).

\subsubsection{Turbulent velocity dispersion --  size relation}
\label{subsec:turbulence}

\begin{figure}[!t]
  \centering
  \includegraphics[width=\linewidth]{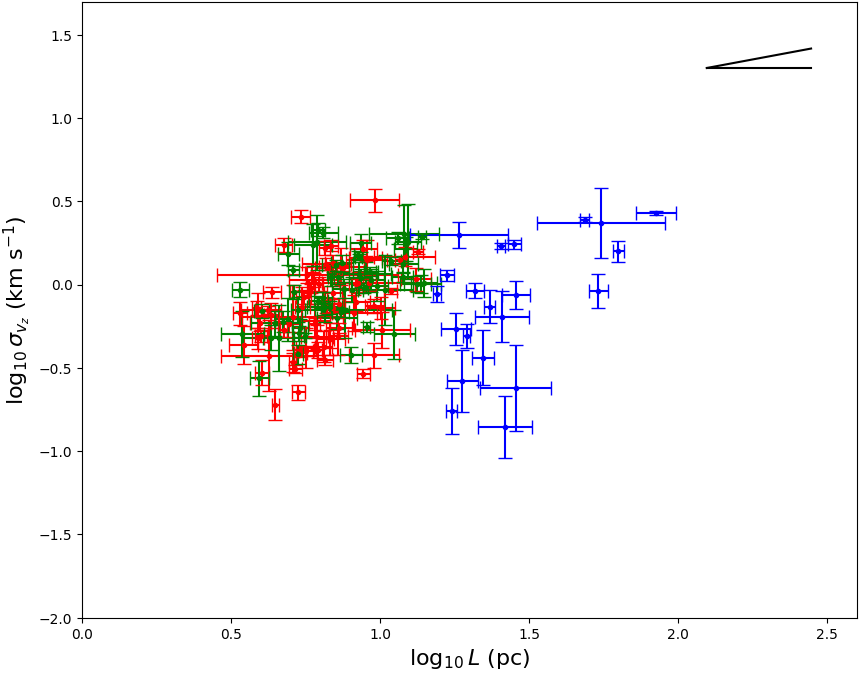}
  \caption{Scatter plot of $\log_{10}\sigma_{v_z}$ vs $\log_{10}L$. Values for structures from CNM$_{\rm F}$, LNM$_{\rm F}$, and WNM$_{\rm F}$ are shown in red, green, and blue, respectively. The Pearson correlation coefficient is 0.27 for the ensemble. The theoretical slope 1/3 of sub/trans-sonic turbulence is illustrated in the upper right corner.}
  \label{fig:dend_sig_vz_size}
\end{figure}

Figure~\ref{fig:dend_sig_vz_size} shows the $\sigma_{v_z}-L$ relation, color coded by phase.  
The two variables are positively correlated, with Pearson correlation coefficient is 0.27.
Although the Pearson coefficient is significantly positive, its value is too low for a reliable determination of the power law exponent.
Given the Mach numbers obtained in Sect.~\ref{subsec:turbulent-properties}, one might expect to see the scaling law of sub/trans-sonic compressible turbulence \citep{kim_density_2005}, close to Kolmogorov's prediction of incompressible turbulence $\sigma_{v_z} \propto L^{1/3}$ \citep{kolmogorov_local_1941}.  This is illustrated in the top right corner of Fig.~\ref{fig:dend_sig_vz_size}.  

\subsubsection{Kinematic viscosity and energy transfer rate -- size relations}
\label{subsec:turbulence}

\begin{figure}[!t]
  \centering
  \includegraphics[width=\linewidth]{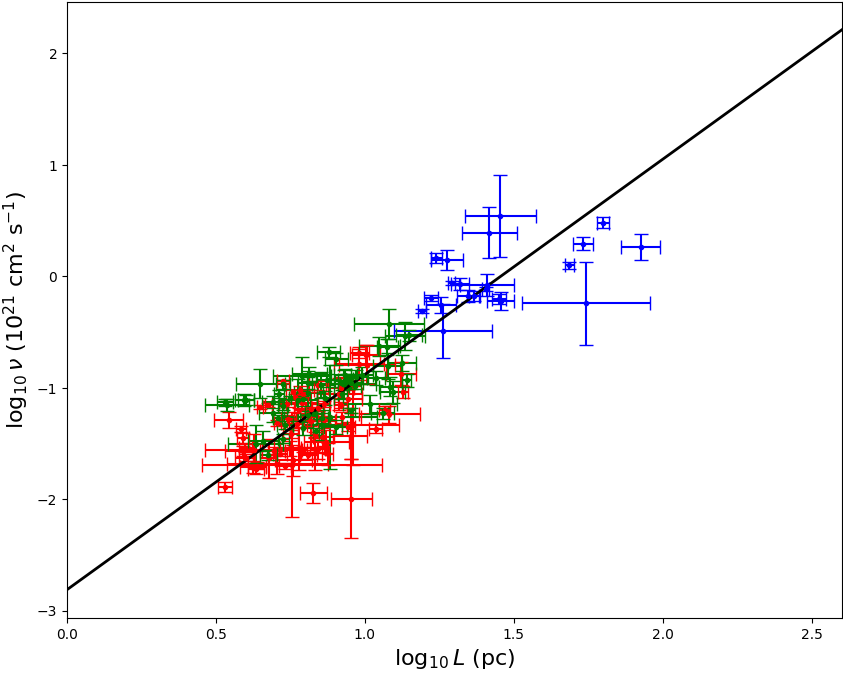}
  \caption{Scatter plot of $\log_{10}\nu$ vs $\log_{10}L$. Values for structures from CNM$_{\rm F}$, LNM$_{\rm F}$, and WNM$_{\rm F}$ are shown in red, green, and blue, respectively. The Pearson correlation coefficient is 0.83 for the ensemble and solid black line shows a power law fit.}
  \label{fig:dend_viscosity_size}
\end{figure}

Figure~\ref{fig:dend_viscosity_size} shows the kinematic viscosity as a function of scale. The Pearson correlation coefficient is 0.83 for the ensemble and a power-law fit using the bisector estimator gives 
exponent $1.9\pm0.1$.
Noting that $\nu \propto T_k^{1/2} n^{-1}$ from Eq.~\ref{eq:visc}, this pronounced scaling is seen to be a direct consequence of the phase transition, which lowers the kinetic temperature and increases the average number of H atoms per unit volume of the fluid as the scale decreases.

The Pearson correlation coefficient of the energy transfer rate -- size scaling law (not plotted) is -0.06, showing that no significant trend is observed.

\subsubsection{Insights into the turbulent cascade}
\label{subsec:turbulent-cascade}

In summary, the physical properties of structures extracted from \env\ F across phases reveal the presence of a sub/trans-sonic turbulent energy cascade. The warm component is the least turbulent phase with $Re=0.62\,(3.9)\times10^4$. 

When the fluid undergoes a phase transition, condensed gas with higher $n$ and lower $T_k$ appears as smaller structures and filaments (see Figs.~\ref{fig:mosaic_field_0_1} (left), \ref{fig:RGB_WCS} (left), \ref{fig:dend_length}, and \ref{fig:dend_aspect_ratio}). This change in the thermodynamic properties of the gas modifies a fundamental property of the turbulence, the kinematic viscosity $\nu$. The scale dependence of $\nu$ (Fig.~\ref{fig:dend_viscosity_size}) increases the relative strength of turbulence from large scales to small scales (see Fig.~\ref{fig:dend_Re_size} for $Re$ and Figs.~\ref{fig:dend_Mach} and \ref{fig:dend_Mach_size} for $\mathcal{M}_s$). There is a progression from subsonic to trans-sonic turbulence, reaching a state characterized by $\mathcal{M}_s=($0.86\,(2.2)$)$ in the coldest phase (Table~\ref{table:turbulence_1}).

Despite this change in the statistical properties of the turbulent cascade, a constant energy transfer rate is observed over scales. This favors a scenario where no energy is injected or dissipated along the energy cascade. In other words, this turbulence appears to be self-similar, even if influenced by a phase transition. 

The traversal time $\tau_L$ decreases from the warm phase to the cold phase (Table~\ref{table:turbulence_1}). This suggests that velocity fluctuations (and therefore density fluctuations) will last longer in the warm phase than in the cold phase. Note that in each phase the traversal time and turbulent crossing time $t_{\rm cross}^{v_z}$ are very similar.

\section{Thermal equilibrium and thermal instability}
\label{sec:discussion}

The following discussion relates to the interpretation of thermal properties of the structures in different phases in \env\ F. While it is difficult to demonstrate unequivocally that the colder and denser condensations that are observed have formed simply by a triggered thermal instability, we find multiple pieces of evidence pointing in this direction.

\subsection{Thermal equilibrium}
\label{subsec:thermal-equilibrum}

\begin{figure*}[!t]
  \centering
  \includegraphics[width=0.49\linewidth]{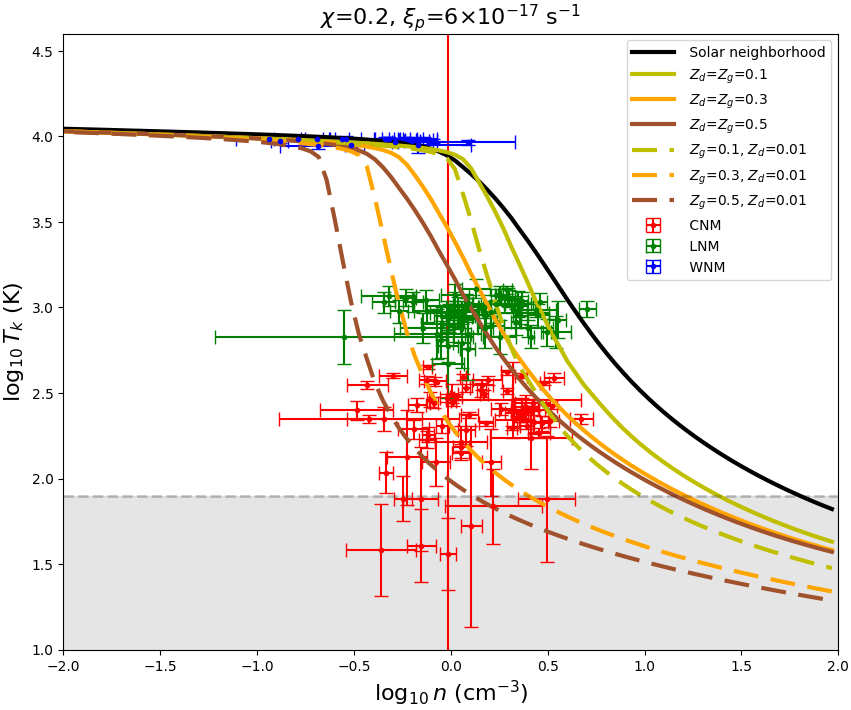}
  \includegraphics[width=0.49\linewidth]{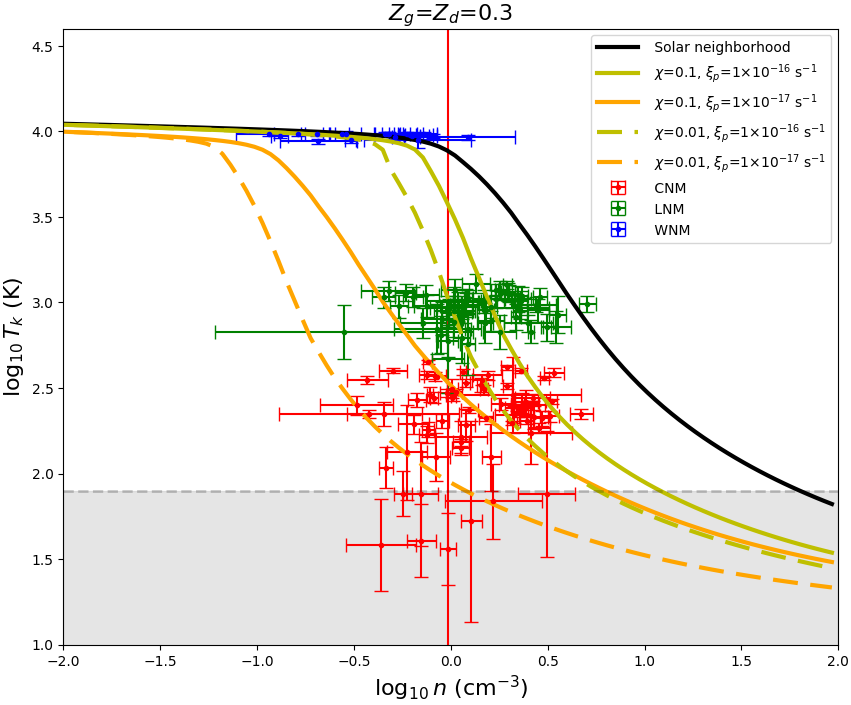}
  \caption{$\log_{10}T_k$ -- $\log_{10}n$ diagram showing values for structures extracted from WNM$_{\rm F}$, LNM$_{\rm F}$, and CNM$_{\rm F}$, in blue, green, and red, respectively. These data are repeated in left and right panels.
  The gray dashed line shows the kinetic temperature corresponding to the channel spacing of the observations.
  Superimposed are theoretical thermal equilibrium curves for a range of model parameters. 
  Left: Colored and dashed lines show curves for varying gas phase metallicity ($Z_g$) and dust-to-gas ratio ($Z_d$), with fixed FUV interstellar radiation field strength ($\chi=0.2$) and the primary CR ionization rate ($\xi_p=6\times10^{-17}$\,s$^{-1}$). The black line shows the equilibrium curve for solar neighborhood conditions ($\chi=1$ and $\xi_p=2\times10^{-16}$\,s$^{-1}$). 
  Right: Colored and dashed lines show curves for varying $\chi$ and $\xi_p$, for a fixed gas phase metallicity and dust-to-gas ratio $Z_g=Z_d=0.3$.
  }
  \label{fig:Zdg_highz_radTn}
\end{figure*}

\subsubsection{Empirical $T_k$ -- $n$ diagram and some caveats}
\label{subsubsec:dataTn}

Figure~\ref{fig:Zdg_highz_radTn} shows the $T_k$ -$n$ diagram for structures extracted from WNM$_{\rm F}$, LNM$_{\rm F}$, and CNM$_{\rm F}$, color coded by phase.  
There are nine structures from CNM$_{\rm F}$ with $T_k$ below the horizontal gray dashed line (see Sect.~\ref{subsec:sepmo}) and these have large uncertainties.

It is instructive to use this figure for comparisons with thermal equilibrium models, but there are some effects that could compromise the comparison in its detail.
First, for each phase the dispersion in $T_k$ is much smaller than that in $n$. For WNM$_{\rm F}$, this is expected because the steep temperature dependence of cooling by collisional excitation of Ly-$\alpha$ constrains the temperature of the warm gas to a narrow range close to $T_k \sim10^4$\,K. However, for the LNM$_{\rm F}$ and CNM$_{\rm F}$, we would not expect the gas to be so closely isothermal for different structures. 
The narrowness of the range of $T_k$ found arises at least in part from our spectral decomposition using {\tt ROHSA}, 
which in enabling phase separation favors a solution with each Gaussian component having a similar Doppler velocity dispersion across the field, as is apparent in Fig.~\ref{fig:heatmap}. 
The non-thermal component of the dispersion is fairly uniform, and so this propagates to the derived uniformity in $T_k$.
Therefore, the ensemble for a given phase, LNM$_{\rm F}$ or CNM$_{\rm F}$, provides a single estimate of the typical temperature for that phase and any temperature variation with density is lost.

Second, the observed $\sigma_{T_b}$ for CNM$_{\rm F}$, which is quite small, might be biased high by two effects, beam smearing (Sect.~\ref{subsubsec:size}) and the finite spectral resolution of the spectrometer.
This bias would propagate such that $T_k$ might then be lower than we have inferred.

Third, as discussed in Sect.~\ref{subsubsec:size}, the limited spatial resolution of the observation is likely to lower the inferred average number of H atoms per unit volume of extracted structures for CNM$_{\rm F}$. Allowing for these last two effects, the red dots for structures from CNM$_{\rm F}$ might tend to be shifted down and/or to the right in Fig.~\ref{fig:Zdg_highz_radTn}.

\subsubsection{Modeling the thermal state in \cib}
\label{subsubsec:modelTn}

We have calculated the thermal state of the gas using the approximation of a static 1D photodissociation region (PDR) model at the location of \cib\ in the Galactic halo. This is admittedly oversimplified, but should be a useful benchmark toward deeper understanding.
In particular, we calculated the thermal equilibrium curve $\mathcal{L}=0$ using the chemical network presented in \cite{gong_2017}, where $\mathcal{L}$ is the net heating and cooling.
In the neutral atomic phase of the ISM,
heating is dominated by photo-electrons from small dust grains and cooling is dominated by collisional excitation of Ly-$\alpha$ and fine structure lines of OI, CII, and CI. 
Input parameters of this PDR model are 
the FUV interstellar radiation field strength $\chi$ 
\citep[in units of the Draine (1978) field strength,][]{draine_1978}, 
the dust abundance $Z_d$ and the gas metallicity $Z_g$ (each relative to the value in the solar neighborhood), and the primary cosmic ray (CR) ionization rate per H atom, $\xi_p$.

To choose plausible values of $\chi$ and $\xi_p$, we first evaluated these parameters at the Galactocentric radius of \cib, $R_{\rm \cib} = 11$~kpc, using table~2 in \citet{wolfire_neutral_2003}.
We then applied the scaling with height $z$ above the Galactic plane using equation~4 in \citet{wolfire_multiphase_1995}, assuming $z=10\,\sin(b_{\rm \cib})$\,kpc. In addition, $\chi$ was multiplied by a factor of 0.6 to lower the midplane intensity to the intensity at the surface of the disk, in order to match the FUV optical depth obtained by \cite{tielens_hollenbach_1985} \citep{wolfire_multiphase_1995}.\footnote{This factor matches observations at the solar Galactocentric radius but could be different at $R_{\rm \cib}$, e.g., due to the Galactic warp.} 
This approach led to $\chi\sim0.2$ and $\xi_p=6\times10^{-17}$\,s$^{-1}$. These values anchor ranges that we have explored.  
We note that an X-ray radiation field that might assume some importance in the Galactic halo is not included in the model. Its potential effects might be mimicked in part by larger values of these two parameters.
 
Three gas phase metal abundances, $Z_g=(0.1,0.3,0.5)$, were selected ranging over estimated metallicities in complex C \citep{gibson_2001,collins_2003,collins_2007,tripp_2003}. The choice of $Z_d$ is more problematic because no dust has been detected in HVCs to date;\footnote{Further unknown is whether the dust size distribution would be the same as in the diffuse ISM.} we explored two possibilities, dust limited by the metallicity, $Z_d=Z_g$, and a much lower value $Z_d=0.01$.

\begin{figure*}[!t]
  \centering
  \includegraphics[width=0.49\linewidth]{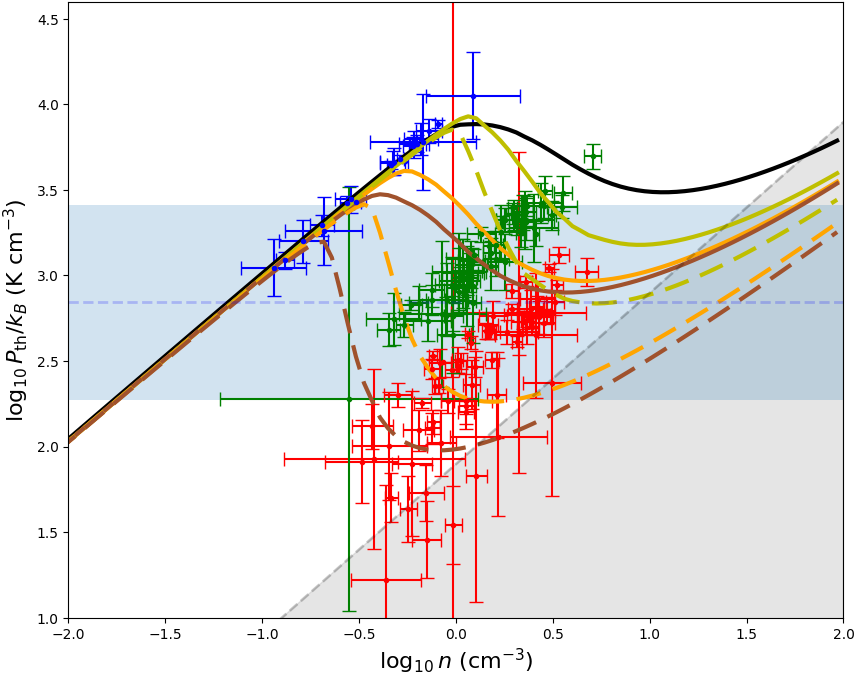}
  \includegraphics[width=0.49\linewidth]{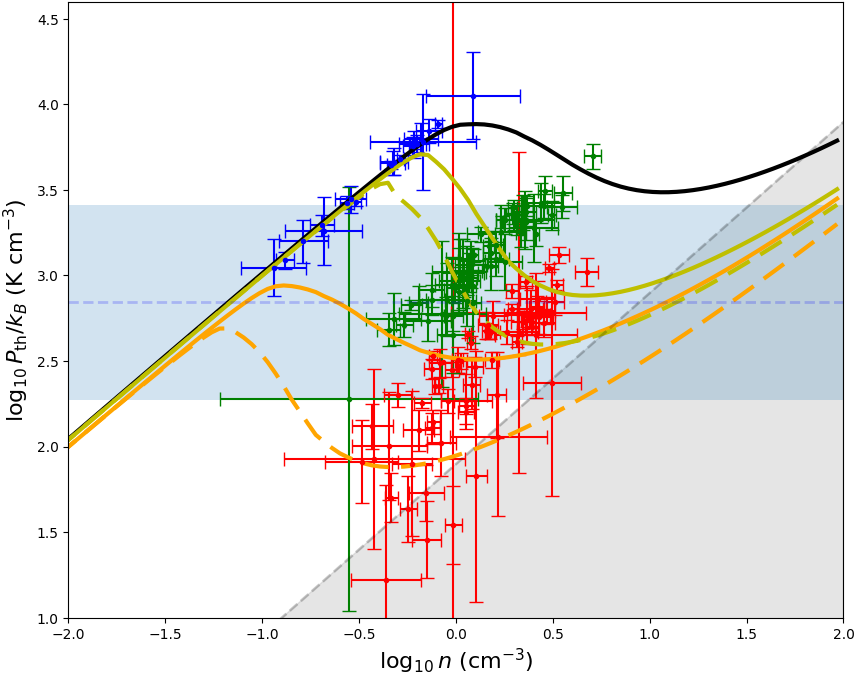}
  \caption{Like Fig.~\ref{fig:Zdg_highz_radTn} but for a $\log_{10}P_{\rm th}/k_B$ -- $\log_{10}n$ diagram.
   The blue horizontal dashed line and surrounding shaded area show the mean and spread of the thermal pressure for the ensemble of structures in \env\ F.
  }
  \label{fig:Zdg_highz_radPn}
\end{figure*}

\subsubsection{Comparison of models and data}
\label{subsubsec:compTn}

Results of our explorations of parameter space are summarized in the two panels in Fig.~\ref{fig:Zdg_highz_radTn}. Any comparisons made with the properties inferred independently from the data (hereafter, data) are subject to the previous caveats.

In the left panel, colored lines and dashed lines show thermal equilibrium curves for varying metallicity and dust abundance, with fixed $\chi=0.2$ and $\xi_p=6\times10^{-17}$\,s$^{-1}$. Solid lines are for different pairs of $Z_g=Z_d= (0.1,0.3,0.5)$. 
The stability of these curves when $Z_g$ and $Z_d$ are decreased simultaneously arises because the photoelectric heating by dust is roughly proportional to $Z_d$ and the gas cooling is roughly proportional to $Z_g$, so that changes in these two processes track each other and the change in $\mathcal{L}$ is small. 
If the dust abundance $Z_d$ is reduced without changing $Z_g$, the photo-electric heating is reduced, leading to lower temperatures (dashed curves). 
In each case, the theoretical thermal equilibrium curve $\mathcal{L} = 0$ is unable to reproduce the data in WNM$_{\rm F}$, LNM$_{\rm F}$, and CNM$_{\rm F}$ simultaneously.
For example, in the case of $Z_g=0.3$ and $Z_d=0.01$ (dashed orange curve), the theoretical $\mathcal{L}=0$ curve shows rough agreement with the average values in CNM$_{\rm F}$ but does not extend to the higher density structures seen in the other phases.

In the right panel, solid and dashed lines show thermal equilibrium curves of varying FUV interstellar radiation field strength ($\chi=0.1$ or $\chi=0.01$) and CR ionization rate ($\xi_p=10^{-16}$\,s$^{-1}$ or $\xi_p=10^{-17}$\,s$^{-1}$), for a fixed metallicity and dust abundance ($Z_g=Z_d=0.3$). 
Reducing $\chi$ directly reduces the photo-electric heating. 
Reducing $\xi_p$ reduces the electron abundance (ionization fraction of the gas), which then also reduces the efficiency of photo-electric heating.
Similar to the left panel, no single theoretical thermal equilibrium curve can reproduce the range of data seen in all three phases. 
The model with $\chi=0.1$ and $\xi_p=10^{-17}$\,s$^{-1}$ (solid orange curve) passing through the cloud of points for CNM$_{\rm F}$, again fails to reproduce the higher densities seen in the other phases.

\subsubsection{Deviation from a single thermal equilibrium curve}
\label{subsubsec:devTn}

Although general trends relating to phase separation are present in static models with heating and cooling mechanism in equilibrium, no single thermal equilibrium curve can reproduce the data. 
There are many failure modes. For example, models that allow a broad range of WNM$_{\rm F}$ densities predict warmer LNM$_{\rm F}$ and CNM$_{\rm F}$ at their inferred densities.

One possible explanation might be local variations in the physical environment ($\chi$ and $\xi_p$) or in the gas ($Z_g$ and $Z_d$).
However, given the relatively low gas surface density ($N_\mathrm{HI} \gtrsim 2\times 10^{20}~\mathrm{cm^{-2}}$, see Fig. \ref{fig:mosaic_field_0_1}) and metallicity, significant variations in $\chi$ and $\xi_p$ from shielding are not expected, except perhaps in the densest parts of CNM$_{\rm F}$ that are not well resolved.
Local variations of metallicities would require a higher $Z_g$ or lower $Z_d$ in LNM$_{\rm F}$ and CNM$_{\rm F}$. Plausible local variations of $Z_g$ might arise if the \HI\ observed in \cib\ is a mixture of the original infalling HVC and the Galactic halo material 
\textcolor{xlinkcolor}{(Heitsch et al. 2021, in preparation)}. 
Local variations of $Z_d$ might be produced by mixing with halo material and/or local destruction of dust due to the systematic motions in the encounter.

Alternatively, the gas might just be out of thermal equilibrium. This possibility is implied by the lower thermal pressure in LNM$_{\rm F}$ and CNM$_{\rm F}$ (see Table~\ref{table:turbulence_1}) and is supported by the fact that the dynamical timescale (thermal crossing times in Table~\ref{table:turbulence_1}) is comparable to the cooling timescale (see also Sect.~\ref{subsec:dynamical}).  
A related caution (or hint) is the widespread presence of the LNM$_{\rm F}$ phase,
which the models suggest is thermally unstable.

\subsection{Thermal instability}
\label{subsec:thermal-instability}

In the ISM, the formation and steady-state presence of non-gravitational cold condensations has been related to the condensation mode of thermal instability \citep{field_thermal_1965}.
To assess the relevance of this physical mechanism to the observed phase transition in EN, we evaluated whether perturbations around the mean thermodynamic state of the gas in \cib\ would allow the condensation mode of thermal instability to develop and then to grow freely. 

\subsubsection{Development of the condensation mode}
\label{subsubsec:devopPn}

In an idealized non-viscous static fluid in thermal equilibrium, the isobaric criterion for development of the condensation mode of thermal instability can be expressed as
\begin{equation}
    \left(\frac{\partial P}{\partial n}\right)_{\mathcal{L} = 0} < 0
    \label{eq:criterion_condensation}
\end{equation}
\citep{field_thermal_1965, wolfire_neutral_1995}.
To assess whether this idealized criterion is satisfied in the mean pressure and density range where the phase separation is observed in EN, the properties extracted from WNM$_{\rm F}$, LNM$_{\rm F}$, and CNM$_{\rm F}$ are presented in the relevant $P_{\rm th}/k_B$ -- $n$ phase diagram in Fig.~\ref{fig:Zdg_highz_radPn}, color coded by phase. 
The mean and spread of the thermal pressure of all structures in \cib\ are indicated by the light blue line and shaded area. This is lower than in standard models of the solar neighborhood ISM \citep[e.g.,][see also the black curves in Fig. \ref{fig:Zdg_highz_radPn}]{wolfire_neutral_1995}. The data also reveal the relevant density range to consider.

Superimposed on the two panels are model thermal equilibrium curves corresponding to those presented in Fig.~\ref{fig:Zdg_highz_radTn}. As in the discussion 
in Sect.~\ref{subsec:thermal-equilibrum}, there is no single model that reproduces the data.
However, in the average number of H atoms per unit volume range observed in EN, where a number of the plausibly relevant models do intersect with the mean pressure in LNM$_{\rm F}$ and CNM$_{\rm F}$ (green and red points), we can see directly that the curves satisfy Eq.~\ref{eq:criterion_condensation}.\footnote{As noted in Sect.~\ref{subsubsec:dataTn}, the apparent distribution of data points from LNM$_{\rm F}$ and CNM$_{\rm F}$ along the diagonal isothermal lines is a consequence of the phase separation performed with \ROHSA. 
Therefore, we can compare the mean pressure and density of LNM$_{\rm F}$ and CNM$_{\rm F}$ but not values among the data points within each phase.
}
On average, from WNM$_{\rm F}$ to LNM$_{\rm F}$ and CNM$_{\rm F}$, the pressure drops and the density increases, also satisfying Eq.~\ref{eq:criterion_condensation}.
Thus, depending on the actual local net cooling of the gas in the \cib\ conditions, pressure perturbations around the average pressure might plausibly trigger the condensation mode of thermal instability, leading to the phase separation observed. 

\subsubsection{Can the condensation mode grow freely?}
\label{subsec:dynamical}

\begin{figure}[!t]
  \centering
  \includegraphics[width=\linewidth]{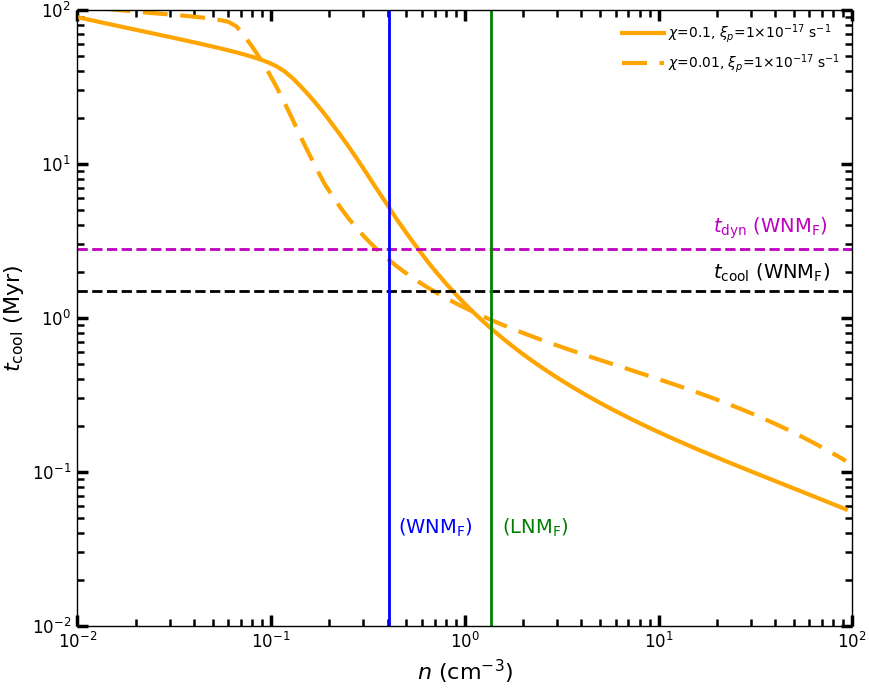}
  \caption{Curves of cooling time versus $n$ for thermal equilibrium models with parameters $\chi=0.2$, $\xi_p=6\times10^{-17}$\,s$^{-1}$, $Z_d=0.01$, $Z_g=0.3$ (orange dashed curve), and $\chi=0.01$, $\xi_p=1\times10^{-17}$\,s$^{-1}$, $Z_d=Z_g=0.3$ (orange solid curve).
  The horizontal black dashed line shows the typical cooling time in the WNM$_{\rm F}$ from Eq.~\ref{eq:tcool}.
  The magenta dashed line shows $t_{\rm dyn}$, the mean thermal crossing time of WNM$_{\rm F}$ (Table~\ref{table:turbulence_1}).
  Vertical lines show the mean density of structures from WNM$_{\rm F}$ and LNM$_{\rm F}$, blue and green respectively.}
  \label{fig:Zdg_tcool}
\end{figure}

A second criterion, i.e., ensuring that the condensation mode can grow freely, involves dynamical and cooling time scales. When the gas experiences a perturbation (i.e., a compression), condensation is possible if the cooling time of the fluid element is shorter than its dynamical time \citep{hennebelle_dynamical_1999}:\footnote{Otherwise, the energy lost by radiation is small when compared to the increase of internal energy and the process can be considered as adiabatic (i.e., no transfer of heat between the fluid element and its surrounding medium).}
\begin{equation}
    \mathcal{I} = t_{\rm cool}/t_{\rm dyn} <  1 \, .
    \label{eq:criterion-dynamical}
\end{equation}
The fact that we observe a phase separation implies that $\mathcal{I} < 1$ is satisfied.  The presence of LNM gas further suggests that the phase transition is ongoing. 

The dynamical time, the typical thermal crossing time of the warm phase WNM$_{\rm F}$, is about 2.8\,Myr (Table~\ref{table:turbulence_1}).  Understanding the balance between time scales requires an assessment of $t_{\rm cool}$ in \cib. 

First, we inferred $t_{\rm cool}$ from
\begin{equation}
    \label{eq:tcool}
    t_{\rm cool}  =  \lambda_{\rm cool} / C_s^{\rm WNM_1} \, .
\end{equation}
where the typical cooling length $\lambda_{\rm cool}$ was estimated in Sect.~\ref{subsec:mass-size-relation} to be about 15\,pc. Using $C_s^{\rm WNM_1}=9.6$\,\kms, we find $t_{\rm cool}\sim1.5$\,Myr. Compared to the dynamical time, $\mathcal{I} < 1$ is satisfied weakly.

Second, we evaluated the theoretical cooling time for the models presented in Sect.~\ref{subsec:thermal-equilibrum}.  A selection of the curves is shown in Fig.~\ref{fig:Zdg_tcool} as a function of $n$.
They correspond to the models with parameters $\chi=0.2$, $\xi_p=6\times10^{-17}$\,s$^{-1}$, $Z_d=0.01$, $Z_g=0.3$ (orange dashed curve as in Fig.~\ref{fig:Zdg_highz_radTn} (left), and $\chi=0.01$, $\xi_p=1\times10^{-17}$\,s$^{-1}$, $Z_d=Z_g=0.3$ (orange solid curve as in Fig.~\ref{fig:Zdg_highz_radTn} right). As discussed above, these models do not reproduce the dispersion observed in the $T_k-n$ diagram but they provide examples of consistent models approximating the average thermodynamic state of the gas in \cib\ and establish the plausibility of thermal instability (Fig.~\ref{fig:Zdg_highz_radPn}).
For comparison, the horizontal black dashed line shows the typical cooling time from Eq.~\ref{eq:tcool}.
The magenta dashed line shows the above estimate of $t_{\rm dyn}$ from the mean thermal crossing time of structures from WNM$_{\rm F}$, which are larger than $\lambda_{\rm cool}$. 
Vertical lines show the mean density of structures from WNM$_{\rm F}$ and LNM$_{\rm F}$, in blue and green respectively (see Table~\ref{table:turbulence_1}), bounding the range of typical densities where condensation is observed.
In that range, these example models have cooling times that are compatible with the Eq.~\ref{eq:tcool} estimate
and also suggestively close enough to $t_{\rm dyn}$ to allow some perturbations to satisfy $\mathcal{I} < 1$.

\section{Origin of substructure in the large-scale context of complex C}
\label{sec:context}

Elongated structures (Fig.~\ref{fig:dend_aspect_ratio}) are a multi-scale property of the flow in \env\ F. Structures and filaments are very well correlated across phases and scales (Figs.~\ref{fig:Cross_SPS_NHI_COMBINED} and \ref{fig:Cross_coeff_NHI}), which indicates that the phase transition is already spatially and anisotropically shaped on large scales. There is no homogeneous warm phase of \HI\ from which the cooler structures condense.

There is some evidence for the importance of thermal instability in the phase transition (Sect.~\ref{sec:discussion}), but thermal equilibrium models are obviously not adequate to describe the complex physical state arising from the interaction of the HVC gas with the Galactic halo.  
Investigation of the mechanisms that are generated at large scales, lead to the observed thermal condensation, and shape the gas will require numerical simulations.

Simulations in turn need to be set up with appropriate geometries and initial conditions. With observations of just small areas like \cib, this would be challenging, even misleading.  
For example, the HVC gas in the N1 field has been classified morphologically to be among the substantial sub-population of compact HVCs that show prominent head-tail structure \citep{bruns_2000}.  However, the observed CNM here occurs in what in that phenomenology is the  ``tail". 
Of more fundamental importance, this gas is not isolated and so it is instructive to consider the context of its place in complex C.

\subsection{Large-scale view from EBHIS}
\label{sec:ebhis}

\begin{figure*}[!t]
  \centering
  \includegraphics[width=0.85\linewidth]{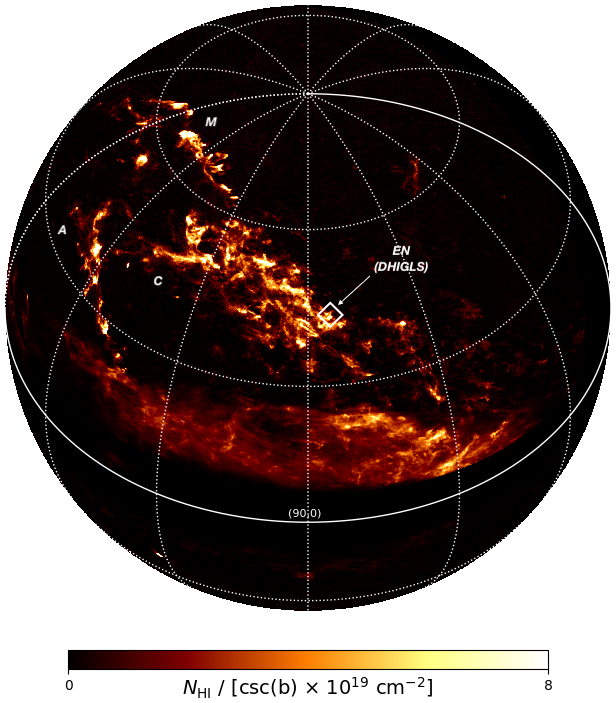}
  \caption{Orthographic projection of \Nh\ of HVC gas selected from EBHIS using $v_{\rm{DEV}} < -80$\,\kms\ ($|v_{\rm{DEV}}| > 80$\,\kms), showing the location of \cib\ along a projected edge. 
  The white box shows the coverage selected from EN similar to Fig.~\ref{fig:NHI_EN_TOT} (right).
  } 
  \label{fig:NHI_EBHIS_nomask_csc_label}
\end{figure*}

\begin{figure*}[!t]
  \centering
  \includegraphics[width=0.85\linewidth]{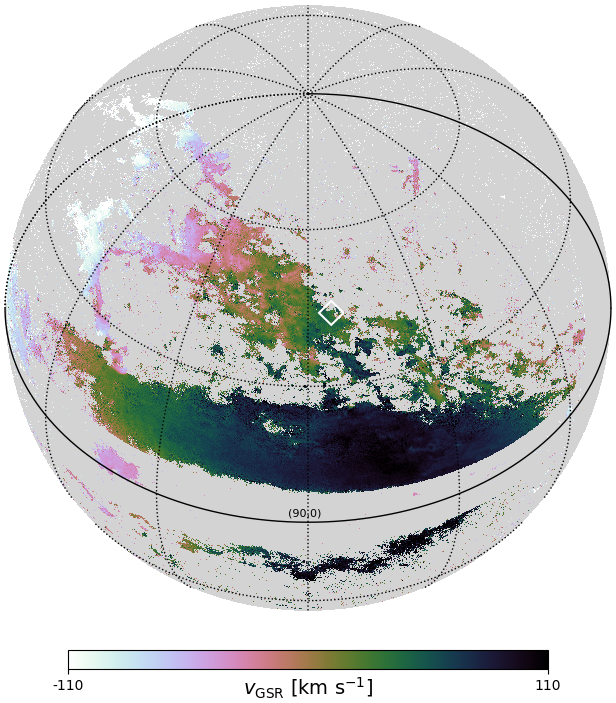}
  \caption{Orthographic projection of the centroid velocity field of the HVC gas in Fig.~\ref{fig:NHI_EBHIS_nomask_csc_label}. Note that the velocity is shown in the Galactic Standard of Rest (GSR). The centroid velocity can be calculated even at low \Nh\ but the gray background corresponds to a mask with $\Nh < 0.5 \times 10^{19}$ cm$^{-2}$.
  The white box shows the coverage selected from EN.
  }
  \label{fig:CV_GSR_EBHIS}
\end{figure*}

This is enabled by the wide sky coverage of the EBHIS data \citep{kerp_2011,winkel_effelsberg-bonn_2016}. Surveyed with the Effelsberg 100-meter radio telescope, EBHIS has a spatial resolution comparable to GHIGLS (10\farcm8 and 9\farcm4 beams, respectively).
To select only the HVC emission, we used the deviation velocity $v_{{\rm DEV}}$ (Appendix~\ref{app:deviation-velocity}), which measures the difference between the observed $v_{\mathrm{LSR}}$ and the predicted velocities of gas throughout a model \HI\ disk.
In complex C and its neighborhood, $v_{\mathrm{LSR}}$ and $v_{\rm{DEV}}$ are negative.  Selecting $v_{\rm{DEV}} < -80$\,\kms\ ($|v_{\rm{DEV}}| > 80$\,\kms) isolates the HVC from the IVC emission (particularly the extended ``Intermediate Velocity Arch'' \citep[IV Arch,][]{kuntz_intermediate-velocity_1996} that would be included at smaller deviations.

Figure~\ref{fig:NHI_EBHIS_nomask_csc_label} shows a map of the column density \Nh\ of the selected HVC emission, in an orthographic projection centred on $(l,\,b)=(90\degree,\, 45\degree)$, close to \cib\ at 
$(l,\, b) = (84\fdg 5,\, 44\fdg 3)$
indicated by the white arrow. The column density is dominated by complex C and its neighbors, complexes A and M. Most of the emission from the Galactic plane has been removed by the $v_{{\rm DEV}}$ cut, but some emission from the Milky-Way extra-planar gas is still visible at $\lvert b\rvert\leq 30\degree$ \citep{marasco_2011}.  The relative prominence of this non-HVC emission is suppressed by imaging $\Nh/\csc b$ here. Also suppressed is the faint extension of complex C to lower $l$ and $b$.\footnote{This has been called the ``tail'' of complex C \citep[][see Sect.~\ref{comparison-previous}]{hsu_2011}.}

As mentioned in Sect.~\ref{subsec:data} and discussed below, \cib\ straddles a projected edge of the main body of complex C. Beyond this edge \Nh\ is very low, about two orders of magnitude less than the brighter features in N1.

Figure~\ref{fig:CV_GSR_EBHIS} maps the centroid velocity (in GSR) of the same selected gas with the same orthographic projection. The centroid velocity can be calculated for low column densities too, leading to a more frothy appearance compared to Fig.~\ref{fig:NHI_EBHIS_nomask_csc_label}. A velocity gradient on large scales is seen across complex C, going from about $50$\,\kms\ at lower longitudes to about $-50$\,\kms\ in the prominent emission adjacent to complexes A and M.

A connection in PPV space between complexes C and A, reflected in Figs.~\ref{fig:NHI_EBHIS_nomask_csc_label} and \ref{fig:CV_GSR_EBHIS}, was reported by \cite{encrenaz_1971} but its faint emission led \cite{wakker_woerden_1991} to catalog them as two distinct entities. In addition to their sub-solar metallicities \citep{kunth_1994,wakker_1996,wakker_distances_2001, gibson_2001,collins_2003,collins_2007,tripp_2003}, their distance estimates of $D_C=10\pm2.5$\,kpc \citep{thom_2008} and $D_A\simeq8-10$\,kpc \citep[][for the high-latitude part of complex A]{wakker_1996,ryans_1997b,van_Woerden_1999,wakker_2003,barger_2012} also support that they could be physically associated.

Curiously, a similar PPV connection apparently links complexes C and M.  However, complex M has a higher, possibly super-solar metallicity \citep{yao_2011} and current estimates place complex M at $D_M \leq 4$ kpc \citep{danly_1993,ryans_1997,yao_2011}, together suggesting that these two complexes are not in fact physically related.  \citet{ryans_1997} suggest that complex M is a part of the Intermediate Velocity Arch.

Clearly, modeling the connections would require further investigation, notably the determination of the 3D orientation of complex C and its neighbors \citep{heitsch_three-dimensional_2016}. A locus of velocities and distances might also constrain the ``orbits" of the gas complexes \citep[e.g.,][for the Smith Cloud]{lockman_2008} and possible related galactic progenitors.

\subsection{The edge of complex C}
\label{subsec:edgeC}

Focusing now on the edge of the main body of complex C probed by the EN data, Figs.~\ref{fig:NHI_EBHIS_nomask_csc_label} and \ref{fig:CV_GSR_EBHIS} show the presence of quasi-periodic scalloping and finger-like structures, of which, importantly, \cib\ is a part. See also the top right panel of Fig.~\ref{fig:NHI_complex_C_zoom_edge_complex_A_M} in Appendix~\ref{app:scalloping-finger-like} for a zoomed view. Note that we deliberately use the terminology ``edge'' to describe what appears in the column density map. This edge corresponds to the projection of the 3D ``boundary'' between complex C and its surrounding environment, which seems devoid of \HI\ and could be warm/hot ionized gas of the Galactic halo or of complex C itself \citep{tufte_1998,haffner_2003,fox_2004}.

These quasi-periodic finger-like structures that stick out beyond the edge strongly resemble the effects expected from hydrodynamic instabilities, i.e., Kelvin-Helmholtz (KH) instabilities or Rayleigh–Taylor (RT) instabilities \citep{chandrasekhar_1961}.
Differentiating between these two possibilities is challenging, requiring the 3D velocity field of both the HVC gas at this edge and that of the adjacent Galactic halo gas (or unseen ionized component of complex C). 
In a simplified model, motion parallel to the boundary in a static halo would favor an interpretation in terms of KH instabilities. On the other hand, a relative velocity perpendicular to the boundary would favor an interpretation as RT instabilities.

A further clue is that other substructure with a similar orientation, appears projected against the eastern body of complex C (see, e.g., the zoom in Fig.~\ref{fig:NHI_complex_C_zoom_edge_complex_A_M}, (top left)), suggesting a complex boundary and interaction in 3D.
Interestingly, similar periodic scalloping and finger-like structures are also seen in complexes A and M (see zooms in Fig.~\ref{fig:NHI_complex_C_zoom_edge_complex_A_M} (bottom) in Appendix~\ref{app:scalloping-finger-like}), again suggesting interaction with warm/hot ionized material. 
Hydrodynamic instabilities are also relevant in producing structure seen in intermediate velocity clouds like the Draco Nebula \citep{miville-deschenes_structure_2017}.
All of these merit further study.

In recent work, \cite{barger_2020} presented a qualitative study of the hydrodynamic instabilities observed in complex A using archival GBT data contained in the GHIGLS NCPL mosaic plus new targeted data.
We had seen similar structures using the EBHIS data (see Fig.~\ref{fig:NHI_EBHIS_nomask_csc_label} and the zoomed in Fig.~\ref{fig:NHI_complex_C_zoom_edge_complex_A_M}, bottom left). The GBT data reach a spatial resolution of 9\farcm1 and a sensitivity of 75\,mK per 0.8\,\kms channel, not unlike EBHIS.
\cite{barger_2020} compared their GBT data to the spatial resolution and sensitivity of the HI4PI survey (16\farcm2 beam, 43\,mK per 1.29\,\kms channel), concluding that targeted observations of HVCs at the higher resolution were needed to resolve hydrodynamic instabilities. However, HI4PI is a combination of EBHIS (10\farcm8 beam) and Galactic All-Sky Survey \citep[GASS; 16\farcm1 beam,][]{mcclure-griffiths_gass:_2009,kalberla_gass:_2010,kalberla_gass:_2015} and for HI4PI the EBHIS data had to be degraded to the resolution of GASS to produce an all-sky survey. Therefore, one should use EBHIS data where available.

Whatever the origin of these large-scale instabilities, they are arguably related to the phase transition found in the EN data. The nature of WNM$_{\rm A}$ and LNM$_{\rm A}$ (see the upper and middle left panel of Fig.~\ref{fig:mosaic_field_0_1}) can now be revisited in this large-scale context. This warm/lukewarm arch follows the local orientation of the edge. Very little CNM is observed along this arch, suggesting that cold structures are preferentially formed along fingers and not in the gas connecting them to the main body. A multiphase analysis using data along the entire edge together with simulations will be useful for understanding the complex details of the connections between scales and the role that instabilities play in the phase transition.

\subsection{Comparison with structures in the ``tail'' of complex C}
\label{comparison-previous}

\cite{hsu_2011} analyzed the physical properties of 79 structures located in what they call the tail\footnote{Here this nomenclature is used in the context of the morphology at large angular scales, not of substructures. It should not necessarily suggest a direction of motion for the complex.}
of complex C, at $(l, \, b) \sim (32\degree, \, 18\degree)$, 
using 4\arcmin-resolution data from the GALFA-HI survey \citep{peek_galfa-hi_2011,peek_galfa-h_2018}.
Assuming the same distance of 10\,kpc \citep{thom_2008}, the masses deduced for these structures spanned $10^{1.1}-10^{4.8}$\,$M_{\odot}$ and the sizes $10^{1.2}-10^{2.6}$\,pc.
Typical line widths of $20-30$\,\kms\ found were characteristic of warm gas. 

Unlike in \cib, where cold structures of both smaller mass and size are found using EN data ($\sim2$\,$M_{\odot}$ and $\sim3$\,pc, respectively), thermal condensation does not seem to be occurring in this part of the complex.
The lower spatial resolution relative to DHIGLS seems unlikely to explain this difference.  CNM structures are identified in \cib\ using GHIGLS N1 data at 9\farcm 4 resolution (Appendix~\ref{app:GHIGLS}) and so GALFA-HI data at 4\arcmin\ resolution ought to have revealed similar cold structures if they were present.

\cite{hsu_2011} suggested that this lack of multiphase structure could be related to the low metallicity of complex C (Sect.~\ref{subsubsec:modelTn}), 
which would increase the cooling time so that it is long compared to the typical lifetime of a structure; if that lifetime is a proxy for the typical dynamical time, then the condensation mode of thermal instability cannot grow freely (Sect.~\ref{subsec:dynamical}).

At least for the conditions found in \cib\ above, where the phase transition is observed, cooling does seem to be sufficiently rapid.  Perhaps the metallicity is even lower in the tail, for example, due to a higher mixing ratio of the original HVC gas with a (perhaps counter-intuitively) lower-metallicity Galactic halo gas 
\textcolor{xlinkcolor}{(Heitsch et al. 2021, in preparation)}. Alternatively, perhaps because of some details of the interaction there is a shorter dynamical time for these structures in the tail preventing the condensation mode from growing freely.
The actual situation is undoubtedly complex and, as we began in the introduction of Sect.~\ref{sec:context}, we conclude by emphasizing the importance of simulations and understanding the basic geometry and environmental context of the underlying interaction.

\section{Summary}
\label{sec:summary}
Our novel study of the multiphase and multi-scale properties of the concentration \cib\ of HVC complex C is based on high spatial resolution \HI\ spectra from DHIGLS. 
\ROHSA\ was used to decompose the spectra and produce maps of the column density, centroid velocity, and velocity dispersion of the multiphase gas. 
In one of two physical regions, \env\ F, the three thermal phases, WNM$_{\rm F}$, LNM$_{\rm F}$, and CNM$_{\rm F}$, are well correlated, associated with the thermal phase transition. 
Multi-scale properties of phases within each region have been quantified by a power spectrum analysis.
We used dendrograms to perform a hierarchical segmentation of column density maps of the phases and analysed physical properties of the ensembles of structures.
As a benchmark of the physical environment at the location of the condensation \cib, we used series of PDR models that compute the chemical and thermal properties of the gas in different environments.  
Building on this, we evaluated whether perturbations around the mean thermodynamic state of the gas allow the condensation mode of thermal instability to develop and grow freely.
Finally, we investigated the large-scale context of \cib\ within complex C using \HI\ data from EBHIS. 

We conclude that there is an ongoing phase transition in \env\ F, located along a pronounced edge of complex C. The thermal condensation proceeds from large to small scales and the cold phase has more small-scale structure:
\begin{itemize}
    \itemsep-0.2em
    \item Values corresponding to the mean and spread of the logarithmic PDFs of the average density and kinetic temperature of structures in the \HI\ phases are 0.41\,(1.9)\,cm$^{-3}$ and  $9.5\,(1.0)\times10^3$\,K (WNM$_{\rm F}$), 1.4\,(1.8)\,cm$^{-3}$ and $0.90\,(1.2)\times10^3$\,K (LNM$_{\rm F}$), and 
    1.3\,(1.9)\,cm$^{-3}$ and $0.20\,(2.1)\times10^3$\,K (CNM$_{\rm F}$).
    Corresponding values for the mass are
    $296\,(3.4)\,M_{\odot}$, 
    $18\,(2.3)\,M_{\odot}$, and 
    $12\,(1.9)\,M_{\odot}$,
    and for the size 
    $28\,(1.6)$\,pc, 
    $7.3\,(1.4)$\,pc, and 
    $6.4\,(1.4)$\,pc.
    \item The angular power spectrum of the WNM$_{\rm F}$ column density map is significantly steeper than for LNM$_{\rm F}$ and CNM$_{\rm F}$. The same trend is observed in the slopes of the mass--size relation of the structures. 
\end{itemize}    

The turbulent energy cascade in the \cib\ gas is well described by compressible sub/trans-sonic turbulence:
\begin{itemize}
    \itemsep-0.2em
    \item From the warm phase to cold phase, both the turbulent Reynolds number and the Mach number increase.
    \item Nevertheless, a constant energy transfer rate is observed over scales, suggesting that energy is neither injected nor dissipated along the energy cascade.
\end{itemize}

Our simplified modeling of the thermal state in \cib\ supports the plausible relevance of the condensation mode of thermal instability:
\begin{itemize}
    \itemsep-0.2em
    \item There could be local variations of gas properties of the physical environment ($Z_g$, $Z_d$, $\chi$, and $\xi_p$), and/or deviation from thermal and dynamical equilibrium.
    \item Nevertheless, the mean thermodynamic properties of the gas across different phases are suggestive of the development of the condensation mode of thermal instability.
    \item The typical scale at which the gas is unstable is about 15\,pc, corresponding to a typical cooling time about 1.5\,Myr.  This is suggestively low enough compared to the mean thermal crossing time in WNM$_{\rm F}$ (2.8\,Myr) to allow the thermal condensations initiated by some perturbations to grow freely.
\end{itemize}

Clues to the triggering of the thermal instability require a large scale context.
The large scale view of complex C suggests that the prominent protrusion in \env\ F, extending from the edge of the complex as the condensation \cib, is the result of a hydrodynamic instability (KH or RT) at the interface. Other similar ``fingers" are observed along the edge in complex C and in other complexes. Understanding the complex and intricate connections between scales and the role that these instabilities play in this phase transition will require further investigation through comparison between observations and numerical simulations.

\acknowledgments

This work was supported by the Natural Sciences and Engineering Research Council of Canada (NSERC).
This work made use of the NASA Astrophysics Data System.
This work began under the program ``The Self-Organized Star Formation Process'' organized and hosted by 
Institut Pascal at Universit\'e Paris-Saclay and the Interstellar Institute.  We thank M.-A. Miville-Desch\^enes, J.E.G. Peek, and other participants for enlightening conversations. MG acknowledges the support of Paola Caselli and the Max-Planck society. We thank the anonymous referee whose comments and suggestions have improved this manuscript.

\software{matplotlib \citep{hunter_2007}, NumPy \citep{van_der_walt_2011}; Astropy\footnote{\url{http://www.astropy.org}}, a community-developed core Python package for Astronomy \citep{astropy_2013, astropy_2018}; \textit{astrodendro}\footnote{\url{http://www.dendrograms.org/}}, a Python package to compute dendrograms of Astronomical data; the HEALPix package \citep{gorski_2005}; and galpy\footnote{\url{http://github.com/jobovy/galpy}} \citep{bovy_2015}.}

\clearpage
\appendix

\section{Decomposition of the N1 field from GHIGLS}
\label{app:GHIGLS}

\begin{figure}[!t]
  \centering
  \includegraphics[width=0.5\linewidth]{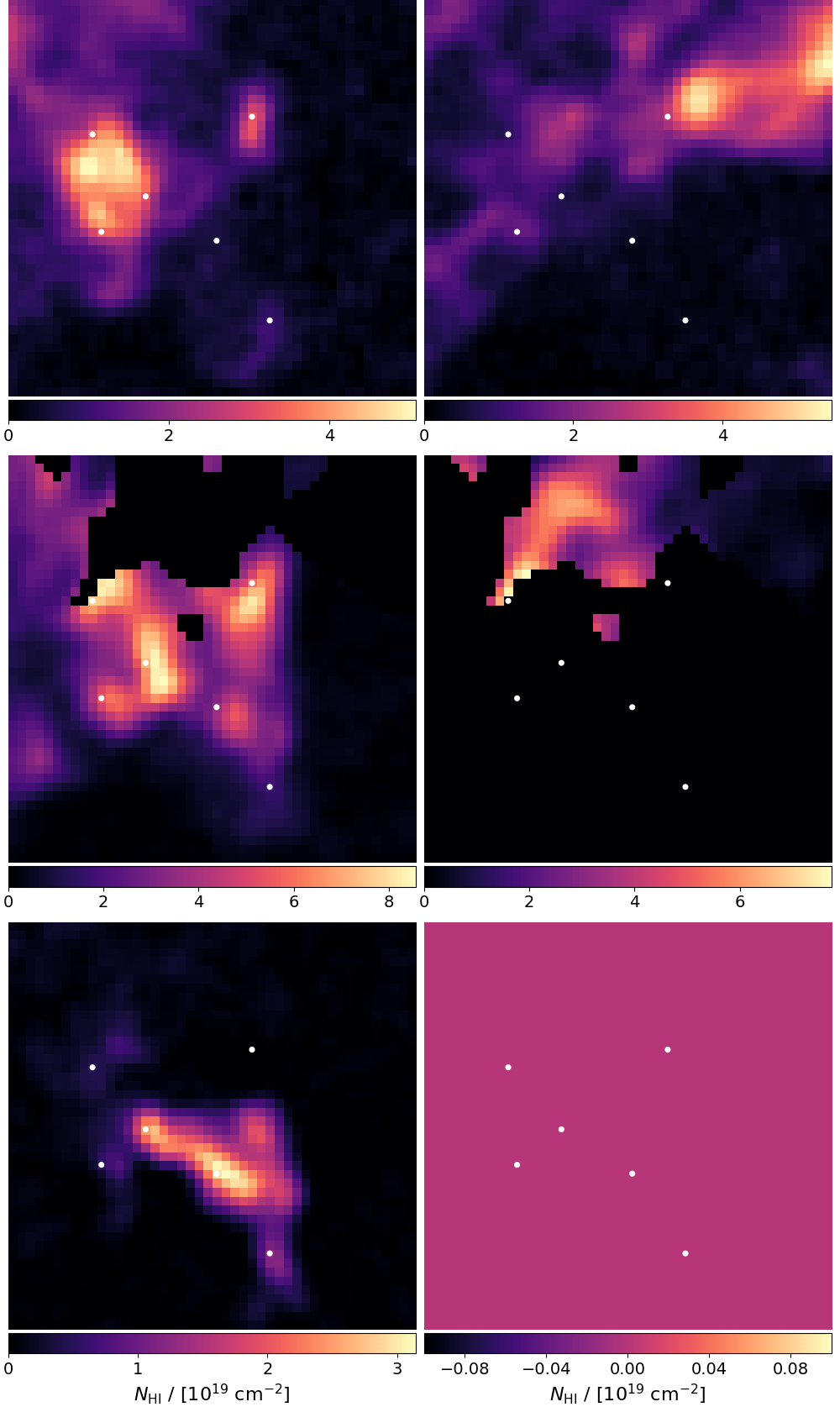}
  \caption{Column density maps of the six HVC phase components
          from \ROHSA\ from N1 data, to be compared with the higher resolution Fig.~\ref{fig:mosaic_field_0_1} from EN data. 
          (left): WNM$_{\rm F}$ (top), LNM$_{\rm F}$ (middle), which is actually quite warm (Table~\ref{table:mean_var_DHIGLS_GHIGLS}), and CNM$_{\rm F}$ (bottom); (right): WNM$_{\rm A}$ (top), LNM$_{\rm A}$ (middle), and CNM$_{\rm A}$ (bottom). Note that color bars have different scales.
          Coordinates (not shown here) are the same as in Fig.~\ref{fig:NHI_EN_TOT} (left).
          White circles indicate the positions of the six spectra shown in Fig.~\ref{fig:mosaic_spectra_all}.
          }  
  \label{fig:mosaic_field_N1}
\end{figure}

As summarized in Sect.~\ref{subsubsec:decomposition-resolution}, to evaluate the impact of the spatial resolution we performed a decomposition of N1 HVC spectra from the GHIGLS survey, using $N=6$ and all hyper-parameters equal to those used for the decomposition of the GHIGLS/N1 data.
In this case, \ROHSA\ also converges toward four components (Table~\ref{table:mean_var_DHIGLS_GHIGLS}). We applied the same procedure as for the DHIGLS/EN data to find the unstable and cold gas associated with regions F and A.

Figure~\ref{fig:mosaic_field_N1} shows column density maps of the six phase components encoding the HVC in \cib\ (N1). Although at lower resolution here, phase structure as seen for the DHIGLS/EN data in  
Fig.~\ref{fig:mosaic_field_0_1} can be recognized.
Importantly, the phase separation in \env\ F shows evidence for cold gas (CNM$_{\rm F}$), albeit at a somewhat higher velocity dispersion because of beam smearing (see Sect.~\ref{subsubsec:decomposition-resolution}, Table~\ref{table:mean_var_DHIGLS_GHIGLS}).
On the other hand, the small amount of emission modeled in CNM$_{\rm A}$ using EN data is not present using N1 data (see lower right panel in Fig.~\ref{fig:mosaic_field_N1}).

\section{Segmentation of \Nh\ maps using dendrograms}
\label{app:dendrogram}

\begin{figure*}[!t]
  \centering
  \includegraphics[width=0.7\linewidth]{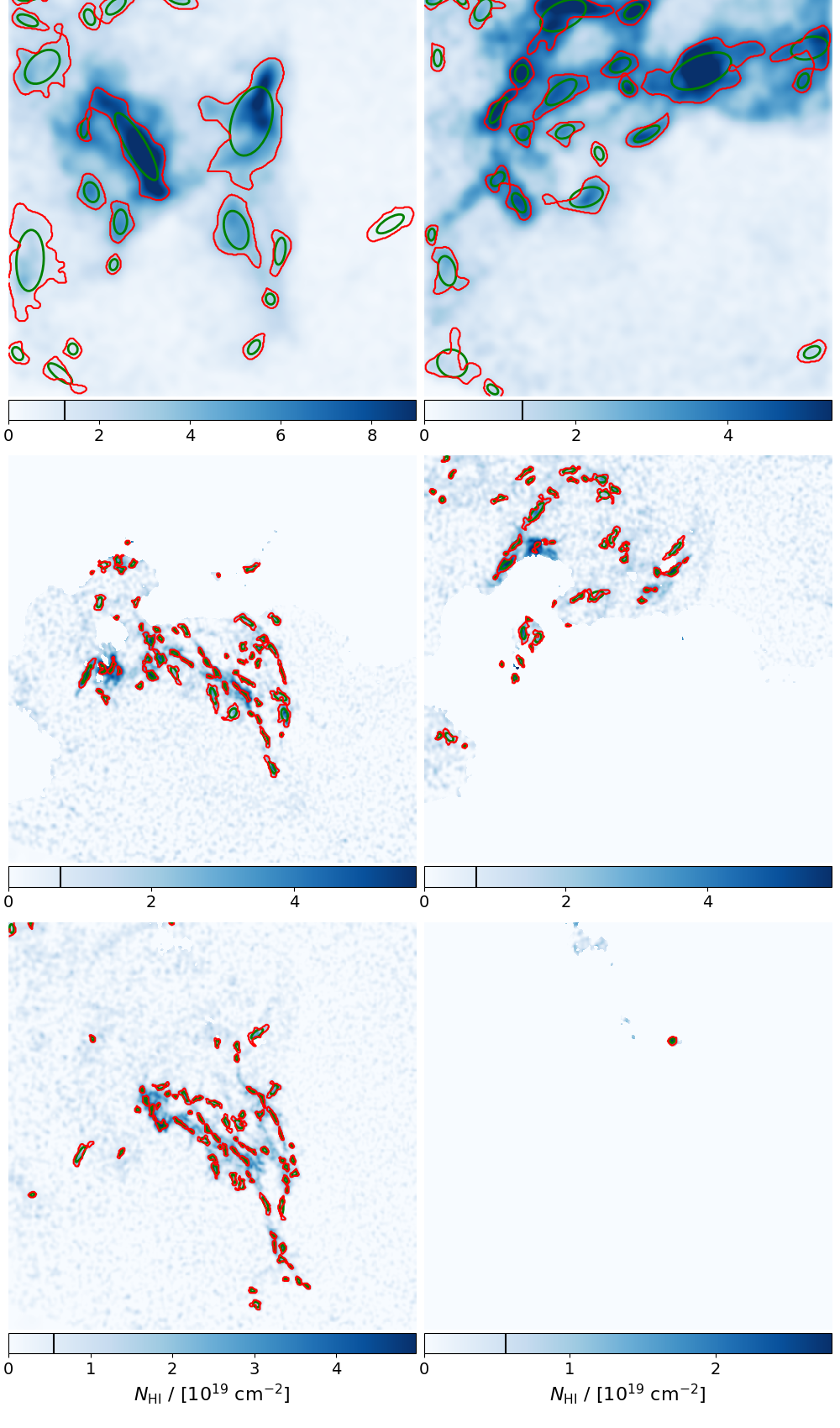}
  \caption{Structures extracted in the three HVC phases in each of two regions identified in \cib\ from EN using \textit{astrodendro}. These are shown over the parent map of \Nh, as red contours over a  blue color map (rather than that in Fig.~\ref{fig:mosaic_field_0_1}) to highlight the clustering results. Corresponding ellipses are superimposed in green.
  (left): WNM$_{\rm F}$ (top), LNM$_{\rm F}$ (middle), and CNM$_{\rm F}$ (bottom); (right): WNM$_{\rm A}$ (top), LNM$_{\rm A}$ (middle), and CNM$_{\rm A}$ (bottom).
  The black mark on the color bars shows the sensitivity limits ($3\sigma$) tabulated in Table~\ref{table:detection_limits}. For WNM$_{\rm F}$ and WNM$_{\rm A}$, this has not been corrected for the convolution applied to suppress noise.
  Coordinates (not shown here) are the same as in Fig.~\ref{fig:NHI_EN_TOT} (left).
  }
  \label{fig:mosaic_field_0_1_dendrograms}
\end{figure*}

Using \textit{astrodendro} we obtained a segmentation via hierarchical clustering in maps of \Nh\ from EN data for  WNM$_{\rm F}$, LNM$_{\rm F}$, and CNM$_{\rm F}$, and for WNM$_{\rm A}$, LNM$_{\rm A}$, and CNM$_{\rm A}$.
To suppress noise, maps of WNM$_{\rm F}$ and WNM$_{\rm A}$ were convolved first to 4\farcm4, four times the native spatial resolution of EN data.
For the six phases, Fig.~\ref{fig:mosaic_field_0_1_dendrograms} shows the structures obtained overlaid on the respective parent \Nh\ map.

Table~\ref{table:astrodendro} summarizes the values of the three user-selected parameters.  For each LNM and CNM phase these are the same:
\textit{min$_{\rm value}$}, the threshold in \Nh\ below which data are ignored, set at twice the sensitivity limit (see Table~\ref{table:detection_limits}); 
\textit{min$_{\rm delta}$}, the minimum height in \Nh\ required for a structure to be retained, set at the detection limit; 
and \textit{min$_{\rm npix}$}, the minimum number of pixels required for an independent structure, set at 16 pixels ($\sim1.2$ times the size of the synthesized beam). 
Due to the convolution applied to WNM$_{\rm F}$ and WNM$_{\rm A}$, lowering the sensitivity limits from those tabulated in Table~\ref{table:detection_limits}, both min$_{\rm value}$ and min$_{\rm delta}$ were chosen manually to ensure a consistent visual clustering of the data. On finding the resulting clustering, the original WNM maps were used to infer the properties of the structures.

The convolution applied to WNM$_{\rm F}$ and WNM$_{\rm A}$, to lower the noise, increases the size of the smallest structure that could be extracted. Although our methodology provides a realistic segmentation of the WNM maps, smaller warm structures ($<15$\,pc) might be missed. However, such structures would be smaller than the larger unstable (LNM) structures extracted and their origin would more likely be attributed to turbulent cascade rather than thermal condensation.

\begin{deluxetable}{lccc}
\tablecaption{User parameters for \textit{astrodendro}}
\label{table:astrodendro}
\tablehead{
\nocolhead{}  & \colhead{\textit{min$_{\rm value}$}} & \colhead{\textit{min$_{\rm delta}$} } & \colhead{\textit{min$_{\rm npix}$}}\\
\nocolhead{}  & \colhead{10$^{19}$ cm$^{-2}$} & \colhead{10$^{19}$ cm$^{-2}$} & \nocolhead{}
}
\startdata
WNM$_{\rm F}$ & 0.5 & 0.5 & 16 \\
LNM$_{\rm F}$ & $2\times\Nh^{\rm lim}$ & $\Nh^{\rm lim}$ & 16 \\
CNM$_{\rm F}$ & $2\times\Nh^{\rm lim}$ & $\Nh^{\rm lim}$ & 16 \\
\hline
WNM$_{\rm A}$ & 1 & 0.5 & 16 \\
LNM$_{\rm A}$ & $2\times\Nh^{\rm lim}$ & $\Nh^{\rm lim}$ & 16 \\
CNM$_{\rm A}$ & $2\times\Nh^{\rm lim}$ & $\Nh^{\rm lim}$ & 16 \\
\enddata
\end{deluxetable}

\section{Properties of structures from segmentation of \Nh\ maps in environment A} \label{app:properties_env_0}

Table~\ref{table:turbulence_1} in Sect.~\ref{sec:dendrogram} summarizes the results from the dendrogram analysis for \env\ F.
Here, for completeness and comparison, Table~\ref{table:turbulence_0} provides a summary for \env\ A.  

\begin{figure}[!t]
  \centering
  \includegraphics[width=0.45\linewidth]{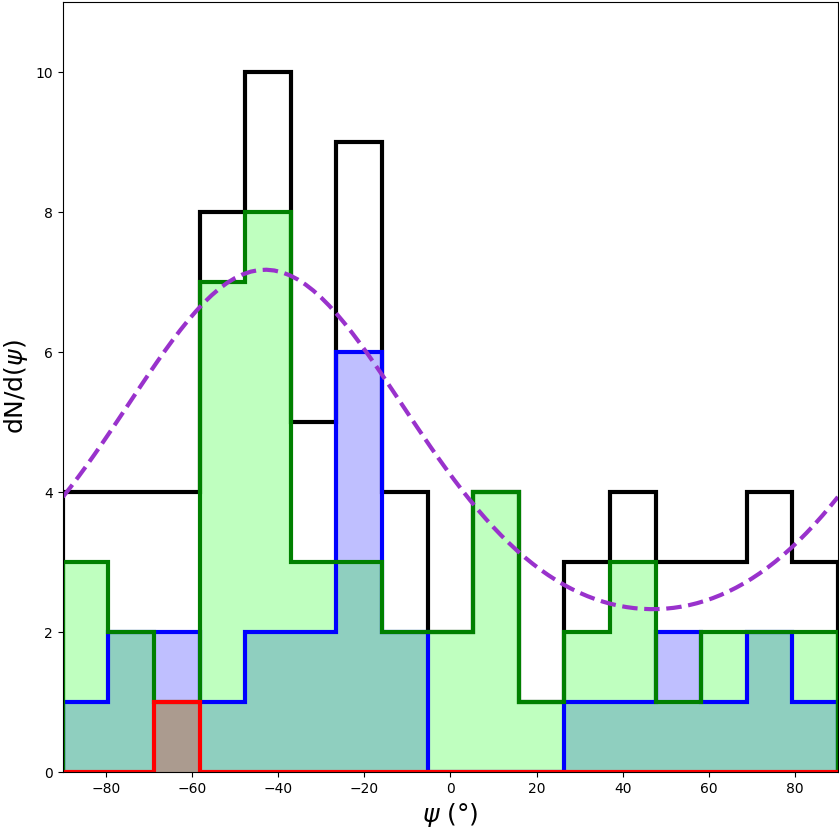}
  \caption{Probability distribution function of the position angle $\psi$ of each structure extracted from WNM$_{\rm A}$ (blue), LNM$_{\rm A}$ (green), and CNM$_{\rm A}$ (red). The total is shown in black. The purple dashed line shows the MLE fit of the Von Mises distribution to the orientations of all structures.}
  \label{fig:dend_orientation_0}
\end{figure}

Figure~\ref{fig:dend_orientation_0} shows the orientation of the dendrogram structures found for \env\ A. In contrast to Fig.~\ref{fig:dend_orientation_1} for \env\ F, where the preferred orientation was $+63\degree$, here the preferred orientation in almost orthogonal, $-43\degree$.   The latter is interestingly similar to the orientation of the edge defined in Sect.~\ref{subsec:data} ($-60\degree$) and the arc of gas in WNM$_{\rm A}$ (Sect.~\ref{subsec:general-description}).

\begin{deluxetable}{lccccccc}
\tablecaption{Properties of structures in \env\ A}
\label{table:turbulence_0}
\tablewidth{0pt}
\tablehead{
\nocolhead{\textbf{Quantity}} & \colhead{\textbf{Symbol}} & \colhead{\textbf{WNM$_{\rm A}$}} & \colhead{\textbf{LNM$_{\rm A}$}} & \colhead{\textbf{CNM$_{\rm A}$}} & \colhead{\textbf{Total$_{\rm A}$}} & \colhead{\textbf{Units}} }
\startdata
 & $N_c$ & 26 & 48 & 1 & 75\\
\hline
\textit{\textbf{Physical}}     &                            &                           & && &                    
\\
Size & $L$ & 26\,(1.5) & 7.2\,(1.5) & 7.9\,(1.0) & 11.3\,(2.1) & pc \\
Aspect ratio & $r$ & 1.7\,(1.3) & 2.0\,(1.4) & 1.1\,(1.0) & 1.9\,(1.4) & \\
Mass & $\mh$ & 260\,(2.6) & 16\,(2.3) & 19\,(1.0) & 42\,(4.9) & M$_{\odot}$  \\
Average number of H atoms per unit volume & $n$ & 0.41\,(1.9) & 1.21\,(2.0) & 1.1\,(1.0) & 0.84\,(2.3) & cm$^{-3}$ \\
\hline
\textit{\textbf{Thermodynamic}}     &  &  &  & & &  \\
Doppler velocity dispersion & $\sigma_{T_b}$ & 9.8\,(1.0) & 3.1\,(1.1) & 2.1\,(1.0) & 4.6\,(1.8) & \kms \\
Turbulent velocity dispersion & $\sigma_{v_{z}}$ & 0.81\,(2.5) & 0.60\,(2.0) & 1.5\,(1.0) & 0.67\,(2.2) & \kms  \\
Thermal velocity dispersion & $\sigma_{\rm th}$ & 9.7\,(1.0) & 2.9\,(1.1) & 1.4\,(1.0) & 4.4\,(1.8) & \kms \\
Kinetic temperature & $T_k$ & 11\,(1.0) & 1.0\,(1.2) & 0.24\,(1.0) & 2.4\,(3.2) & 10$^{3}$\,K \\
Sound speed & $C_s$ & 11\,(1.0) & 3.2\,(1.1) & 1.6\,(1.0) & 4.8\,(1.8) & \kms \\
Thermal crossing time & $t_{\rm cross}^{\rm th}$ & 2.4\,(1.5) & 2.2\,(1.6) & 5.1\,(1.0) & 2.3\,(1.6) & Myr \\
Turbulent crossing time & $t_{\rm cross}^{v_z}$ & 32\,(2.1) & 12\,(1.8) & 5.1\,(1.0) & 17\,(2.2) & Myr \\
Thermal pressure & $P_{\rm th}/k_B$ & 4.8\,(1.9) & 1.3\,(1.9) & 0.3\,(1.0) & 2.0\,(2.5) & 10$^{3}$\,K\,cm$^{-3}$ \\
Turbulent pressure & $P_{\rm v_z}/k_B$ & 0.1\,(6.7) & 0.2\,(3.8) & 1.0\,(1.0) & 0.1\,(4.9) & 10$^{1}$\,K\,cm$^{-3}$ \\
Total pressure & $P_{\rm tot}/k_B$ & 5.6\,(1.9) & 1.7\,(1.9) & 1.3\,(1.0) & 2.6\,(2.3)  & 10$^{3}$\,K\,cm$^{-3}$ \\
\hline
\textit{\textbf{Turbulent cascade}} &                            & &&&&                             \\
Turbulent sonic Mach number & $\mathcal{M}_s$ & 0.13\,(2.5) & 0.32\,(2.0) & 1.7\,(1.0) & 0.24\,(2.5) & \\
Mean free path  & $\lambda$ & 0.78\,(1.9) & 0.27\,(2.0) & 0.30\,(1.1) & 0.39\,(2.3) & 10$^{-3}$\,pc \\
Kinematic molecular viscosity & $\nu$ & 1.1\,(1.9) & 0.11\,(2.0) & 0.06\,(1.0) & 0.24\,(3.6) & 10$^{21}$\,cm$^2$\,s$^{-1}$ \\
Knudsen number  & $K_n$ & 3.0\,(1.6) & 3.7\,(1.5) & 3.7\,(1.0) & 3.4\,(1.6) & 10$^{-5}$ \\
Reynolds number & $Re$  & 0.56\,(3.4) & 1.1\,(2.4) & 5.8\,(1.0) & 0.89\,(2.9) & 10$^4$   \\
Dissipation scale & $\eta$ & 4.1\,(2.2) & 0.68\,(2.0) & 0.21\,(1.0) & 1.2(3.1) & 10$^{-2}$\,pc \\
Dissipation time & $t_{\eta}$ & 4.7\,(3.8) & 1.3\,(2.6) & 0.24\,(1.0) & 2.0\,(3.6) & 10$^{-1}$\,Myr \\
Convective time & $t_{L}$ & 35\,(2.1) & 13\,(1.8) & 6\,(1.0) & 18\,(2.2) & Myr \\
Traversal time & $\tau_L$ & 36\,(2.1) & 13\,(1.8) & 6\,(1.0) & 18\,(2.3) & Myr \\
Energy transfer rate & $\epsilon$ & 0.25\,(12) & 0.35\,(6) & 5.3\,(1) & 0.32\,(8) & 10$^{-5}$\,L$_{\odot}$\,M$_{\odot}^{-1}$ \\
\enddata
\tablecomments{Values correspond to the mean and spread from the logarithmic PDF.}
\end{deluxetable}

\begin{figure*}[!t]
  \centering
  \includegraphics[width=0.42\linewidth]{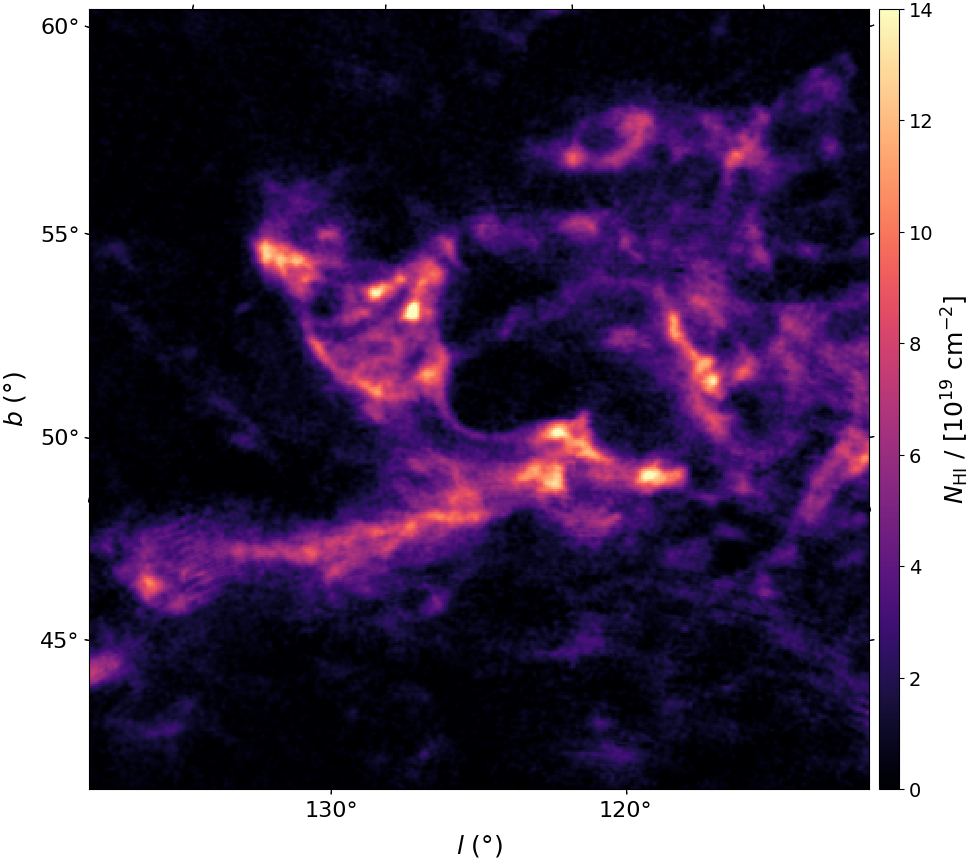}
  \includegraphics[width=0.42\linewidth]{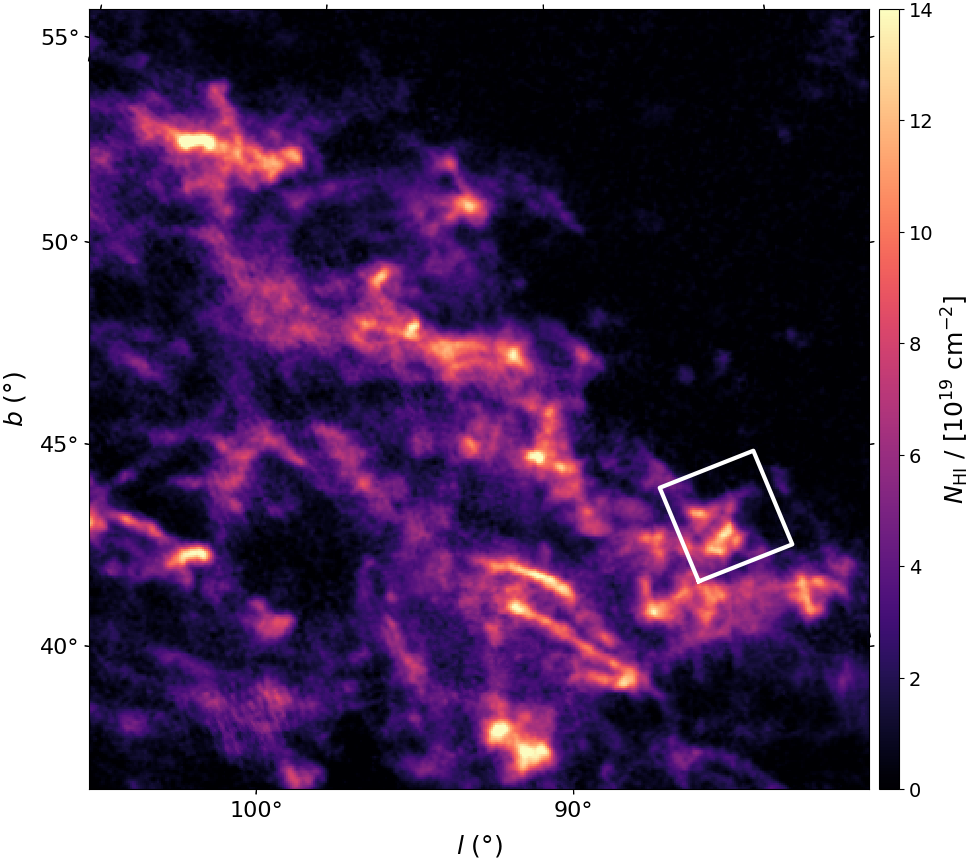}
  \includegraphics[width=0.42\linewidth]{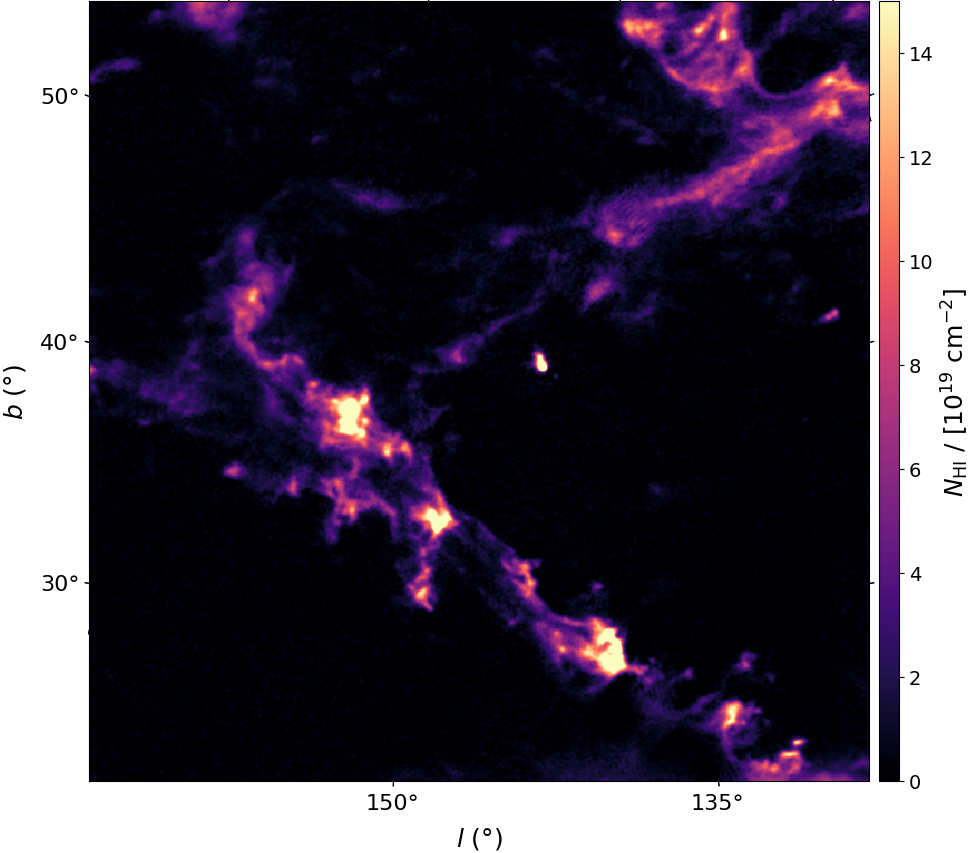}  \includegraphics[width=0.42\linewidth]{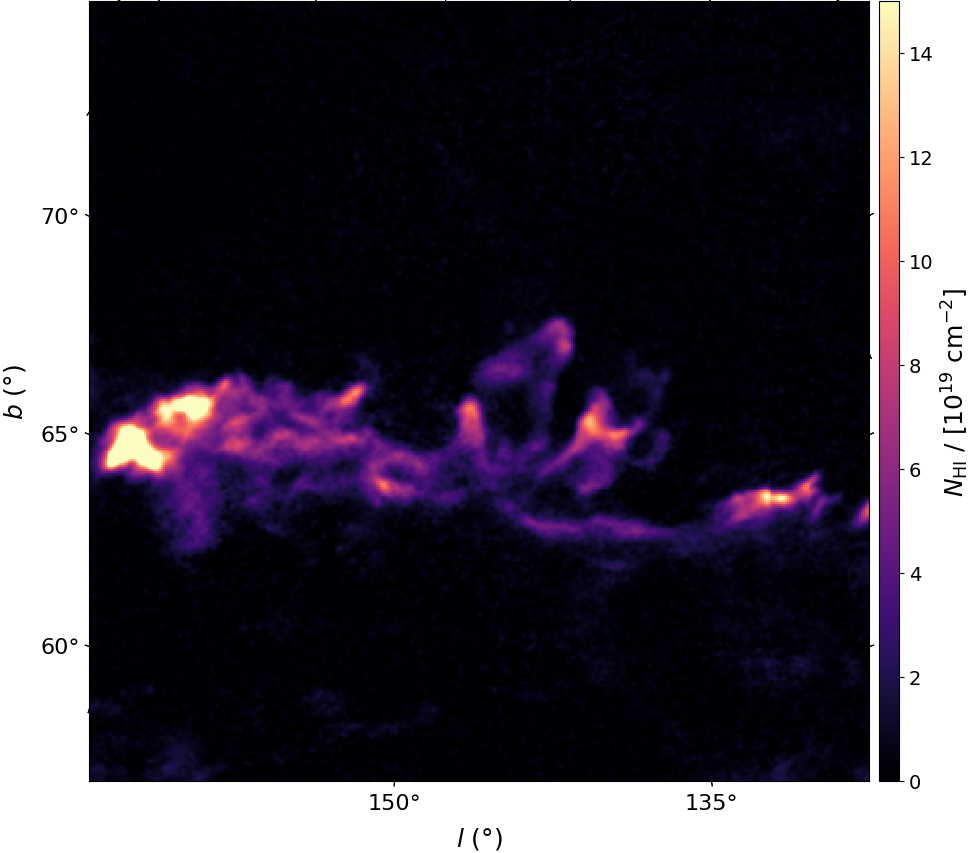}
  \caption{
  Zoomed images from Fig.~\ref{fig:NHI_EBHIS_nomask_csc_label} highlighting quasi-periodic scalloping and finger-like structures in \Nh\ toward complex C and its neighbors. The apparent orientation of the features depends on the projection in the different maps.
  Top left: Emission toward the eastern end of complex C.
  Top right: Main body of complex C with its pronounced upper edge containing \cib\ (right). The white box shows the coverage selected from the EN data.
  Bottom left: Complex A and its bridge to complex C in the image above.
  Bottom right: The western part of complex M, adjacent to upper east end of complex C in projection.
  }
  \label{fig:NHI_complex_C_zoom_edge_complex_A_M}
\end{figure*}

\section{Deviation velocity}
\label{app:deviation-velocity}

The deviation velocity for a given line of sight measures the difference between the observed $v_{\mathrm{LSR}}$ and the predicted velocities of \HI\ gas throughout a differentially-rotating model Galactic disk \citep{Wakker_1991a}. It can be used to identify gas in the halo that does not share the motion expected from the disk.

To evaluate the model, we used the rotation curve $v(R)$ at Galactocentric radius $R$ from the galpy ``MWPotential2014'' \citep{bovy_2015} in
\begin{equation}
\label{eq:wherer}
  v_{\mathrm{LSR,\,model}}=\left(\frac{R_{0}}{R} v(R)-v\left(R_{0}\right)\right) 
  \sin l \cos b \,.
\end{equation}
We adopted the simple model of the \HI\ disk from \citet{wakker_2004}, which has radius $R_{\rm{max}}$ = 26\,kpc and flared edges $\pm z_{\max }$ in the vertical direction given by 
\begin{equation}
\label{eq:zmodel}
    z_{\max} = z_{1} \,\, (R < R_{0}) \text { and }  z_{\max} = z_{1}+(z_2 - z_1) \times \frac{\left(R / R_{0}-1\right)^{2}}{4} \,\,  (R>R_{0})\, ,
\end{equation}
where $z_1$ and $z_2$ are 1 and 3 kpc, respectively.
At any distance $d$ along the line of sight,
\begin{equation}
\label{eq:whered}
   R = \sqrt{R_{0}^2 + (d \cos b)^2 -2 R_{0} (d \cos b) \cos l} \text { and } z = d \sin b\, .
\end{equation} 
For any direction $(l,b)$, $v_{\mathrm{LSR,\,model}}$ can be evaluated using Eqs.~\ref{eq:wherer} -- \ref{eq:whered} over the range of $d$ inside the model \HI\ disk ($R < R_{\rm{max}}$ and $|z| < z_{\max}$) to find the minimum and maximum, $v_{\rm{min,\, model}}$ and $v_{\rm{max,\, model}}$, respectively.  The deviation velocity is defined as
\begin{equation}
\label{eq:vdev}
    v_{\mathrm{DEV}} = v_{\mathrm{LSR}} -  v_{\rm{min,\, model}} \, (v_{\mathrm{LSR}} < 0) \text { and } v_{\mathrm{DEV}} = v_{\mathrm{LSR}} -  v_{\rm{max,\, model}} \, (v_{\mathrm{LSR}} > 0)\, .
\end{equation}

\section{Quasi-periodic scalloping and finger-like structures in complexes C, A, and M}
\label{app:scalloping-finger-like}

In support of the discussion in Sect.~\ref{subsec:edgeC}, Fig.~\ref{fig:NHI_complex_C_zoom_edge_complex_A_M} shows \Nh\ maps of the selected HVC emission from EBHIS for areas dominated by complex C and its neighbors, complexes A and M.  The relative positions of the panels on the sky can be judged by using the coordinates on the axes, reference to Fig.~\ref{fig:NHI_EBHIS_nomask_csc_label} (albeit in a different projection), and the caption.
These zoomed views highlight quasi-periodic scalloping and finger-like structures not only at obvious projected edges but also against the main body of gas, presumably because of projection of the complex boundary in 3D.

\bibliography{myref}{}
\bibliographystyle{aasjournal}

\end{document}